\documentclass[11pt,a4paper,fleqn,twoside]{report}
\usepackage[latin1]{inputenc} 
\usepackage[T1]{fontenc} 
\usepackage{amsmath}
\usepackage{amssymb}
\usepackage{latexsym}
\usepackage[dvips]{graphicx}
\usepackage[german,english]{babel}
\usepackage[clearempty]{titlesec}
\usepackage{hyperref}

\showhyphens{temperature}

\newcommand{\pd}{\partial}
\newcommand{\oh}{\frac{1}{2}}
\newcommand{\rar}{\rightarrow}
\newcommand{\munu}{{\mu\nu}}
\newcommand{\sss}[1]{\scriptscriptstyle{#1}}
\newcommand{\sst}{\scriptscriptstyle}
\newcommand{\hyphen}{\text{-}}
\newcommand{\M}{\mathcal{M}}
\newcommand{\x}{\ensuremath{\mathbf{x}}}
\newcommand{\ve}[2]{\left(\begin{array}{c}#1\\#2\end{array}\right)}
\newcommand{\vve}[3]{\left(\begin{array}{c}#1\\#2\\#3\end{array}\right)}
\newcommand{\cmb}{\mathrm{SU(2)}_{\mbox{\tiny CMB}}}
\newcommand{\order}{\ensuremath{\mathrm{O}}}
\renewcommand{\i}{\ensuremath{\mathrm{i}}}

\begin{document}
\rm 
\pagenumbering{roman}
\pagestyle{headings}

\begin{titlepage}
  \centering \renewcommand{\baselinestretch}{1.5} \sc
  \huge Fakult\"at f\"ur Physik \\
  \LARGE Universit\"at Karlsruhe (TH)\\
  \vfill \rm \large {\bf
    Diploma thesis \\
    in Physics \\
    submitted by \\
    Julian P.~Moosmann \\
    born in Bretten, Germany. \\
    2008 }
\end{titlepage}

\thispagestyle{empty} \cleardoublepage

\begin{titlepage}
  \centering \renewcommand{\baselinestretch}{1.5} \vspace*{3cm} \sc
  \huge
  Solitonic fermions\\
  in the confining phase\\
  of SU(2) Yang-Mills theory \vfill \rm \large {\bf
    This diploma thesis has been carried out by \\
    Julian P.~Moosmann \\
    at the \\
    Institut f\"ur Theoretische Physik \\
    under the supervision of \\
    Priv.-Doz.~Dr.~Ralf Hofmann}
\end{titlepage}

\thispagestyle{empty} 
\cleardoublepage

\setlength{\oddsidemargin}{0cm} 
\setlength{\evensidemargin}{0.85cm} 
\pagenumbering{roman}

\begin{abstract}
  
  \noindent We consider spatial coarse-graining in statistical
  ensembles of non-selfintersecting and one-fold selfintersecting
  center-vortex loops as they emerge in the confining phase of SU(2)
  Yang-Mills thermodynamics.  This coarse-graining is due to a noisy
  environment and described by a curve shrinking flow of center-vortex
  loops locally embedded in a two-dimensional flat plane.  The
  renormalization-group flow of an effective `action', which is
  defined in purely geometric terms, is driven by the curve shrinking
  evolution.


  In the case of non-selfintersecting center-vortex loops, we observe
  critical behavior of the effective `action' as soon as the
  center-vortex loops vanish from the spectrum of the confining phase
  due to curve shrinking.  This suggest the existence of an asymptotic
  mass gap.

  An entirely unexpected behavior in the ensemble of one-fold
  selfintersecting center-vortex loops is connected with the
  spontaneous emergence of order.  We speculate that the physics of
  planar, one-fold selfintersecting center-vortex loops to be relevant
  for two-dimensional systems exhibiting high-temperature
  superconductivity.

  \vspace{12mm} 
\selectlanguage{german}
  \begin{center} 
{\bf Zusammenfassung}
  \end{center}
  \vspace{1.3mm}

  \noindent Die Anregungen der konfinierten Phase in der
  thermodynamischen Behandlung der SU(2) Yang-Mills Theorie sind
  Zentrumsvortexschlaufen welche aufgrund der Wechselwirkung mit einer
  rauschenden Umgebung Schrumpfungsprozess unterliegen.  Wir
  betrachten statistische Ensemble von Zentrumsvortexschlaufen ohne
  und mit einfachen Schnittpunkt welche in einer flachen
  zweidimensionalen Ebene lokal eingebettet sind.  Der
  Schrumpfungsprozess von eingebetteten Zentrumsvortexschlaufen wird
  durch eine Diffusionsgleichung beschrieben.  Der
  Renormierungsgruppenfluss einer in rein geometrischen Gr\"o\ss en
  definierten effektiven \glqq Wirkung\grqq\ wird durch die Evolution
  schrumpfender Kurven bestimmt.

  Im Falle von Zentrumsvortexschlaufen ohne Schnittpunkt beobachten
  wir ein kritisches Verhalten der effektiven \glqq Wirkung\grqq\
  sowie die Vortexschlaufen aufgrund des Schrumpfungsprozesses aus dem
  Spektrum der konfinierten Phase verschwinden.  Dies legt die
  Existenz eines asymptotischen Massen-Gaps nahe.

  Ein vollkommen unerwartetes Verhalten im Ensemble von
  Zentrumsvortexschlaufen mit einfachem Schnittpunkt steht in engem
  Zusammenhang mit dem spontanen Auftreten von Ordnung.  Wir vermuten,
  dass die Physik ebener Zentrumsvortexschlaufen relevant ist f\"ur die
  Beschreibung zweidimensionaler Systeme, welche die Eigenschaft
  der Hochtemperatursupraleitung aufweisen.

\end{abstract}

\selectlanguage{english} 
\thispagestyle{plain} 
\cleardoublepage
\setcounter{tocdepth}{3} 
\tableofcontents
\thispagestyle{plain} 
\cleardoublepage

\chapter{Introduction}
\label{chap:Intro}
\pagenumbering{arabic}

The importance of Yang-Mills theories in mathematical and theoretical
physics is generally acknowledged.  Yang-Mills gauge theories are the
cornerstone of quantum field theories in the Standard Model of
Particle Physics: Besides gravity, all fundamental interactions are
incorporated as gauge symmetries in the Standard Model.  Although it
has been examined in the framework of perturbation theory due to the
enormous complexity implied in the full story of (especially
non-Abelian) gauge theories, the Standard Model has produced a lot of
striking results and predictions.  There are many examples, such as
the explanation of the anomalous magnetic moment of the electron, the
feature of asymptotic freedom of Quantum Chromodynamics
in the high energy limit, or the prediction of flavor-changing neutral
currents in electroweak processes \cite{NeutralCurrents}.  However,
there are still a number of unsolved mathematical problems and
unexplained experimental observations.  Among those are: The necessity
of an asymptotic mass gap and a rigorous proof of color confinement in
pure Yang-Mills theory \cite{JaffeWitten}.  In the Standard Model, the
assumption of a zero rest mass of the neutrino is refuted by the
observation of neutrino oscillations \cite{PDG} and the double $\beta$
decay \cite{KKG}.  These observations indicate a small, finite rest
mass that also cannot be excluded by recent experiments measuring the
spectrum of the single $\beta$ decay of tritium nuclei near the
endpoint \cite{Lobashev1999,Kraus2005}.  Furthermore, the Standard
Model does not provide for an explanation of Dark Matter and Dark
Energy that account for about 96\% of the energy density in the
present universe, and the predicted Higgs particle has evaded
experimental detection so far.  Moreover, the perturbation series of
four-dimensional quantum field theories is most likely an asymptotic
series; the fact that a perturbative calculation of the
thermodynamical pressure cannot be driven beyond order $g^5$ in the
coupling constant due to the weak screening of the magnetic sector
causing infrared instabilities \cite{Linde}, could be shown for
Quantum Chromodynamics at finite temperature.

Since perturbation theory is an expansion in powers of a necessarily
small coupling constant about a trivial a priori estimate for the
vacuum of the theory, it fails to describe strongly coupled physics as
well as the according nontrivial vacuum state.  This vacuum is
certainly composed of finite-action solitonic solutions of the
classical Yang-Mills action.  The so called instantons are
topologically nontrivial objects in pure Yang-Mills theory that
describe tunneling processes between topological distinct vacua,
e.g.~\cite{Ryder}.  Their weight possesses an essential zero at
vanishing coupling, and thus instanton contributions to the partition
function of the theory are completely ignored by perturbation theory.
Instantons at finite temperature are called calorons.

Therefore, we are advised to consider a nonperturbative approach to
gauge theories.  Such a treatment has already been proven successful
in terms of an effective theory for superconductivity
\cite{MullerBednorz1986}.  An analytical and nonperturbative approach
to SU(2) Yang-Mills thermodynamics was developed in
\cite{Hofmann2005}.  In this approach the basic idea is to subject the
highly complex dynamics of the topologically nontrivial field
configurations to a spatial coarse graining that leads to the
emergence of macroscopic scalar fields, and pure gauges.  Due to
nontrivial (thermal) ground states, the fundamental gauge symmetry is
broken successively as temperature decreases.  As a consequence,
Yang-Mills thermodynamics occurs in three phases: the deconfining, the
preconfining and the confining phase.  The latter, in which we are
primarily interested in this thesis, exhibits three unexpected
results.  These are the exact vanishing of the energy density and the
pressure of the ground state at zero temperature, the Hagedorn
character of the preconfining-confining phase transition and the
spin-1/2 nature of the massless and massive excitations in the
confining phase.

\begin{figure}
  \centering
  \includegraphics[width=0.9\textwidth]{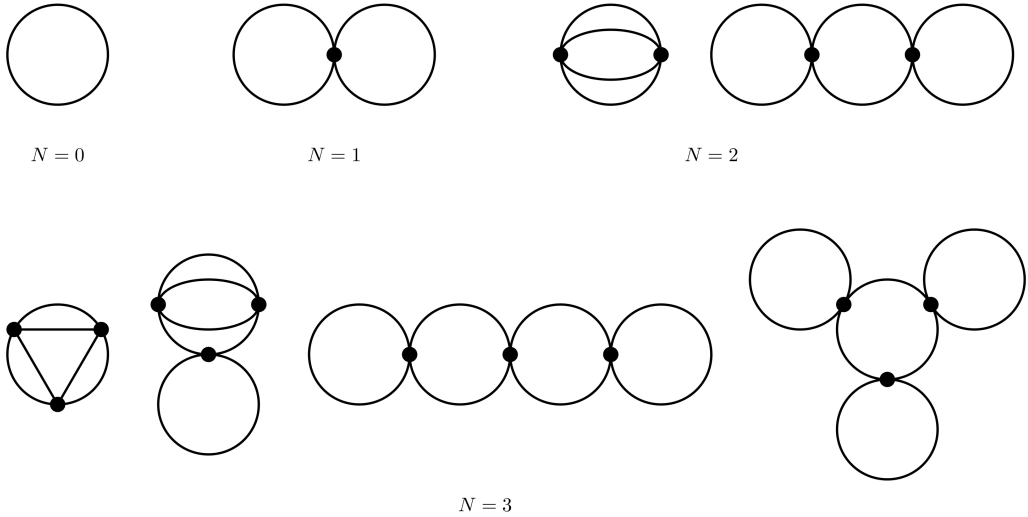}
  \caption{The topologies of solitonic excitations with up to $N=3$
    selfintersections for an SU(2) Yang-Mills theory in the confining
    phase.  A magnetic monopole of charge $+1$ or $-1$ is located at
    each point where center-flux lines intersect.  Solitons with $N=0$
    are unstable in presence of a noisy environment, whereas solitons
    with $N=1$ are always stable.  Excitations with $N>1$ are unstable
    if subjected to mixing with theories possessing propagating gauge
    fields.}
  \protect{\label{fig:Excitations}}
\end{figure}

The ground state of the deconfining phase is composed of interacting
calorons and anticalorons and exhibits negative pressure.  The
propagating excitations within that phase are two massive gauge modes
- due to the dynamically broken SU(2) - and one massless gauge
mode.  
As temperature decreases, the likeliness for calorons and anticalorons
to dissociate into (BPS saturated) magnetic monopoles and
antimonopoles increases strongly in the vicinity of a critical
temperature.  The ground state of the preconfining (or magnetic) phase
starts to form by the pairwise condensation of monopoles and
antimonopoles.  Excitations in that phase are propagating dual gauge
modes of mass $m_{\sst D}$ (dynamically broken U(1)$_D$).  Unstable
defects of the magnetic ground state are closed magnetic flux lines of
finite core size $d$ that collapse as soon as they are created.  This
is because, as long as $d>0$, the pressure inside the vortex loop is
more negative than outside, thus leading to the contraction of the
vortex loop.  The magnetic phase exhibits negative pressure.  At the
Hagedorn transition towards the confining (also called center) phase,
a complete decoupling of the gauge fields takes place.  To put it more
precisely, by the decay of the magnetic ground state into
selfintersecting and non-selfintersecting center-vortex loops the mass
of the dual gauge field diverges and the core size of center-vortex
loops vanishes, see also \cite{NO1973,tHooft1978}.  As a result of $d
\rightarrow 0$, the negative pressure $P$ is confined to the vanishing
vortex core.  This implies that center-vortex loops become stable
particle-like excitations with $P = 0$.  These solitons are classified
according to their center charge and the number of selfintersections
$N$, see Fig.~\ref{fig:Excitations}.  The mass of an $N$-fold
selfintersecting soliton is $N\Lambda_C$, where $\Lambda_C$ is the
Yang-Mills scale.  Topologically, solitons with non-vanishing $N$ are
stable in the absence of external gauge modes coupling to the charges
at the intersection points.  On the other hand, for $N=0$, there is no
topological reason for stability.

Now consider a situation where a planar center-vortex loop, which is a
(local) embedding of a center-vortex loop into a two-dimensional flat
and spatial plane, still has non-vanishing core size $d > 0$ and the
mass of the dual gauge field $m_{\sst D}$ is still finite due to a
noisy environment which locally resolves the otherwise infinitely thin
vortex.  In this case, the pressure $P$ is locally nonzero and the
soliton starts shrinking.  Such a situation is described by a curve
shortening flow in the (dimensionless) parameter $\tau$.  Here, $\tau$
is a variable measuring the decrease of externally provided resolving
power applied to the system.  There is a functional dependence of
$\tau$ on the corresponding resolution $Q$ (momentum transfer).  For
an isolated SU(2) theory the role of the environment is played by the
sectors with $N>0$.  If the confining SU(2) is part of a world with
additional gauge symmetries, then a portion of such an environment
arises from the mixing between the corresponding gauge groups.  Either
way, a center-vortex loop acquires a finite core size and as a
consequence, a finite mass for the $N=0$ soliton by frequent
interaction with the environment after it was generated by a process
that was subject to an inherent, finite resolution $Q_0$.

Knot-like structures are relevant in a number of chemical, biological
and physical systems \cite{FaddeevNiemiKnots}, e.g. in polymer
physics, particularly in molecular biology, in type-II superconductor,
where string-like vortices confine magnetic fields to the cores of the
vortex-like structures, in superfluid helium ($^4\mathrm{He}$), as
well as in liquid crystals.  As early as 1897 Lord Kelvin proposed
that elementary particles - at that time atoms were considered to be
elementary by Kelvin and others - should be described as knotted lines
of vortex tubes in a medium (the aether) \cite{Kelvin}.  As we know
now, the point particle interpretation of Quantum Mechanics appears to
be a much more elegant and efficient framework to describe the physics
of atoms and molecules.  But at the same time, the notion of an
electron as a spinning point particle, albeit an excellent description
in a bulk of physical situations in atoms, colliders and condensed
matter systems, causes theoretical and experimental inconsistencies.
On the one hand, there is the problem of diverging classical
self-energy of the electron.  On the other hand, the unexpected
explosive behavior in recent high-temperature plasma experiments
\cite{Sandia,ZPinch} and the strong correlations of electrons in
two-dimensional planar systems \cite{MullerBednorz1986} are
indications of non-local effects possibly related to the extended
spatial structure of the electron.  Also, recent theoretical
developments revive Kelvin's description of elementary particles as
non-local knot-like entities.  In
\cite{FaddeevNiemiKnots,FaddeevNiemi}, the argument is that confining
strings, tied into stable knotted solitons, exist when decomposing the
gauge field in the low-energy domain of four-dimensional SU(2)
Yang-Mills theory.

\begin{figure}
  \centering
  \includegraphics[width=0.80\textwidth]{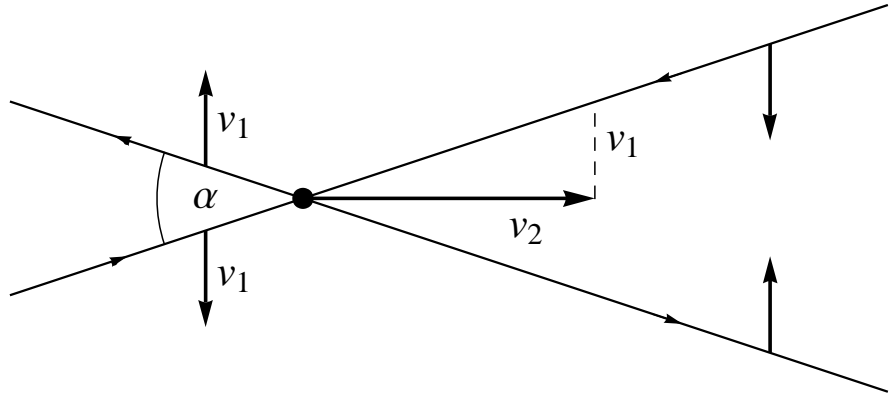}
  \caption{\protect{\label{fig:SuperluminalMotion}}Points on the center
    flux lines moving oppositely on a line perpendicular to the
    bisecting line of the angle $\alpha$ with velocity modulus $v_1$.
    For sufficiently small $\alpha$ the velocity modulus $v_2$ of the
    intersection point is superluminal:
    $v_2=v_1\cot{\frac{\alpha}{2}}$.}
\end{figure}

According to the approach in \cite{Hofmann2005}, we tend to interpret
one-fold selfintersecting center-vortex loops as electrons and
accordingly non-selfintersecting center-vortex loops as neutrinos.
This implies that the Yang-Mills scale $\Lambda_C$ must be set equal
to the electron mass $m_e = 511$~keV.  The spin-1/2 nature of a
center-vortex loop is a consequence of its two-fold degeneracy with
respect to the direction of flux which is lifted in the presence of an
electric or magnetic background field.  It should be noticed for a
selfintersecting center-vortex loop that, as long as both wings of
center flux are of finite size, a spatial shift of the intersection
point requires a negligible amount of energy only.  In particular, if
the inner angle $\alpha$ between in- and out-going center-flux at the
intersection is sufficiently small, then a motion of points on the
vortex line that is directed perpendicularly to the bisecting line of
the angle $\alpha$ easily generates a velocity of the intersection
point which exceeds the speed of light,
e.g.~Fig.~\ref{fig:SuperluminalMotion}.  Here, it should be considered
that the path-integral formulation of Quantum Mechanics admits such
superluminal motion in the sense that the according trajectories
contribute to transition amplitudes \cite{Feynman}.

The purpose of this thesis is to treat the behavior of $N=0$ and $N=1$
center-vortex loops under curve shrinking as a Wilsonian
renormalization-group flow governed by an effective `action'. The term
`action' is slightly misleading since we do not aim at a time evolution
of the system by demanding stationarity of the `action' under curve
variation.  We consider resolution dependent statistical ensembles in
the presence of an environment represented by a parameter $\sigma$.  The
corresponding weight-functional for the members of the ensemble, written
as the exponential of an `action', is defined in purely geometric terms.
In turn the resolution dependence of the `action' is determined by the
curve shrinking flow.  The `action' possesses a natural decomposition into
a conformal and a non-conformal factor.  We consider the partition
function of a given ensemble of planar curves to be invariant under the
condition of changing the resolution.  Once the evolution of the
weight-functional is determined, we are able to compute the resolution
dependence of `observables' as ensemble averages of (local or non-local)
operators.

What we observe is that the $N=0$ sector becomes unresolvable from a
finite resolution $Q_*$ downward.  That is, as a consequence of a noisy
environment planar $N=0$ center-vortex loops shrink to points with
circular limiting shape within a finite decrease of resolving power
and thus disappear from the spectrum of the confining phase of SU(2)
Yang-Mills theory for a resolution smaller than the critical $Q_*$.
Since center-vortex loops with $N>0$ have finite mass this generates an
asymptotic mass gap.  We show that the observed transition to the
conformal limit of vanishing curve length is a critical phenomenon with
a mean-field exponent of the coefficient associated with the
non-conformal factor.
For the $N=1$ sector we observe the unexpected behavior that, starting
from a finite value, the entropy of the system decreases to an almost
zero value as the resolving power is lowered: the ensemble evolves
into a highly ordered state in a sense that only a single curve
survives the process of coarse-graining.

The thesis is organized as follows: Chapter~\ref{chap:YMT} gives a
brief outline of the effective theory of thermalized SU(2) Yang-Mills
dynamics in all of its phases as it is developed in
\cite{Hofmann2005}.  Chapter~\ref{chap:MP} provides prerequisites for
the mathematics of curve shrinking flows in two and three space
dimensions.  Chapters \ref{chap:0CVL} and \ref{chap:1CVL} investigate
the $N=0$ sector and respectively, the $N=1$ sector.  In Sections
\ref{sec:0WRGF} and \ref{sec:1WRGF}, we explain our philosophy
underlying the statistics of geometric fluctuations and how the
renormalization-group flow of the effective `action' is driven by the
curve-shrinking evolution of the members of a given ensemble of $N=0$
and respectively, $N=1$ center-vortex loops.  In
Sec.~\ref{sec:0Simulation}, we explain our numerical analysis
concerning the computation of the effective action and the variance of
the `center of mass' which is compared to Heisenberg's uncertainty
relation.  In Sec.~\ref{sec:1Simulation}, we elucidate our numerical
analysis concerning the computation of the effective `action', the
variance of the location of the selfintersection, and the
resolution-dependent entropy associated with a given ensemble.
Chapter~\ref{chap:Applications} deals with electrons which are
interpreted as center-vortex loops with one selfintersection.  In
Sec.~\ref{sec:Solitons}, we give reasons for this interpretation, and
in Sec.~\ref{sec:HTS}, we consider strongly correlated systems of
electrons in cuprates exhibiting high-temperature superconductivity
and the new class of recently discovered iron-based high-temperature
superconductors.  Chapter~\ref{chap:Summary} gives a short summary of
our findings.

\cleardoublepage


\chapter{Brief review of SU(2) Yang-Mills thermodynamics}
\label{chap:YMT}

In this section, we give a short outline of the analytical and
nonperturbative approach to SU(2) Yang-Mills thermodynamics as it is
developed in \cite{Hofmann2005}.  The basic idea is to subject the
highly complex dynamics of topologically nontrivial field
configurations to a spatial coarse-graining that is described by
emergent macroscopic scalar fields, one for each phase.  Conceptually,
this approach is similar to the macroscopic Landau-Ginzburg theory of
superconductivity.  Although we are only concerned with the confining
phase in this work, we start our outline of \cite{Hofmann2005} in the
deconfining phase at high temperatures which leads us by consecutive
phase transitions to the confining phase.

\section{Basics of thermal Yang-Mills theory}
\label{sec:ThermalYM}

Yang-Mills theories are non-Abelian gauge theories whose Lagrangian is
demanded to be invariant under local gauge transformations.  In this
thesis, we restrict ourselves to the case of SU(2) gauge
transformations.  In pure Yang-Mills theory, only gauge field terms
appear in the fundamental Lagrangian while matter fields are absent.
Wick-rotating to Euclidean signature by $t \rightarrow - \i \tau$ and
moving to finite temperature $T$, which corresponds to imaginary time
compactified on a circle with circumstance $\beta = \frac{1}{T}$, the
gauge-invariant action is given by
\begin{equation}
  \label{eq:pureYMaction}
  S\equiv \frac{1}{2g^2} \mathrm{tr}
  \int_0^\beta\mathrm{d}\tau\int\mathrm{d}^3 x\,F_{\mu\nu}F_{\mu\nu},
\end{equation}
where $g$ denotes the dimensionless coupling constant and
$\mathrm{tr}$ the trace operation.  It holds that
$(x_1,x_2,x_3,x_4)\in\mathbb{R}^4$.  The Yang-Mills field strength
tensor is defined as\footnote{The gauge coupling $g$ is absorbed in
  the definition of the gauge fields.}
\begin{equation}
  \label{eq:YMfieldstrength}
  F_{\munu} = \pd_\mu A_\nu - \pd_\nu A_\mu - \i [A_\mu,A_\nu],
\end{equation}
with the Lie-algebra valued gauge fields in the adjoint representation
\begin{equation}
  A_{\mu} \equiv A_{\mu}^a \frac{\sigma^a}{2},
\end{equation}
where the generators $\sigma^a$ are given by the Pauli matrices.  The
action density $\frac{1}{2g^2}\mathrm{tr}\,F_{\munu}F_{\munu}$ is
invariant under local SU(2) gauge transformations
\begin{equation}
  \label{eq:gaugetrafo}
  A_{\mu}(x)\overset{\Omega}{\rightarrow}
  \Omega(x) A_{\mu}(x) \Omega^{\dagger}(x) +i\Omega(x)\pd_{\mu}\Omega(x),
\end{equation}
where $\Omega$ is an element of $SU(2)$.

Instantons are localized finite-action classical solutions in
Euclidean field theory.  The BPST (Belavin-Polyakov-Schwartz-Tyupkin)
instanton is an (anti)selfdual, that is BPS
(Bogomol'nyi-Prasad-Sommerfield) saturated, configuration solving the
Euler-Lagrange equations $D_{\mu}F_{\mu\nu} = 0$ subject to the action
(\ref{eq:pureYMaction}) \cite{BPST1975}.  For the
covariant derivative $D_{\mu}$ of the field $\phi$ in the adjoint
representation we have
\begin{equation}
  \label{eq:YMcovderivativ}
  D_{\mu}\phi = \pd_{\mu}\phi - \i [A_{\mu},\phi].
\end{equation}
The (anti)selfduality condition reads
\begin{equation}
  \label{eq:selfduality}
  F_{\mu\nu} = \pm \tilde F_{\mu\nu},
\end{equation}
where the dual field strength is defined as $\tilde F_{\mu\nu} \equiv
\oh \epsilon_{\mu\nu\kappa\lambda} F_{\kappa\lambda}$,
$\epsilon_{\mu\nu\kappa\lambda}$ being the total antisymmetric tensor
with $\epsilon_{1234}=1$.  (Anti)selfdual configurations saturate the
BPS bound on the action which therefore is minimal (in a given
topological sector) and of value
\begin{equation}
  \label{eq:BPSaction}
  S=\frac{8\pi^2}{g^2}|Q|,  
\end{equation}
where the Pontryagin index $Q$ is a topological invariant (charge) and
defined as
\begin{equation}
  \label{eq:topcharge}
  Q \equiv
  \frac{1}{32\pi^2} \int_0^{\beta}\mathrm{d}\tau\int\mathrm{d}^3x\,
  F_{\munu}^a\tilde F_{\munu}^a.
\end{equation}
BPS saturated field configurations $A_{\mu}$ have \emph{vanishing}
energy-momentum tensor.

(Anti)calorons are BPS saturated, periodic-in-$\tau$ configurations at
finite temperature with finite action and topological charge $Q=\pm
1$.  They are classified according to the eigenvalues of their
Polyakov loop (time-like Wilson loop evaluated in periodic gauge) at
spatial infinity.
An (anti)caloron is said to be of trivial holonomy, if its Polyakov
loop, evaluated at spatial infinity, is an element of the center of
the gauge group.  Otherwise it is said to have nontrivial holonomy.
The Harrington-Shepard (HS) (anti)caloron is a periodic-in-$\tau$
instanton in singular gauge with topological charge $Q = \pm 1$ and
trivial holonomy, whereas the Lee-Lu-Kraan-van Baal (LLKvB)
(anti)caloron is of nontrivial holonomy. Descriptively, trivial
holonomy means that the caloron has no substructure.  The LLKvB
(anti)caloron contains BPS magnetic monopoles constituents which, by
virtue of quantum corrections \cite{Diakonov2004}, are subject to an
attractive interaction in the case of small holonomy and to a
repulsive interaction for large holonomy.  In the case of large
holonomy, the repulsion leads to a dissociation of the caloron into a
pair of a screened magnetic monopole and antimonopole. On the other
hand, for small holonomy, the (anti)caloron collapses back to the
stable configuration of a HS (anti)caloron by annihilation of their
BPS monopole constituents.  Thus, single LLKvB calorons are unstable
under quantum deformations.

\section{The deconfining phase}
\label{sec:DePhase}

The complex microscopic dynamics in Yang-Mills theory does not seem to
allow for a direct analytic calculation of macroscopic quantities in
terms of the fundamental gauge fields.  A spatial coarse-graining, that
is the computation of a spatial average over the sector of topologically
nontrivial, BPS saturated 
field configurations of trivial holonomy turns out to be a feasible and
thermodynamically exhaustive approach.  The coarse-graining procedure is
described in terms of a macroscopic adjoint field $\phi$.  In order to
characterize the macroscopic ground state, $\phi$ has to satisfy for
following conditions:

\noindent (i) due to spatial isotropy and homogeneity $\phi$ must be a
Lorentz scalar;

\noindent (ii) homogeneity of the ground state implies that the modulus
of $\phi$ is independent of space and time.  A dynamically generated
Yang-Mills scale $\Lambda$ enters this modulus as a parameter;

\noindent (iii) $\phi$ is a composite of local fields and therefore
has to transform under the adjoint representation, because in pure
Yang-Mills theory all local fields transform under the adjoint
representation of the gauge group;

\noindent (iv) only the color orientation of $\phi$ in a given gauge,
also referred to $\phi$'s phase, depends on $\tau$.  Since $\phi$ is
constructed from (anti)calorons, which are periodic in Euclidean time,
$\phi$'s phase is also periodic in $\tau$, and since the classical
caloron action $S=\frac{8\pi}{g^2}$ is independent of temperature
$\phi$'s phase is not explicitly time dependent.  The computation of the
phase of $\phi$ does not require any information about the Yang-Mills
scale.

\noindent Consequently, the field can be written as
\begin{equation}
  \label{eq:phi}
  \phi^a = |\phi|(\Lambda_E,\beta) 
  \frac{\phi^a}{|\phi|}\left(\frac{\tau}{\beta}\right).
\end{equation}
%
In \cite{Hofmann2005,GH2006,HerbstHofmann2004} equations of motion for
the phase and modulus of the spatially homogeneous, composite, emergent
adjoint scalar field $\phi$ obeying the above conditions are derived.
The (non-perturbatively) temperature dependent modulus is given by
\begin{equation}
  \label{eq:modulusphi}
  |\phi|(\Lambda_E,\beta) =\sqrt{\frac{\Lambda_E^3\beta}{2\pi}} 
  =  \sqrt{\frac{\Lambda_E^3}{2\pi T}}.
\end{equation}
The corresponding action is found to be
\begin{equation}
  \label{eq:ElectricAction}
  S_{\phi} = \mathrm{tr}\int_0^{\beta}\mathrm{d}\tau\int\mathrm{d}^3x\; 
  ((\pd_{\tau}\phi)^2 +\Lambda_E^6\phi^{-2}),
\end{equation}
where $\phi^{-1} \equiv \frac{\phi}{|\phi|^2}$.  The field $\phi$ turns
out to be quantum mechanically and statistically inert\footnote{This can
  be checked by direct computation but also is implied by the fact that
  a spatial average over non-propagating gauge fields must generate a
  composite that itself is not propagating.}.  It serves as a spatially
homogeneous background for the topologically trivial ($Q=0$) sector of the
coarse-grained, propagating gauge fields $a_{\mu}$.  In
Eq.~(\ref{eq:ElectricAction}) interactions between calorons are not yet
included.  This is done via minimal coupling by substituting
\begin{equation}
  \pd_{\mu}\phi \rightarrow D_{\mu}\phi =\pd_{\mu}\phi +\i e[\phi,a_{\mu}].
\end{equation}
The interactions are mediated by the topologically trivial fields that
change the holonomy of the (anti)calorons and subsequently induce
interactions between the magnetic monopole constituents of nontrivial
holonomy (anti)calorons.  The action for the minimally coupled fields is
given by
\begin{equation}
  \label{eq:IntElectricAction}
  S = \mathrm{tr} \int_0^{\beta}\mathrm{d}\tau\int\mathrm{d}^3x \; (\oh
  G_{\mu\nu}\,G_{\mu\nu} +(D_{\mu}\phi)^2 +\Lambda^6\phi^{-2}),
\end{equation}
where the field strength is $G_{\mu\nu} \equiv
\frac{\sigma^a}{2}(\pd_{\mu}a_{\nu}^a -\pd_{\nu}a_{\mu}^a -e f^{a b
  c}a_{\mu}^b a_{\nu}^c)$ and $e$ denotes the effective gauge coupling
which determines the strength of interaction between topologically
trivial gauge field fluctuations and the macroscopic field $\phi$.
Due to Lorentz invariance, gauge invariance, perturbative
renormalization, and the inertness of $\phi$ the action
(\ref{eq:IntElectricAction}) is \emph{unique}.  The topologically
trivial sector is written as a decomposition
\begin{equation}
  \label{eq:adecom}
  a_{\mu} = a_{\mu}^{gs} + \delta a_{\mu},  
\end{equation}
where $a_{\mu}^{gs}$ is a pure-gauge solution of the equations of motion
for $a_{\mu}$ following from action (\ref{eq:IntElectricAction}), and
$\delta a_{\mu}$ is a (periodic) finite-curvature propagating
fluctuation.  The pressure $P^{gs}_E$ and energy density $\rho^{gs}_E$
of the ground state, following from the energy-momentum tensor, read
\begin{equation}
  \label{eq:decgsequ}
  P^{gs}_E =-\rho^{gs}_E =-4\pi \Lambda_E^3 T.
\end{equation}
Microscopically, the negative ground state pressure arises from the
creation and annihilation of BPS monopoles and antimonopoles within
small-holonomy (anti)calorons.  The emergence of the macroscopic adjoint
scalar field $\phi$ breaks the fundamental gauge group SU(2) down
dynamically to the subgroup U(1).  Due to the adjoint Higgs mechanism,
two out of three gauge modes $\delta a_{\mu}^{(1,2)}$ acquire a
temperature dependent mass, while the third remains massless,
\begin{equation}
  \label{eq:gaugefielmass}
  m_1 = m_2 = 2 e(T) |\phi| = 2e(T)\sqrt{\frac{\Lambda_E^3}{2\pi T}}
  \qquad \textnormal{and} \qquad  m_3 = 0.
\end{equation}
Evaluating the Polyakov loop in a different (unitary) gauge gives rise
to the conclusion that the ground state is two-fold degenerated with
respect to the (broken) global electric $Z_2$ symmetry.  Thus, the
electric phase is deconfining.
The temperature evolution of the effective gauge coupling $e$ is derived
from the demand for thermodynamical self-consistency, it reaches a
plateau value rapidly as temperature increases ($T\gg T_{c,E}$) and
diverges logarithmically for $T\searrow T_{c,E}$,
\begin{equation}
  \label{eq:eevolution}
  e(T) \propto -\log(T-T_{c,E}).
\end{equation}
\begin{figure}
  \centering
  \includegraphics[width=0.80\textwidth]{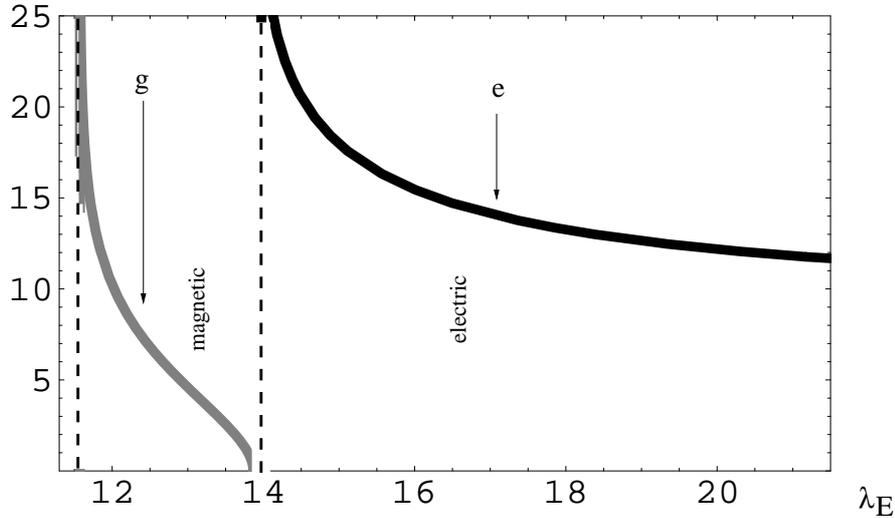}
  \caption{Temperature evolution of the effective gauge couplings $e$
    and $g$ as a function of the dimensionless temperature
    $\lambda_E=\frac{2\pi T}{\Lambda_E}$.  The figure is taken from
    \cite{Axion}.}
  \protect{\label{fig:Couplings}}
\end{figure}
Therefore, the massive gauge $\delta a_{\mu}^{(1,2)}$ modes become
infinitely heavy and decouple at $T_{c,E}$.
%
%

The ground state in the deconfining phase is composed of interacting
calorons and anticalorons of topological charge-modulus one and trivial
holonomy.  Screened magnetic BPS saturated monopoles are spatially
isolated defects in the electric phase.  Screening occurs due to
short-lived magnetic dipoles provided by intermediary small-holonomy
LLKvB (anti)calorons and due to all other stable and screened
(anti)monopoles.


\section{The preconfining phase}
\label{sec:PrePhase}

At $T_{c,E}=13.87\frac{\Lambda_E}{2\pi}$, the effective gauge coupling
$e$ diverges.  Thereby, magnetic monopoles and antimonopoles, which
are generated by the dissociation of large-holonomy calorons, become
massless and condense pairwise, thus terminating the deconfining
phase.  Note that for $T<T_{c,E}$, the average caloron-anticaloron
holonomy gradually increases with decreasing temperature.  After a
spatial coarse-graining, the thermal ground state of the Bose
condensate of interacting monopoles and antimonopoles is entirely
described by a macroscopic complex scalar field $\varphi$ and a pure
gauge $a_{\mu}^{\sss{D},gs}$; only gauge fields transforming under
U(1) survive the electric-magnetic phase transition.  The macroscopic
complex scalar field $\varphi$ turns out to be quantum mechanically
and statistically inert.  Interactions between monopoles mediated by
pure gauges generate isolated but unstable defects.  These defects are
closed magnetic flux lines that are composed of magnetic monopoles
moving oppositely directed to each other in the vortex core along the
flux lines.  The closed flux lines collapse as soon as they are
created, thereby inducing a negative pressure.  It should be noticed
that the magnetic flux lines need to be closed due to the absence of
isolated magnetic charges in the monopole condensate.  The spatially
homogeneous and BPS saturated complex scalar field $\varphi$ breaks
the dual gauge symmetry $\mathrm{U(1)}_{D}$ dynamically: the stable
and propagating excitations in the magnetic phase are massive dual
gauge modes.

In \cite{Hofmann2005} equations for phase and modulus of the
macroscopic complex scalar field are derived.
The modulus of the field $\varphi$ is found to be
\begin{equation}
  \label{eq:modulusvarphi}
  | \varphi|(\Lambda_M,\beta) =\sqrt{ \frac{\Lambda_M^3\beta}{2\pi}}
  =\sqrt{\frac{\Lambda_M^3}{2\pi T}}.
\end{equation}
The effective action for $\varphi$ reads
\begin{equation}
  \label{eq:MagneticAction}
  S_{\varphi} =\int_0^{\beta}\mathrm{d}\tau\int\mathrm{d}^3x\; 
  \left(\oh\,\overline{\pd_{\tau}\varphi}\,\pd_{\tau}\varphi 
    +\oh\frac{\Lambda_M^6}{\bar\varphi\varphi}\right),
\end{equation}
where interactions between (screened) monopoles are absent and
$\Lambda_M$ is an externally provided Yang-Mills scale.  Interactions
are accounted for in analogy to Sec.~\ref{sec:DePhase}: the
topologically trivial sector $a_{\mu}^{\sst D}$ is decomposed into
\begin{equation}
  \label{eq:dualgaugedecomposition}
  a_{\mu}^{\sst D} =a_{\mu}^{\sss{D},gs} + \delta a_{\mu}^{\sst D}
\end{equation}
and is minimally coupled to $\varphi$.  The unique effective action
including interaction reads
\begin{equation}
  \label{eq:IntMagneticAction}
  S=\int_0^{\beta}\mathrm{d}\tau\int\mathrm{d}^3x \;
  \left(\frac{1}{4} F^{\sst D}_{\mu\nu}\,F^{\sst D}_{\mu\nu} 
    +\oh\overline{\mathcal{D}_{\mu}\varphi}\,\mathcal{D}_{\mu}\varphi 
    +\oh\frac{\Lambda_M^6}{\bar\varphi\varphi}\right),
\end{equation}
where the Abelian field strength of the dual gauge field is given by
\begin{equation}
  \label{eq:dualfieldstrength}
  F_{\mu\nu}^{\sst D} = (\pd_{\mu}a^{\sst D}_{\nu} -\pd_{\nu}a^{\sst D}_{\mu}),
\end{equation}
and the covariant derivative involving the effective magnetic coupling
$g$ by
\begin{equation}
  \label{eq:dualcovderivative}
  \mathcal{D}_{\mu} =\pd_{\mu} +\i g a_{\mu}^{\sst D}.  
\end{equation}
A pure-gauge solution $a_{\mu}^{\sss{D},gs}$ to the equations of
motions for the dual gauge-field in the background of $\varphi$ is
found.

The evaluation of the Polyakov loop suggests that the electric $Z_2$
degeneracy, as occurred in the electric phase, no longer exists in the
magnetic phase: the ground state of the magnetic phase is unique and
confines fundamental, heavy and fermionic test charges.  Nevertheless,
massive gauge modes still propagate because the Polyakov loop does not
vanish entirely.  
Therefore, the magnetic phase is called preconfining.  The dual gauge
group U(1)$_D$ is dynamically broken due to the emergence of the
macroscopic scalar field $\varphi$.  As a consequence, the dual gauge
excitation $\delta a_{\mu}^{\sst D}$ becomes Mei\ss{}ner massive via
the dual Abelian Higgs mechanism,
\begin{equation}
  \label{eq:mdual}
  m_{\sst D} = g(T) \varphi.
\end{equation}
The evolution of the temperature dependent effective gauge coupling
$g$ is predicted by thermodynamical self-consistency.  The coupling
vanishes for $T\nearrow T_{c,E}$ and diverges logarithmically for
$T\searrow T_{c,M}$:
\begin{equation}
  \label{eq:gdivergence}
  g \propto - \log(T - T_{c,M}),
\end{equation}
where $T_{c,M}$ denotes the temperature where the transition to the
center phase takes place.  The typical energy of a
non-selfintersecting center-vortex loop (CVL) is $\propto\;g^{-1}$.

\begin{figure}
  \centering
  \includegraphics[width=0.80\textwidth]{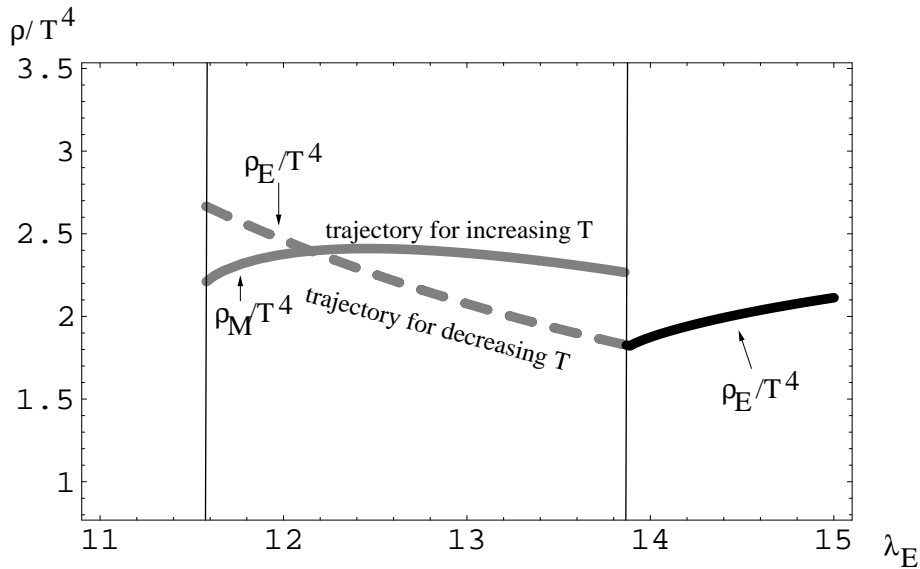}
  \caption{Ratio of the energy density $\rho$ and temperature $T^4$
    across the electric-magnetic phase transition as a function of the
    dimensionless temperature $\lambda_E=\frac{2\pi T}{\Lambda_E}$.
    The dashed line represents the continuation of the energy $\rho_E$
    of the electric phase (solid black line) for decreasing
    temperature $T<T_{c,E}$ (supercooled state, $m_{\sst D}=0$).  The
    solid grey line represents the energy density $\rho_M$ in the
    magnetic phase for increasing temperature ($m_{\sst D}>0$).  As
    long as no additional energy is available, the system remains in a
    supercooled state until a temperature $\lambda_E=12.15$ is
    reached.  The figure is taken from \cite{Axion}.}
  \protect{\label{fig:EnergyDensity}}
\end{figure}

During the electric-magnetic phase transition, the number of
polarizations of the `photon' jumps from two to three, thereby
inducing a discontinuity in the energy density.  The negative pressure
of the ground state arises due to an equilibrium between vortex-loop
creation by dissociation of large-holonomy calorons and the
annihilation of vortex loops by contraction.  The non-vanishing
pressure $P^{gs}_M$ and energy density $\rho^{gs}_M$ of the ground
state evaluate as
\begin{equation}
  \label{eq:precongsequ}
  P^{gs}_M = - \rho^{gs}_M = -\pi\Lambda_{M}^3T.
\end{equation}
Across the electric-magnetic phase transition at $T_{c,E}$, where
$e=\infty$ and $g=0$, the pressure is continuous (see
Fig.~\ref{fig:EnergyDensity}) relating the scales $\Lambda_E$ and
$\Lambda_M$:
\begin{equation}
  \label{eq:EMScaleRelation}
  \Lambda_E = \left(\frac{1}{4}\right)^{1/3}\Lambda_M.
\end{equation}

The magnetic phase is not detected by finite-size lattice simulations,
since the monopoles condensate posses infinite correlation length
($\propto(M_{m+a})^{-1}$), where $M_{m+a}$ is the sum of the monopole
and antimonopole mass after screening:
\begin{equation}
  \label{eq:ScreenedMonopoleMass}
  M_{m+a} = \frac{8\pi}{e\beta}.
\end{equation}

%

\section{The confining phase}
\label{sec:ConPhase}

First we provide some facts on the Abrikosov-Nielsen-Olesen (ANO)
vortex.  When embedded in three space dimensions, a point-like
two-dimensional ANO vortex becomes a  vortex line.  A mesoscopic
description of a static ANO vortex is given by the action of
Eq.~(\ref{eq:IntMagneticAction}) where the potential is absent.  A BPS
saturated solution to the equation of motions, following from action
(\ref{eq:IntMagneticAction}), can be found that carries one unit of
magnetic flux ($\frac{2\pi}{g}$) and has vanishing core size.  Outside
the vortex core the pressure $P_v(r)$, which is isotropic in the $x_1
\hyphen x_2$ plane, reads
\begin{equation}
  \label{eq:ANOpressure}
  P_v(r) =-\oh\frac{\Lambda_M^3\beta}{2\pi}\frac{1}{g^2}{r^2},\qquad(r>0),
\end{equation}
where $r$ is the radial vector in the $x_1 \hyphen x_2$ plane.  Notice
the minus sign on the right hand side of Eq.~(\ref{eq:ANOpressure}).
For a finite energy, the length of the ANO vortex line must be finite.
The configuration is static as long as it possesses cylindric
symmetry, but as soon as the vortex is bend the configuration becomes
unstable: the pressure inside the vortex loop is more negative than
outside.  Thus, the vortex collapses as soon as it is created at
finite coupling $g$.  Notice that in the limit where $g$ diverges the
pressure vanishes.  This implies that the formerly unstable vortex
loop becomes a stable and massless particle-like excitation for
temperatures below $T_{c,M}$.  The typical core size $d$ and energy
$E_v$ of a CVL are given by
\begin{equation}
  \label{eq:CoreEnergy}
  d \propto \frac{1}{m_{\sst D}} =
  \frac{1}{g} \sqrt{\frac{\Lambda_M^3}{\beta}} 
  \qquad\textnormal{and}\qquad
  E_v \propto \frac{\pi}{g} \sqrt{\frac{\Lambda_M^3\beta}{2\pi}}.  
\end{equation}

By the collective dissociation of large-holonomy calorons and
anticalorons in the preconfining phase, isolated and closed magnetic
flux lines start to form.  At $T_{c,M}$, where the magnetic coupling
$g$ diverges logarithmically, the dual gauge field becomes infinitely
heavy.  Thus, a complete decoupling of the dual gauge modes takes
place at the magnetic-center phase transition.  As a result, only
contact interactions between center-vortex-loops are possible.

%
The decay of the monopole-antimonopole condensate and the subsequent
formation of the (Bose)condensate of CVLs is described by a
macroscopic complex scalar field $\Phi$ in a potential $V_C(\Phi)$.
The expectation of $\Phi$ is proportional to the expectation of the
't Hooft loop operator which is a dual order parameter for confinement.
CVLs in the magnetic phase are created by phase jumps of $\Phi$ and an
increase in the modulus of $\Phi$.  This process continues until
$\Phi$ relaxes to one of the $Z_2$ degenerated, energy and pressure
free minima of a potential $V_C$.  The phase of $\Phi$ is given by a
line integral of the dual gauge field $a_{\mu}^{\sst D}$ along a
spatial circle of infinite radius $S^{R=\infty}_1$ measuring the
(quantized) magnetic flux through the minimal surface
$\mathcal{M}_{S_1^{R=\infty}}$.  The creation of a CVL now proceeds by
an infinitely thin flux line and its flux reversed partner traveling
in from infinity and intersecting with the $S_1^{R=\infty}$, thereby
piercing the surface $\mathcal{M}_{S_1^{R=\infty}}$.  The energy
needed to create a single center-vortex loop is provided by the
potential $V_C$.
Selfintersecting and therefore massive CVLs come into existence when
generated single CVLs that move fast enough to convert some of their
kinetic energy into mass collide and merge, thus creating
selfintersections.
\begin{figure}
  \centering
  \includegraphics[width=0.60\textwidth]{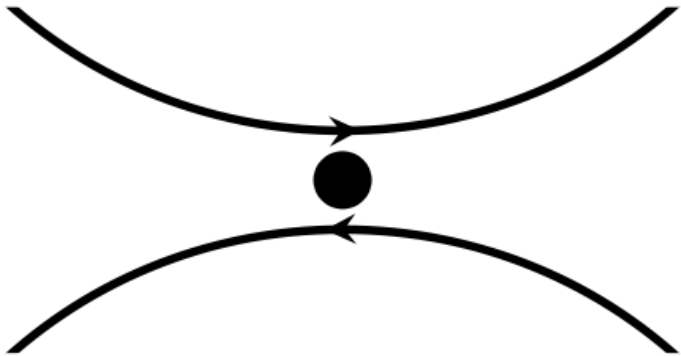}
  \caption{The oppositely directed center fluxes in the core of the
    intersection of a selfintersecting center-vortex loop generate an
    eddy where an isolated magnetic $Z_2$ monopole is located.}
  \protect{\label{fig:Intersection}}
\end{figure}
Each intersection point carries one unit of magnetic charge, see
Fig.~\ref{fig:Intersection}, where each sign is equally likely.  The
spectrum of excitations is equidistant since the mass of a soliton
with $N$ self-interactions is given by $N\Lambda_C$, $\Lambda_C$ being
the Yang-Mills scale.  Modulo the charge multiplicities the number of
distinct topologies of $N$-fold selfintersecting solitons is given by
the number of distinct topologies of connected bubble diagrams with
$N$ vertices in a scalar $\lambda\phi^4$ quantum field theory.  In
Fig.~\ref{fig:Excitations} the topologies of CVLs with up to $N=3$
selfintersections are shown.  If subjected to mixing with theories
possessing propagating photons the only stable excitations are
non-selfintersecting and one-fold selfintersecting CVLs.  This is due
to the repulsive or attractive forces between the charges of CVLs with
more than one selfintersection.  CVLs without selfintersection,
however, are unstable in the presence of a noisy environment, see
Sec.~\ref{sec:MbyC}.

In \cite{Hofmann2005} the potential $V_C$ for the macroscopic field
$\Phi$ is found to be
\begin{equation}
  \label{eq:CenterPotential}
  V_C=\overline{\left(\frac{\Lambda_C^3}{\Phi}-\Lambda_C\Phi\right)}
  \left(\frac{\Lambda_C^3}{\Phi}-\Lambda_C\Phi\right),
\end{equation}
satisfying the following properties (see also \cite{SHS2007}):

\noindent (i) $V_C$ is invariant under center jumps
$\Phi\rar\exp(\i\pi)\Phi$ only;

\noindent (ii) it allows for the creation of spin-1/2 fermions by a
forward- and backward tunneling which corresponds to local center
jumps of $\Phi$'s phase;

\noindent (iii) the degenerated minima of $V_C$ have zero energy
density and are related by local center jumps;

\noindent (iv) a mass scale $\Lambda_C$ occurs to parameterize the
potential $V_C$;

\noindent (v) $V_C$ needs to be real.

The process of relaxation of $\Phi$ to one of the minima of $V_C$ is
described by the action
\begin{equation}
  \label{eq:CenterAction}
  S=\int\mathrm{d}x^4\; 
  \left(\oh \overline{\pd_{\mu}\Phi}\,\pd_{\mu}\Phi -\oh V_C \right).
\end{equation}
Once $\Phi$ has reached $V_C$'s minima, quantum fluctuation
$\delta\Phi$ are absent because every potential fluctuation would be
harder than the maximal resolution.

Neglecting contact interactions between and internal degrees of
freedom within solitons as well as long-range interactions between
charges mediated by photons, the naive series for the total pressure
$P_C$ at temperature $T$ represents an asymptotic expansion in powers
of a suitably defined coupling coupling constant $\lambda \equiv
\exp(-\Lambda_C/T)$.  That is, the sum
\begin{equation}
\label{eq:CenterPressure}
P_C=\sum_{N=0}^{\infty}P_{C,N}
\end{equation}
over partial pressures $P_{C,N}$ of spin-1/2 states arising from
solitons with $N$ selfintersections seems to converge up to a
critical, temperature-dependent value $N_c(T)$, but converges when
including higher contributions.  This signals that the assumption that
solitons with arbitrary $N$ are stable breaks down to hold for $N>N_c$
as a consequence of contact interactions which increase due to the
higher density of intersection points and vortex lines.  Though
formally divergent, the sum over partial pressures $P_{C,N}$ turns out
to be Borel summable for negative (unphysical) values of $\lambda$.
The inverse Borel transformation is meromorphic\footnote{A meromorphic
  function is holomorphic on an open subset of the complex plane
  except for a set of isolated poles.} in the entire $\lambda$-plane
except for a branch cut along the positive-real axis.  Continuation to
the physical region $\lambda > 0$ leads to a sign-indefinite imaginary
part which is smaller than the real part for sufficiently small
temperatures.  Complex admixtures to the pressure become manifest as
turbulence-like phenomena in the plasma and thus violate thermal
equilibrium.  At zero temperature, the pressure of the ground state is
precisely \emph{nil}.  Because of the over-exponential rise of
spin-1/2 fermion states with increasing temperature, the imaginary
part starts to dominate the pressure and the thermodynamical
description of the system begins to fail (violation of spatial
homogeneity).  That is, at temperature $T_H\sim\Lambda_C$, the entropy
wins over the Boltzmann suppression in energy and the partition
function diverges.  This is an indication for the Hagedorn transition
to the preconfining phase.  For details see \cite{YMlow}.  Similar
behavior is observed for the expansion of the energy density
\cite{ConfiningPhase}
\begin{equation}
  \label{eq:CenterEnergyDensity}
\rho_C=\sum_{N=0}^{\infty}\rho_{C,N},  
\end{equation}
where $\rho_{C,N}$ is the energy density of soliton states with $N$
selfintersections and mass $N\Lambda_C$.

Demanding for continuity of the negative pressure across the
magnetic-center phase transitions yields a relation between
$\Lambda_M$ and $\Lambda_C$:
\begin{equation}
  \label{eq:MCScaleRelation}
  \Lambda_M \propto 2^{1/3}\Lambda_C.
\end{equation}

The question may arise whether there are stable selfintersecting
vortex-loops in the magnetic phase.  By the decay of the macroscopic
ground state in the magnetic phase its energy density is used to
create selfintersecting CVLs.  An $N$-fold twisted CVL possesses a
mass $N \Lambda_C$, where the Yang-Mills scale is about
$\Lambda_C\propto T_H$.  For this reason, the potential in the magnetic
phase cannot provide enough energy density to create a
selfintersection in the magnetic phase.

\section{\texorpdfstring{The postulate $\cmb = \mathrm{U(1)}_Y$}{The
    postulate SU(2) = U(1)}}
\label{sec:SU2CMB}

We have mentioned in Sec.~\ref{sec:DePhase} that the spatial
coarse-graining over the topologically nontrivial sector leads to the
emergence of a macroscopic adjoint Higgs field which breaks the
fundamental gauge group SU(2) down dynamically to the subgroup U(1).
Thereby, two out of three gauge bosons become massive.  At
$T_{c,E}=13.87\frac{\Lambda_C}{2\pi}$, where the electric-magnetic phase
transition takes place, the mass of these two gauge field diverges and
the massless mode remains exactly massless because radiative corrections
are absent due to the decoupling from its heavy partners.

Now consider the U(1)$_Y$ factor of the electroweak gauge group
$\mathrm{SU(2)}_W\times \mathrm{U(1)}_Y$ of the present Standard Model
of Particle Physics (SM).  In Quantum Electrodynamics (QED), the photon
is observed to be unscreened and practically massless
($m_{\gamma}<10^{-14}$~eV) \cite{WilliamsFallerHill}.  It is described
by the gauge group U(1) the progenitor of which is the U(1)$_Y$ factor.
As stated above, there is only a single point in the phase diagram of
SU(2) Yang-Mills thermodynamics that exhibits a precisely massless gauge
mode: the deconfining-preconfining phase transition at $T_{c,E}$.
Therefore, in \cite{Hofmann2005,Axion,Hofmann2006,GHS2006}, the
postulate was pushed forward that the U(1)$_Y$ factor of the electroweak
gauge group is the unbroken subgroup of an SU(2) Yang-Mills theory with
a scale comparable to the temperature of the cosmic microwave background
(CMB) $T_{\mbox{\tiny{CMB}}} =2.728$~eV.  This group is denoted $\cmb$.
The photon $\gamma$ in the SM has to be identified with the massless
gauge mode of $\cmb$.  In analogy to the $W^{\pm}$ gauge bosons of the
SM, the remaining two infinitely massive and thus undetectable gauge
modes of $\cmb$ are denoted $V^{\pm}$ ($m_{V^{\pm}}=2e\phi$ with
$e=\infty$ at $T_{c,E}$).  Furthermore, the average temperature of the
universe $T_{\mbox{\tiny CMB}}$ is identified with the critical
temperature $T_{c,E}$ of $\cmb$.  This fixes the only free parameter of
the theory, the Yang-Mills scale $\Lambda_{E}$, to
$\Lambda_E=\frac{2\pi}{13.87}T_{\mbox{\tiny CMB}}=1.065\times
10^{-4}$~eV.  For temperatures much above $T_{c,E}$, the effects of
$V^{\pm}$ are completely negligible, whereas for temperatures a few
times of $T_{\mbox{\tiny CMB}}$, these lead to a visible modification of
the black-body spectrum at low frequencies (spectral gap)
\cite{GHS2006,Gap}.  The spectral gap could also provide for an
explanation why old (estimated age $\sim$~50 million years), cold (mean
brightness temperature $\sim 20$~K) and dilute (number density $\sim
1.5$~cm$^3$) clouds in between the spiral arms of the outer galaxy are
composed of atomic instead of molecular hydrogen, and why these clouds
are stable \cite{Clouds}.

Regarding the transition towards the preconfining phase, the postulate
$\cmb=\mathrm{U(1)}_Y$ implies that the photon will acquire a Mei\ss ner
mass because of the coupling to the newly emerging superconducting
ground state (condensate of magnetic monopoles).  The system, however,
remains in a supercooled state down to $T=12.15\frac{\Lambda_E}{2\pi}$
due to the shift in energy density at $T_{c,E}$ (additional degree of
freedom), see also Fig.~\ref{fig:EnergyDensity}.
In \cite{Axion} an upper bound for the time the photon remains massless
was estimated to be $\sim 2.2$ billion years.  The observed
intergalactic magnetic fields can possibly be explained by the
electric-magnetic phase transition.  Conventional superconductors
consist of a Cooper-pair condensate of \emph{electric} charges and expel
\emph{magnetic} fields from their interior (Mei\ss ner effect).  If the
occurrence of intergalactic magnetic fields is addressed to the
emergence of a superconducting ground state, this leads to the
conclusion that a magnetically charged object in the gauge group $\cmb$
is interpreted as an electrically charged object with respect to
U(1)$_Y$.  Therefore, the ground state of the magnetic phase is a
condensate of electrically charged monopoles and antimonopoles with
respect to U(1)$_Y$, and thus generates intergalactic magnetic fields.

  
\section{Motion by curvature}
\label{sec:MbyC}

Here, we would like to illustrate how the curve shrinking process is
induced by the curvature of a CVL.  Recall that the vortex loop is
generated by the bending of a straight ANO vortex line which exhibits
isotropic pressure perpendicular to its symmetry axis.
\begin{figure}[t!]
  \centering
  \includegraphics[width=0.9\textwidth]{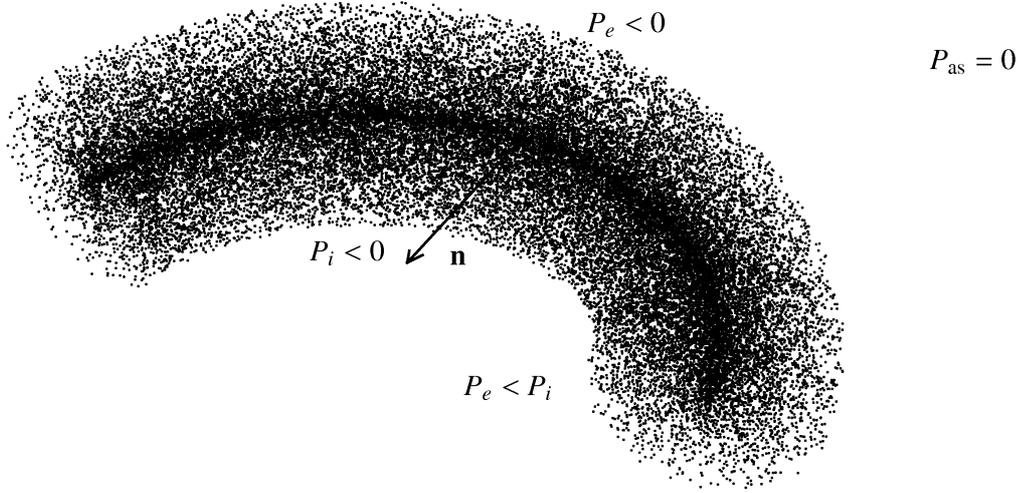}
  \caption{Highly space-resolved snapshot of a segment of a
    center-vortex loop. The pressure $P_i$ in the region pointed to by
    the normal vector $\mathbf{n}$ is more negative than the pressure
    $P_e$ thus leading to a motion of the segment along $\mathbf{n}$.}
  \protect{\label{fig:Pressure}}
\end{figure}
Now consider a situation where a CVL of an isolated SU(2) Yang-Mills
theory is (locally) embedded into a flat two-dimensional surface at
$m_{\sst D} <\infty$ and $d>0$.  Then, a hypothetical observer
measuring a positive (negative) curvature of a segment of the vortex
line experiences more (less) negative pressure in the intermediate
vicinity of this curve segment (see Sec.~\ref{sec:ConPhase}) leading
to its motion towards (away from) the observer, see
Fig.~\ref{fig:Pressure}.  The (inward directed) velocity of a point in
the vortex core will be a monotonic function of the curvature at this
point.  On average, this shrinks the CVL.  Alternatively, one may
\emph{globally} consider the limit $m_{\sst D}\to\infty$, $d\to 0$,
that is the confining phase, but now taking into account the effects
of an environment that \emph{locally} relaxes this limit (by
collisions) and thus also induces curve shrinking.  This situation is
described by a curve shrinking flow in the dimensionless parameter
$\tau$
\begin{equation}
  \label{eq:csfSigma}
  \pd_\tau \vec{x}(\xi,\tau) = \frac{1}{\sigma}\,\pd^2_{\xi} 
  \vec{x}(\xi,\tau),
\end{equation}
where $\vec{x}$ is a point on the planar CVL, $\xi$ is arc length, and
$\sigma$ a string tension effectively expressing the distortions
induced by the (noisy) environment.  After a rescaling to
dimensionless variables,
\begin{equation}
  \label{eq:rescaling}
  \x \equiv \sqrt{\sigma}\vec{x}\qquad \textnormal{and} \qquad 
  s = \sqrt{\sigma}\xi,
\end{equation}
flow equation~(\ref{eq:csfSigma}) assumes the form:
\begin{equation}
  \pd_{\tau}\x(s,\tau) = \pd_s^2\x(s,\tau).
\end{equation}
In the following sections, we will resort to the dimensionless flow
equation.

\cleardoublepage


\chapter{Mathematical Prerequisites: Curve shrinking flow}
\label{chap:MP}

In the 1970s, William Thurston developed a program for the
classification of three-dimen\-sion\-al manifolds.  It had a great
impact in the field of three-dimensional topology and revealed a very
strong connection between low-dimensional topology and differential
geometry, especially between hyperbolic geometry and Kleinian groups
\cite{ThurstonLectures,Leeb}.

Now consider a smooth closed (that is, compact and without boundary)
manifold $\M$ equipped with a smooth time-dependent Riemannian metric
$g(\tau)$.  A (topological) manifold is a topological space which
is locally homeomorphic to a Euclidean space, but with an generally
more complicated global structure.  A manifold equipped with a
Riemannian metric $g$ is a real differentiable manifold $\M$, in which
each tangent space is endowed with an inner product $g$ in a manner
that varies smoothly from point to point.  It should be noted that not
every manifold admits a geometry.  The Ricci flow is a means of
processing the metric $g$ by the evolution of $g$ under the following
partial differential equation (PDE)
\begin{equation}
  \label{eq:RicciFlow}
  \frac{\pd}{\pd \tau} g(\tau) = -2 \mathrm{Ric}(g),
\end{equation}
where Ric is the Ricci curvature.  In local coordinates the
coefficients $R_{ij}$ of the Ricci curvature tensor are given by a
contraction of the Riemannian curvature tensor $R^i{}_{jkl}$, $R_{ij}
= R^k{}_{ikj}$.  Roughly speaking, the Ricci-flow contracts regions of
positive curvature and expands those of negative curvature, thereby
smoothing out irregularities in the metric.  In this spirit, it is
formally analogous to the diffusion of heat that describes how an
irregular temperature distribution in a given region tends to become
more homogeneous over time.  An example of its application is the
proof of the two-dimensional uniformization theorem, which states that
any surface admits a Riemannian metric of constant Gaussian curvature.
Here, the (suitably renormalized) Ricci flow is used to conformally
deform a two-dimensional metric on $\M$ into one of constant curvature
\cite{Topping}.  Richard Hamilton introduced the Ricci flow with the
intention to gain insight into the geometrization conjecture proposed
by William Thurston in 1980 \cite{Thurston}.  The geometrization
conjecture is the analogue for three-manifolds of the uniformization
theorem for surfaces and implies several other conjectures, such as
Thurston's elliptization conjecture or the Poincar\'e conjecture.  Let
us first consider the Poincar\'e conjecture, which was originally
posed as a question at the end of an article by Henri Poincar\'e in
1904.  In its standard form, it states that every simply connected,
compact three-manifold without boundary is homeomorphic to the
three-sphere.
A more precise phrasing is that the fundamental group of a closed
three-manifold $\M$ is trivial, if and only if $\M$ is homeomorphic to
the three-sphere.  Now, the geometrization conjecture concerns the
topological classification of three-dimensional smooth manifolds.  The
original phrasing of Thurston goes as follows \cite{Thurston}: ``The
interior of every compact 3-manifold has a canonical decomposition
into pieces which have geometric structures''\footnote{A geometric
  structure is defined to be a space modeled on a homogeneous space
  $(X,G)$, where $X$ is a manifold and $G$ is a group of
  diffeomorphisms of $X$ such that the stabilizer of any point $x\in
  X$ is a compact subgroup of $G$ \cite{Thurston}.  For every $x$ in
  $X$, the stabilizer subgroup of $x$ (also called the isotropy group
  or little group) is defined as the set of all elements in $G$ that
  fix $x$: $G_x=\{g\in G|g\cdot x =x\}$}.
In three dimension there are precisely eight geometric structures
called the eight Thurston (model) geometries (involving the spherical
geometry $S^3$, the Euclidean geometry $\mathbb{R}^3$, the hyperbolic
geometry $\mathbb{H}^3$, the geometry of $S^2\times\mathbb{R}$, the
geometry of $\mathbb{H}^2\times\mathbb{R}$, the geometry of the
universal cover of $SL_2(\mathbb{R})$, the nil geometry and finally
the sol geometry).  The canonical decomposition is carried out in two
steps.  In the first stage, also referred to as the prime
decomposition, one cuts a three-manifold $\M$ along two-spheres
embedded in $\M$ such that neither of the obtained manifolds is a
three-ball, then one glues three-balls to the resulting boundary
components.  This decomposition is unique up to the sequence and
additional three-balls.  The second stage involves cutting along
certain tori that are nontrivially embedded in $\M$ obtaining a
three-manifold the boundary of which consists of tori.  Hamilton's
basic idea was to place an arbitrary metric $g$ on a given smooth
manifold $\M$ and to dynamically deform $\M$ by the Ricci flow to
yield one of Thurston's geometric structures.  Hamilton succeeded in
proving that a closed three-dimensional manifold, which carries a
metric of positive Ricci curvature, is a spherical space form that
acts like an attractor under the Ricci flow \cite{Hamilton}.  This is
known as the Hamilton theorem.  However, in general, the Ricci flow
can be expected to develop a singularity in finite time.  Then, in a
series of eprints starting in 2002, Grigori Perelman sketched a proof
for the geometrization conjecture \cite{Perelman}.  Thereby, Perelman
modified Hamilton's program to prove Thurston's geometrization
conjecture by stopping the Ricci flow once a singularity has been
formed, then carefully performing `surgery' on the evolved manifold,
systematically excising singular regions before continuing the flow.
This is called Ricci flow with surgery.

The results obtained in this thesis heavily depend on the important
work on the curve shortening flow done by Gage and Hamilton
\cite{GageHamilton}, and Grayson \cite{Grayson,GraysonII}.  The curve
shortening flow, also known as heat equation on immersions\footnote{An
  immersion is a local embedding.}, is the one-dimensional analogue to
the Ricci flow and originally inspired Hamilton in the development of
the Ricci flow.  Let us now consider the properties of curve shrinking
flows in two and three space dimensions.

\section{Embedded curves without selfintersection}
\label{sec:EC}

\subsection{Planar curves}
\label{sec:Planar}

\begin{figure}
  \centering
  \includegraphics[width=0.60\textwidth]{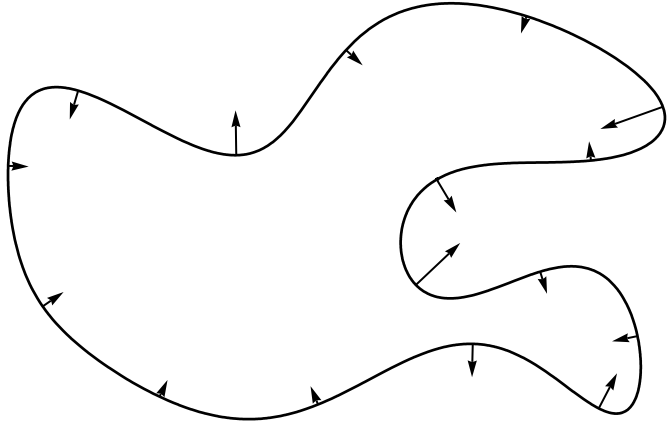}
  \caption{\protect{\label{fig:Elastic}}The Euclidean curve shortening
    flow.  The arrows point towards the unit normal $\mathbf{n}$ and
    the length of the arrows is proportional to the curvature $k$.}
\end{figure}

Consider a family of smooth, closed curves $\x(s,\tau)$ of length $L$
embedded\footnote{An embedding is a map $f:X\rar Y$ between
  differentiable manifolds $X$ and $Y$ where the map $f$ is a
  homeomorphism between $X$ and its image $f(X)$.} in a
two-di\-men\-sion\-al flat plane $\mathbb{R}^2$, where $\x$ is a point
along the curve, $s\in[0,L]$ is the arc length that is unique only up
to a constant and $\tau\in[0,T]$ the flow parameter which parametrizes
the family.  The initial curve $\x(s,0)$ evolves as a function of
`time' $\tau$ to $\x(s,\tau)$.  The Euclidean curve shortening flow is
defined as
\begin{equation}
  \label{eq:csf}
  \pd_{\tau}\x(s,\tau) 
  = \pd_s^2\x(s,\tau) 
  \equiv k(s,\tau)\,\mathbf{n}(s,\tau),
\end{equation}
where the derivative $\pd_{\tau}:=\frac{\pd}{\pd \tau}$ is taken along a fixed value of $s$.
This is a parabolic, nonlinear second-order partial differential
equation, where $\mathbf{n}$ is the inward-pointing Euclidean unit
normal and $k$ the scalar curvature, defined as
\begin{equation}
  \label{eq:curvature}
  k(s,\tau) = |\pd_s^2\x(s,\tau)| 
  = \det(\pd_s\x(s,\tau),\pd_s^2\x(s,\tau)),
\end{equation}
with $|\mathbf{v}|\equiv\sqrt{\mathbf{v}\cdot\mathbf{v}}$,
$\mathbf{v}\cdot\mathbf{w}$ denoting the Euclidean scalar product, and
$\det(\cdot,\cdot)$ denotes the determinant of the $2\times 2$ matrix
created by two $2\times 1$ vectors.  It is a standard result for
parabolic equations that solutions exist for a short time and are
unique.  In the curve shortening flow, the curve $\x(s,\tau)$ is
deformed along its unit normal $\mathbf{n}(s,\tau)$ at a rate that is
proportional to its curvature $k(s,\tau)$.  This flow deserves the
attribute curve shortening, because its flow lines in the space of
closed curves are tangent to the gradient for the curve length
functional, see Eq.~(\ref{eq:Lvar}).  For the remainder of this
section, we assume that a solution to Eq.~(\ref{eq:csf}) exists on the
maximal time interval $[0,T)$.  A more visual description of this flow
is the evolution of an elastic band in a viscous medium.  If the
tension in the elastic is kept constant then its behavior is
approximately determined by Eq.~(\ref{eq:csf}), see also
Fig.~\ref{fig:Elastic}.
Since motion normal to the curve affects arc length, $s$ is not
preserved under curve shrinking.  Thus, $s$ and $\tau$ are not
independent and commute according to the following rule
\begin{equation}
  \label{eq:stcommutator}
  \pd_{\tau}\pd_s = \pd_s\pd_{\tau} + k^2\pd_s.
\end{equation}
Therefore, we introduce the curve parameter $u$ (modulo $2\pi$)
related to $s$ by
\begin{equation}
  \label{eq:su}
  \mathrm{d}s = \left|\pd_u\x\right|\mathrm{d}u.
\end{equation}
The quantity $|\pd_u\x|$ can also be thought of as an arc length
density.  Actually, Eq.~(\ref{eq:su}) defines $s$.  The operator
$\pd_s$ then writes as
\begin{equation}
  \label{eq:operator}
  \pd_s=\frac{1}{\left|\pd_u\x\right|}\pd_u.
\end{equation}
In the following, we resort to a slight abuse of notation by using the
same symbol $\x$ for the functional dependence on $u$ or $s$.  Let us
now introduce coordinates in $\mathbb{R}^2$,
$\x(u,\tau)=(x(u,\tau),y(u,\tau))^T$ (where $T$ denotes the
transpose).  The tangent vector to the curve is given by $\pd_u \x$,
and thus we define the unit tangent $t$ vector as
\begin{equation}
  \label{eq:unittangent}
  \mathbf{t}(u,\tau) :=\frac{\pd_u\x}{|\pd_u\x|} 
  =\frac{1}{|\pd_u\x|}\ve{\pd_u x}{\pd_u y}.
\end{equation}
The unit normal is then given by
\begin{equation}
  \label{eq:unitnormal}
  \mathbf{n}(u,\tau) :=\frac{1}{|\pd_u\x|} \ve{-\pd_u y}{\pd_u x}.
\end{equation}
The unit tangent and normal vectors are written in terms of arc length
$s$ as
\begin{equation}
  \mathbf{t}(s,\tau) =\ve{\pd_s x}{\pd_s y}, \qquad
  \textnormal{and}\qquad
  \mathbf{n}(s,\tau) = \ve{-\pd_s y}{\pd_s x}.
\end{equation}
So we can write the Frenet-Serret formulas, which describe the
kinematic properties of a point (particle) that moves along the planar
curve $\x$ as
\begin{equation}
  \label{eq:Frenet}
  \frac{\pd}{\pd s} \ve{\mathbf{t}}{\mathbf{n}} = \left(
    \begin{array}{cc}
      0 & k \\
      -k & 0
    \end{array}
  \right)  \ve{\mathbf{t}}{\mathbf{n}},
\end{equation}
where the curvature, when expressed in coordinates, is
\begin{equation}
  \label{eq:kincoords}
  k(s,\tau) =\pd_s x\,\pd_s^2y-\pd_s^2x\,\pd_s y.
\end{equation}
The circumference of the curve $L$ at time $\tau$ is defined as
\begin{equation}
  \label{eq:length}
  L(\tau) \equiv \int_0^{L(\tau)}\mathrm{d}s = \int_0^{2\pi}
  \mathrm{d}u\,\left|\pd_u\x(u,\tau)\right|.
\end{equation}
The evolution of $L$ under the flow is given by
\begin{equation}
  \label{eq:Levol}
  \dot L(\tau) := \frac{\mathrm{d}L(\tau)}{\mathrm{d}\tau} 
  \equiv -\int_0^{L(\tau)} \mathrm{d}s\,k^2 
  =-\int_0^{2\pi}\mathrm{d}u\left|\pd_u\x\right|k^2.
\end{equation}
For the area $A$ enclosed by the curve we have
\begin{equation}
  \label{eq:area}
  A(\tau) \equiv \frac{1}{2}\left|\int_0^{L(\tau)}\mathrm{d}s\;
    \x(s,\tau)\cdot\mathbf{n}(s,\tau)\right|.  
\end{equation}
Surprisingly, the time derivative of the enclosed area remains
constant under curve shrinking,
\begin{equation}
  \label{eq:Aevol}
  \dot A(\tau) := \frac{\mathrm{d}A(\tau)}{\mathrm{d}\tau} = -2\pi.
\end{equation}
For planar curves, the decreasing integral $\int_0^L\mathrm{d}s \,|k|$
measures the total change in angle.  In the special case of convex
planar curves, $\int_0^L\mathrm{d}s \,|k| = \int_0^L \mathrm{d}s \,k$
measures the winding number of the curve and is an invariant of the
flow (until a singularity develops).

In \cite{GraysonI}, Grayson stated that under the flow
Eq.~(\ref{eq:csf}) `the curve is shrinking as fast as it can using
only local information'.  Let us see how this statement can be
understood.  Consider the curve length
$L(\tau)=\int_0^{2\pi}\mathrm{d}u\,|\pd_u\x|$.  To take the time
derivative of $L$ we differentiate $|\pd_u\x|^2$ with respect to
$\tau$ and obtain
\begin{equation}
  \label{eq:v2der}
  \pd_{\tau}|\pd_u\x| = \frac{1}{|\pd_u\x|}\pd_u\x\cdot\pd_{\tau}\pd_u\x.
\end{equation}
Substituting this into $\dot L(\tau)$ and integrating
by parts, we obtain the following expression for the rate of decrease
of curve length
\begin{equation}
  \label{eq:Lvar}
  \dot L(\tau) = -\int_0^{L(\tau)}\mathrm{d}s\,k\,
  \mathbf{n}\cdot\pd_{\tau}\x.
\end{equation}
Therefore, Eq.~(\ref{eq:csf}) expresses the local condition that the
rate of decrease of $L(\tau)$ is maximal with respect to a variation
of the direction of the velocity $\pd_{\tau}\x$ of a given point on
the curve at fixed magnitude $|\pd_{\tau}\x|$ \cite{Smith}.  However,
the magnitude $|\pd_{\tau}\x|$ is not in general the speed which
maximizes $\dot L(t)$.

Setting $A(\tau=0)\equiv A_0$, the solution to Eq.~(\ref{eq:Aevol}) is
\begin{equation}
  \label{eq:Aevolexpl}
  A(\tau)=A_0-2\pi\,\tau.
\end{equation}
By virtue of Eq.~(\ref{eq:Aevolexpl}) the critical value $T$, where
$A$ and with it the curve vanishes, is related to $A_0$ as
\begin{equation}
  \label{eq:A0}
  T=\frac{A_0}{2\pi}.
\end{equation}
The isoperimetric ratio is defined as $\frac{L^2}{A}$, and the
isoperimetric inequality states that
\begin{equation}
  \label{eq:isoinequ}
  \frac{L(\tau)^2}{A(\tau)}\ge 4\pi.
\end{equation}
Equality is achieved if and only if the curve is a circle.  Therefore,
one can consider it a measure of `how circular' the curve is.

In 1983, Gage showed that when a smooth convex curve evolves according
to Eq.~(\ref{eq:csf}), the isoperimetric ratio $\frac{L^2}{A}$
decreases, so that if $A\rightarrow 0$, then $L\rightarrow 0$ and the
curve shrinks to a point \cite{Gage1983}.  In 1984, Gage showed that a
convex curve is becoming circular and $\frac{L^2}{A}$ approaches
$4\pi$, as the enclosed area approaches zero, provided that the
curvature does not blow up prematurely, that is the curve does not
form a cusp \cite{Gage1984}.  As a consequence, the ratio
$\frac{R_{\mathrm{out}}}{R_{\mathrm{in}}}$ of the circumscribed ratio
to the inscribed ratio converges to unity.  This can be considered a
$C^0$-convergence to the circle.  Hence, in the absence of
singularities, a strictly convex and embedded curve remains convex and
embedded under the evolution.

In 1986, Gage and Hamilton showed that for convex curves the curvature
does not blow up prematurely for $\lim_{\tau\rightarrow T} A(\tau) =
0$ \cite{GageHamilton}.  Thus, the curve remains convex and becomes
circular, as it shrinks to a point for $\tau\nearrow T$, where
$0<T<\infty$.  The curve shrinks to a circle in the sense that:

\noindent (i) the ratio $\frac{R_{\mathrm{out}}}{R_{\mathrm{in}}}$
approaches unity;

\noindent (ii) the ratio of the maximum curvature to the minimum
curvature $\frac{k_{\mathrm{max}}}{k_{\mathrm{min}}}$ approaches unity
($C^2$-convergence);

\noindent (iii) the higher order derivatives of the curvature $k$
converge to zero uniformly ($C^{\infty}$-convergence).

\noindent In 1987, Grayson showed that embedded (non-selfintersecting)
planar curves become convex before $T$ without developing
singularities \cite{GraysonI}.  Thus, this completes the proof of the
well known Gage-Hamilton-Grayson theorem that curve shortening
determined by Eq.~(\ref{eq:csf}) shrinks embedded plane curves
smoothly to points, with round limiting shape.  It is important to
note that some planar curves, which are immersed but not embedded will
surely develop singularities, e.g. the figure-eight of
Sec.~\ref{sec:IC} or the Lima\c{c}on of Pascal.

Consider the set of all Euclidean transformations in $\mathbb{R}^2$,
that is the set of all rotations, translations and reflections of a
figure in $\mathbb{R}^2$.  Such a transformation
$ET:\mathbb{R}^2\rar\mathbb{R}^2$ is a function of the form
\begin{equation}
  \label{eq:ET}
  ET(\x)= U\x+a,
\end{equation}
where $U$ is an orthogonal $2\times 2$ matrix and $a\in\mathbb{R}^2$.
The Euclidean curve shortening flow is defined in terms of the
Euclidean curvature $k$ and the Euclidean unit normal $\mathbf{n}$
that are invariant under Euclidean transformations $ET$.

\subsection{Space curves}
\label{sec:Space}

There are several possibilities of generalizing the curve shortening
flow.  One is the mean curvature flow, which is the generalization of
Eq.~(\ref{eq:csf}) for hypersurfaces.  In this case, the results of
Sec.~\ref{sec:Planar} continue to hold for convex curves, but for
non-convex curves they do not \cite{ChouZhu}.

For our purposes, it is more interesting to look at the extension of
the curve shortening flow for curves embedded in the three-dimensional
Euclidean space $\mathbb{R}^3$.  Consider a continuous, differentiable
(and not necessarily closed) space curve $\x(s,\tau)$ embedded in
$\mathbb{R}^3$.  The tangent, normal and binormal unit vectors are
denoted $\mathbf{t}$, $\mathbf{n}$ and $\mathbf{b}$, respectively,
also called Frenet-Serret frame, and defined as follows:

\noindent $\mathbf{t}$ is the unit vector tangent to the curve,
pointing in the direction of motion: $\mathbf{t} = \pd_s \x$;

\noindent $\mathbf{n}$ is the normalized derivative of $\mathbf{t}$
with respect to the arc length $s$ of the curve: $\mathbf{n}
=\frac{\pd_s\mathbf{t}}{|\pd_s\mathbf{t}|}$;

\noindent $\mathbf{b}$ is the cross product of $\mathbf{t}$ and
$\mathbf{n}$: $\mathbf{b}=\mathbf{t}\times\mathbf{n}$.

\noindent The Frenet-Serret formulas for a point on the space curve
are given by
\begin{equation}
  \label{eq:FrenetSpace}
  \frac{\pd}{\pd s}\vve{\mathbf{t}}{\mathbf{n}}{\mathbf{b}} 
  = \left(
    \begin{array}{ccc}
      0 & k & 0 \\
      -k & 0 & t \\
      0 & -t & 0
    \end{array}
  \right) \cdot \vve{\mathbf{t}}{\mathbf{n}}{\mathbf{b}},
\end{equation}
where $k$ is the curvature and $t$ the torsion.  The Frenet-Serret
formulas effectively define the curvature and torsion of a space
curve.  It should be noted that the existence of a Frenet-Serret frame
requires $|k\mathbf{n}|^2>0$.  That is, a particle traveling along the
curve must experience acceleration.  The evolution equation for space
curve assumes the same form as Eq.~(\ref{eq:csf}),
\begin{equation}
  \label{eq:Spacecsf}
  \pd_{\tau} \x(s,\tau) = k(s,\tau)\mathbf{n}(s,\tau).
\end{equation}
The unit normal $\mathbf{n}$ is not always defined, though
$k\mathbf{n}$ always makes sense.  It was shown by Altschuler and
Grayson that solutions to the space curve flow exist until the
curvature becomes unbounded.  However, space curves may not remain
embedded in general, and singularities will develop in the case of
closed curves.  A phenomenon of space curve evolution is that
inflection points ($k=0$) may develop during a time interval on which
the curvature is bounded.  When this happens the curvature becomes
zero and the torsion infinite at a point.  Nevertheless the curve
remains embedded since the flow ignores these types of singularities
in the torsion \cite{Altschuler,AltschulerGrayson}.  A rather
surprising property of space curve evolution is, that the formation of
a singularity is a planar phenomenon.  A space curve is said to be
planar at a point $(s',\tau')$ if the ratio of torsion and curvature
vanishes, $\frac{t}{k}(s',\tau')=0$.  In \cite{Altschuler}, Altschuler
showed that if a space curve develops a singularity at $(s',\tau')$,
then $\lim_{(s,\tau)\rar (s',\tau')} \frac{t}{k}(s,\tau)=0$.
Furthermore, Altschuler showed that the space curve is either
asymptotic ($\tau\rar\infty$) to a planar solution which moves by
homothety (self-similarity), or a rescaling of the solution along the
singularity converges in $C^{\infty}$ to a limiting solution
$\x(s,\tau=\infty)$ \cite{Altschuler}, where $\x(s,\tau=\infty)$ is
the family of planar, convex curves.  The most trivial case of a curve
moving by homothety is the circle shrinking down to a point.  It
should be noted that the conjecture due to Grayson, that singularity
formation is a planar phenomenon, can be proven without using the
language of rescalings.

\section{Immersed curves with one selfintersection}
\label{sec:IC}

When a closed curve immersed in a plane evolves by its curvature
according to Eq.~(\ref{eq:csf}), it remains smooth until its curvature
blows up.  From Sec.~\ref{sec:Planar}, we know that an embedded closed
curve cannot develop a singularity until it shrinks to a point, where
the limiting shape of the curve converges in $C^{\infty}$ to circle.
In marked contrast to this behavior, it was shown by Grayson that an
immersed curve can evolve by the curvature flow such that its area
vanishes, but its isoperimetric ratio converges to $\infty$.  Such a
curve, namely a figure-eight, was investigated in \cite{GraysonII}.  A
figure-eight is the simplest non-embedded curve and is defined to be a
smooth immersion into the plane with exactly one double point, and a
total rotation number zero,
\begin{equation}
  \label{eq:trn}
  \int_{0}^{L}\mathrm{d}s\,k=0.
\end{equation}
Here, $s$ is arc length, $L$ the curve length and $k$ the scalar
curvature.  Such a curve divides the plane into three disjoint areas
two of which are finite and denoted (the unsigned areas) $A_1$ and
$A_2$.  Let $\x(s,\tau=0)$ be figure-eight which evolves to
$\x(s,\tau)$ according Eq.~(\ref{eq:csf}) for $0\le\tau<T$.  The
curvature is unbounded as $\tau\rar T$.  The main result of
\cite{GraysonII} is that the isoperimetric ratio $\frac{L^2}{A}$
converges to $\infty$ as $\tau\rar T$ if and only if the loops bound
regions of equal area, $A_1(0)=A_2(0)$.  This in turn implies that
$\frac{L^2}{A}$ for a curve with unequal-area loops is bounded as
$\tau\rar T$.

Since for immersed curves the number of double points is a
non-increasing function of time \cite{GraysonIII}, a figure-eight
remains a figure-eight until one of its loops collapses or the flow
encounters a singularity.  The curve stays smooth and the flow
continues until $A_1$ or $A_2$ converge to zero.  For the total area
$A$ of a figure-eight we have
\begin{equation}
  \label{eq:totalA}
  A(\tau) = A_1(\tau) + A_2(\tau).
\end{equation}
The time derivative of the area enclosed by one of loop of the curve
is equal to $-|\int \mathrm{d}s\,k|$ over the loop.  Unlike the case
of a non-selfintersecting curve, the rate of change of the total area
 is not longer constant, but constrained as
\begin{equation}
  -4\pi\le\frac{\mathrm{d} A(\tau)}{\mathrm{d}\tau}\le -2\pi.
\end{equation}
However, we have the nice property of figure-eights that the
difference of areas bounded by the two loops of $\x(s,\tau)$ remains
constant under the flow evolution:
\begin{equation}
  \label{eq:diffA}
  A_1(\tau)-A_2(\tau)=\mathrm{const}.
\end{equation}

Aside from a number of applications in differential geometry, curve
shortening flows are also used in multi-agent systems, such as mobile
autonomous robots \cite{Smith}, in image processing where the flow
provides an efficient way to smooth curves representing the contours
of objects, or in computer vision.  For a complete account of many of
the results of curve shrinking see \cite{ChouZhu,Cao}.

In the following, we suppress the functional dependence on $u$ in the
argument of $\x$ and $\mathbf{n}$ and write $\x(\tau):=\x(u,\tau)$ and
$\mathbf{u}(u,\tau):=\mathbf{n}(\tau)$

\cleardoublepage


\chapter{Non-selfintersecting center-vortex loops}
\label{chap:0CVL}
%

We apply curve shrinking to the $N=0$ sector in the sense of
Sec.~\ref{sec:MbyC}.  It should be noticed that the restriction of the
motion of a CVL to a two-dimensional flat plane is a major assumption
which needs to be supplemented by additional physical arguments for
its validity.

\section{Wilsonian renormalization-group flow}
\label{sec:0WRGF}

In this section, we exploit the concept of renormalization-group
transformations to yield an effective `action' that enables us to
compute statistical quantities.  The renormalization group allows one
to investigate the change in the physical parameters of a system which
is associated with the change in scale (energy or resolution) and
necessary to keep the physics constant.  In our case, the change of
scale corresponds to a change of the resolution $Q$ used to probe the
system.  Here, the resolution $Q$ is a strictly monotonic decreasing
function of the flow parameter $\tau$.  The change in parameters of
the effective `action' is implicitly determined by a
renormalization-group flow in $\tau$.


\subsection{Geometric partition function}
\label{sec:0GPF}

Let us now interpret the process of curve shrinking determined by
Eq.~(\ref{eq:csf}) as a re\-norm\-al\-iz\-a\-tion-group transformation
of a statistical ensemble made up of planar $N=0$ CVLs.  A partition
function, which is the sum over suitable defined weights of the
members in the ensemble, is considered to be invariant under a
decrease of resolution $Q$ determined by the flow parameter $\tau$.
Physically, $\tau$ is a monotonically decreasing function of $Q/Q_0$,
where $Q$ ($Q_0)$ are mass scales associated with the actual (initial)
resolution applied to the system.  The role of $Q$ can also be played
by the finite temperature of a reservoir that is coupled to the
system.

To define a suitable weight, we devise an ansatz for the effective
`action' $S=S[\x(\tau)]$ in geometric terms of the curves in the
ensemble, since these are the only accessible quantities in the system
of isolated non-interacting CVLs.  The `action' as a functional of $\x$
is expressible in terms of integrals over local densities in
$s$. 
Furthermore, we take advantage of the following symmetries the action
should possess:

\noindent (i) scaling symmetry $\x\to \lambda\x,\
\lambda\in\mathbb{R}_+$: for both conformal limits, $\lambda\to\infty$
and $\lambda\to 0$, where the curves at fixed $L$ gets unobservable
since $\lambda L\to\infty$ and $\lambda L\to 0$, the `action' $S$ should
be invariant under further finite rescalings (decoupling of the fixed
length scale~$\sigma^{-1/2}$);

\noindent (ii) Euclidean point symmetry in $\mathbb{R}^2$, that is the
group of all rotations, translations and reflections of a figure (curve)
in the plane: sufficient but not necessary for this is a representation
of $S$ in terms of integrals over scalar densities with respect to these
symmetries.  That is, the `action' density should be expressible as a
expansion in series involving products of Euclidean scalar products of
$\frac{\pd^n}{\pd s^n}\x,\ n\in\mathbb{N}^+,$ or constancy.  However,
scalar integrals can be constructed which involve non-scalar densities.
For instance, consider the area $A$ enclosed by curve and given by
\begin{equation}
  \label{eq:area2}
  A(\tau)=   \oh \left|\int_0^{L(\tau)}\mathrm{d}s\;
    \x(\tau)\cdot\mathbf{n}(\tau)\right|.
\end{equation}
The density $\x\cdot\mathbf{n}$ in this expression is not a scalar
under translations.

We now decompose the effective `action' into a conformal and a
non-conformal factor
\begin{equation}
  \label{eq:action}
  S=F_c\times F_{nc}\;,
\end{equation}
where in addition to Euclidean point symmetry $F_c$ is invariant under
$\x\to \lambda\x$, whereas $F_{nc}$ is not.  In principle, infinitely
many operators can be defined to contribute to $F_c$.  Since the
evolution generates circles for $\tau\nearrow T$ and thus homogenizes
the curvature, higher derivatives of $k$ with respect to $s$ rapidly
converge to zero \cite{GageHamilton}.  We expect this to be true also
for Euclidean scalar products involving higher derivatives
$\frac{\pd^n}{\pd s^n}\x$.  To yield conformally invariant expressions
such integrals need to be multiplied by powers of $\sqrt{A}$ and/or $L$
or the inverse of integrals involving lower derivatives.  At this stage,
we are not able to constrain the expansion in derivatives by additional
physical or mathematical arguments.  To be pragmatic, we simply set
$F_c$ equal to the isoperimetric ratio:
\begin{equation}
  \label{eq:Fc}
  F_c(\tau)\equiv\frac{L(\tau)^2}{A(\tau)}\,.
\end{equation}
We consider the non-conformal factor $F_{nc}$ in $S$ as a formal Taylor
expansion in inverse powers of $L$ or $A$ due to the conformal
invariance of the curve for $L,A\rightarrow\infty$ and
$L,A\rightarrow0$.  Since the renormalization-group evolution of the
effective `action' is driven by the curve shortening flow of each member
in the ensemble, we allow for an explicit $\tau$ dependence of the
coefficient $c$ of the lowest nontrivial power $\frac{1}{L}$.  The idea
is to include the contribution of higher-order operators, that do not
exhibit an explicit $\tau$ dependence, into a resolution dependence of
the coefficient of the lower-dimensional operators.  Thus, we make the
following ansatz
\begin{equation}
  \label{eq:Fnc}
  F_{nc}(\tau)=1+\frac{c(\tau)}{L(\tau)}\,.  
\end{equation}
The initial value $c(\tau=0)$ is determined from a physical boundary
condition such as the mean length $\bar{L}$ at $\tau=0$ which determines
the mean mass $\bar{m}$ of a $N=0$ CVL as $\bar{m}=\sigma\bar{L}$.  We
have also considered a modified factor
\begin{equation}
  \label{eq:Fncmod}
  F_{nc}(\tau)=1+\frac{c(\tau)}{A(\tau)}  
\end{equation}
in the ansatz for the `action' in Eq.~(\ref{eq:action}).

For later use, we investigate the behavior of $F_{nc}(\tau)$ for
$\tau\nearrow T$ for an ensemble consisting of a single curve only and
require the independence of the `partition function' under changes in
$\tau$.  Using Eq.~(\ref{eq:Aevolexpl}) in the vicinity of $\tau=T$,
where the limiting of curve is a circle with radius $R$, we have
\begin{equation}
  \label{eq:limL}
  L(\tau)=2\pi R=\sqrt{8}\pi\,\sqrt{T-\tau}\,.  
\end{equation}
Since $F_c(\tau\nearrow T)=4\pi$, independence of the `partition
function' under the flow in $\tau$ implies that
\begin{equation}
  \label{eq:limc}
  c(\tau)\propto\sqrt{T-\tau}\,.  
\end{equation}
That is, $F_{nc}$ approaches a constant value for $\tau\nearrow T$
which brings us back to the conformal limit by a finite
renormalization of the conformal part $F_c$ of the effective `action'.
In this parametrization of $S$, the coefficient $c(\tau)$ can thus be
regarded as an order parameter for conformal symmetry with a
mean-field critical exponent.

\subsection{Effective `action'}
\label{sec:0EA}

We now want to derive an effective `action' $S[\x(\tau)]$ resulting
from a partition function $Z$ for a nontrivial ensemble $E$.  The
partition function $Z_M$ is defined as the average
\begin{equation}
  \label{eq:pf}
  Z_M=\sum_{i=1}^M \exp\left(-S[\x_i(\tau)]\right)
\end{equation}
over the ensemble $E=\{\x_1,\dots\,\x_M\}$.  $E_M$ denotes an ensemble
consisting of $M$ curves where $E_M$ is obtained from $E_{M-1}$ by
adding a new curve $\x_M(u,\tau)$.  We are interested in a situation
where all curves in $E_M$ shrink to a point at the same value
$\tau=T$.  Because of $T=A_0/(2\pi)$, we demand that at $\tau=0$ all
curves in $E_M$ have the same initial area $A_0$.  The effective
`action' $S$ in Eq.~(\ref{eq:action}) (when associated with the
ensemble $E_M$ we will denote it as $S_M$, and the corresponding
coefficient $c_M$) is determined by the function $c_M(\tau)$, compare
with Eq.~(\ref{eq:Fnc}), the flow of which follows from the
requirement of $\tau$-independence of $Z_M$:
\begin{equation}
  \label{eq:renflow}
  \frac{d}{d\tau}Z_M=0\,.
\end{equation}
This is an implicit, first-order ordinary differential equation for
$c_M(\tau)$, which is in need for an initial condition
$c_{0,M}=c_M(\tau=0)$.  An obvious choice of initial condition is to
demand that the statistic mean length $\bar{L}(\tau)$, defined as
\begin{equation}
  \label{eq:statmeanL}
  \bar{L}_M(\tau)\equiv\frac{1}{Z_M(\tau)}\sum_{i=1}^M
  L[\x_i(\tau)]\,\exp\left(-S_M[\x_i(\tau)]\right),  
\end{equation}
coincides with the geometric mean length $\tilde{L}_M(\tau)$ defined
as
\begin{equation}
  \label{eq:geommeanL}
  \tilde{L}_M(\tau)\equiv\frac{1}{M}\sum_{i=1}^M L[\x_i(\tau)]
\end{equation}
at $\tau = 0$:
\begin{equation}
  \label{eq:ic}
  \bar{L}_M(0)=\tilde{L}_M(0).
\end{equation}
From this initial condition a value for $c_{0,M}$ follows.  In the
case of the modified `action' in Eq.~(\ref{eq:Fncmod}), the choice of
initial condition $\bar{L}_M(\tau=0)=\tilde{L}_M(\tau=0)$ leads to
$F_{nc}(\tau)\equiv 0$ which is equivalent to a uniform distribution.
This is because initial condition~(\ref{eq:ic}) is identically
fulfilled for the modified `action' if $c(0) = -A_0$ is chosen, then
setting $c(\tau) = -A(\tau)$ solves $\mathrm{d}Z_M/\mathrm{d}t=0$
trivially.  While the geometric effective `action' is thus profoundly
different for such a modification of $F_{nc}(\tau)$, physical results
such as the evolution of the variance of the position of the `center
of mass' agree remarkably well, see Sec.~\ref{sec:0COM}.  We conclude,
that the geometric effective `action' itself has no physical
interpretation in contrast to quantum field theory and conventional
statistical mechanics where the action in principle is related to the
physical properties of a given member of the ensemble.  Rather, going
from one ansatz for $S_M$ to another describes a particular way of
redistributing the weight in the ensemble which seems to have no
significant impact on the physics.

\section{Results of simulation}
\label{sec:0Simulation}

\subsection{Preparation of ensemble}
\label{sec:0POE}

\begin{figure}[th]
  \centering
  \includegraphics[width=0.95\textwidth]{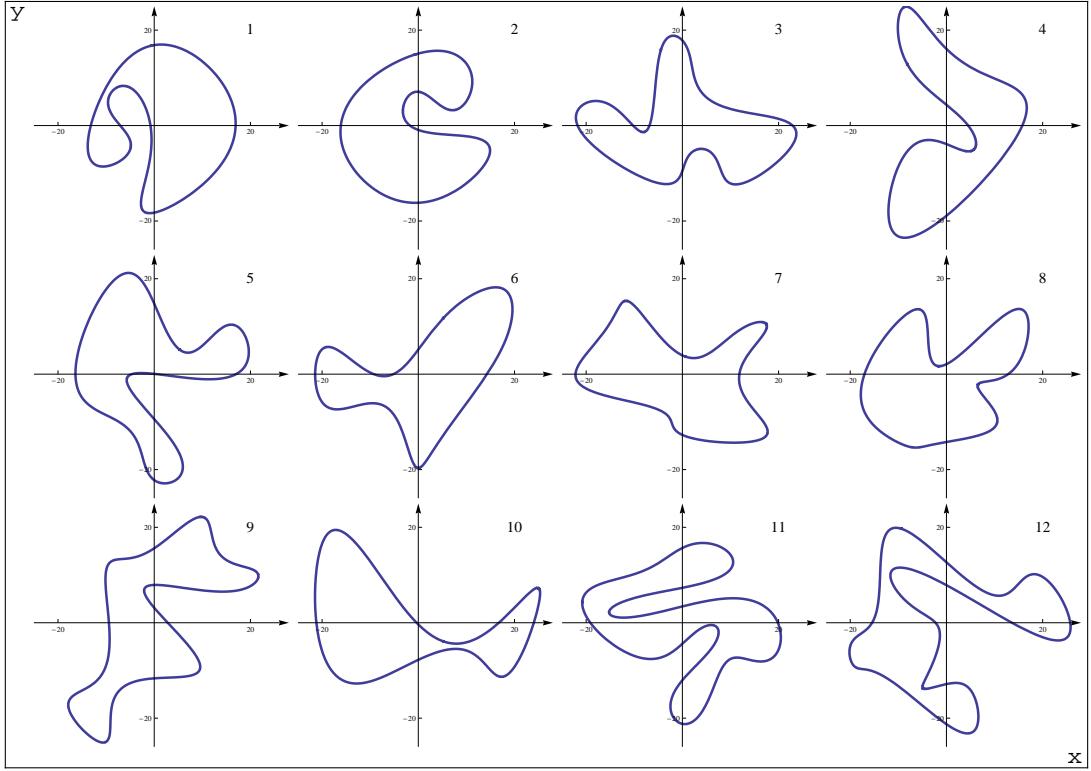}
  \caption{Initial curves contributing to the ensembles $E_M$. The
    positions of the `center of mass' coincide with the origin, and all
    curves have the same area $200\,\pi$.}
  \protect{\label{fig:0InitialCurves}}
\end{figure}

\begin{figure}
  \centering
  \includegraphics[width=0.95\textwidth]{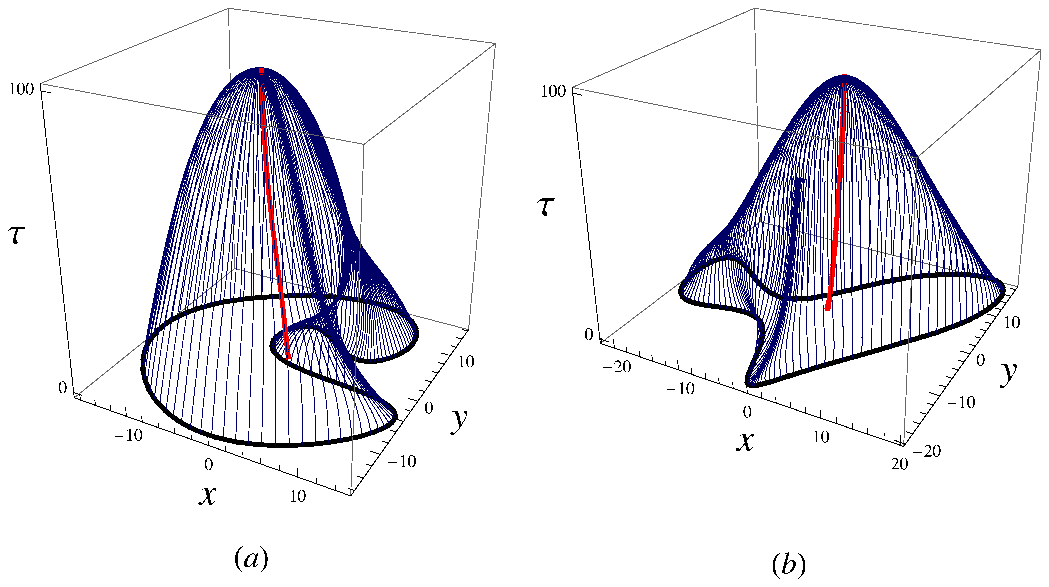}
  \caption{\protect{\label{fig:0EvolutionPlots}}Plots of the evolution
    of planar N=0 CVLs (curve 2 and 6 
    of Fig.~\ref{fig:0InitialCurves}) under the curve shortening
    flow. The thick central lines depict the trajectories of the
    `center of mass' (see Sec.~\ref{sec:0COM}) which coincides with
    the origin at $\tau=0$. The flow is started at $\tau=0$ and
    stopped at $\tau=100$.}
\end{figure}

For the curves depicted in Fig.~\ref{fig:0InitialCurves}, we make the
convention that $A_0\equiv 2\pi\times 100$.  It then follows that
$T=100$ by virtue of Eq.~(\ref{eq:A0}).  Furthermore, we have prepared
the ensembles such that the position of `center of mass' (COM)
coincides with the origin.  It should be recalled that such a
translation does not alter the effective `action' (Euclidean point
symmetry).  Also note that we use the same notation $E_M$ for the
primed and the unprimed ensemble.
In Fig.~\ref{fig:0EvolutionPlots}, the evolution of two different
initial curves under curve shrinking is shown.

\subsection{Numerical procedure}
\label{sec:0NI}

The initial curves depicted in \ref{fig:0InitialCurves} are generated
as follows.  First, we chose a list of points in the $ x\hyphen y$
plane such that the initial curve $\x(u,\tau=0)$ consecutively passes
the points, where initial and final point coincide.  A spline, which
is made of piecewise third-order polynomials with $C^1$ continuity,
interpolates each point in the list.  Since Mathematica's Spline\-Fit
sets the second derivatives of the spline at the endpoints to zero,
the first two points are appended to the end of the list.  The
analogue holds for the last two points of the original list.  To yield
a smooth curve with more than $C^1$ continuity each coordinate of the
curve is fitted by trigonometric functions up to order eight in its
Fourier-expansion.  Area, length and centroid of the initial curve are
computed numerically with NIntegrate, where the latter is given by
\begin{equation}
  \label{eq:centroid}
  \x_{\mbox{\tiny COM}} = \frac{1}{L} 
  \int_0^{2\pi}\mathrm{d}u\, \left|\pd_u\x\right|\,\x.
\end{equation}
Now we can prepare the ensembles as described in Sec.~\ref{sec:0POE}.

To simulate the flow evolution of the initial curves one searches for
solutions to the second-order partial differential equation
\begin{equation}
  \label{eq:flownum}
  \pd_{\tau}\x(u,\tau) = \frac{1}{|\pd_u\x(u,\tau)|}\pd_u 
  \frac{1}{|\pd_u\x(u,\tau)|}\pd_u \x(u,\tau)  
\end{equation}
subject to periodic boundary conditions in the curve parameter,
$\x(u=0,\tau)=\x(u=2\pi,\tau)$, and for the initial conditions
$\x(u,\tau=0)$ depicted in Fig.~\ref{fig:0InitialCurves}.  This is
done using the Numerical Method of Lines.  This is a technique for
solving PDEs by discretizing in all but one dimension, and then
integrating the semi-discrete problem as a system of coupled ordinary
differential equations (ODEs) or differential-algebraic equations.
Here, we partially discretize the flow equation Eq.~(\ref{eq:flownum})
on a uniform grid in the parameter $u$ yielding an ODE initial value
problem in $\tau$ that was solved by the ODE integrators in
Mathematica's NDSolve.  Fig.~\ref{fig:0EvolutionPlots} indicates why
this technique is called the method of lines.  As one can also see
from Fig.~\ref{fig:0EvolutionPlots}, a set of discrete points on the
curve, although remaining equidistant in $u$, may evolve under the
flow such that the spatial distances between adjacent points falls
below the numerical precision.  The flow then encounters a purely
numerically and thus virtual singularity (not to be confused with the
earlier mentioned non-virtual singularities at $\tau=T$).  Therefore,
the execution of NDSolve is broken up into several basic steps which
are carried out separately.  These steps are:

\noindent (i) equation processing and method selection,

\noindent (ii) method initialization,

\noindent (iii) numerical solution,

\noindent (iv) solution processing.

\noindent The low-level functions that are used in Mathematica to
break up these steps are ND\-Solve\`{}Proc\-ess\-Equations (i,ii),
NDSolve\`{}It\-er\-ate (iii) and NDSolve\`{}Proc\-ess\-So\-lu\-tions
(iv).  NDSolve\`{}Proc\-ess\-E\-qua\-tions classifies the differential
system into an initial-value problem, boundary-value problem,
differential-algebraic problem, partial-differential problem, etc.  It
also chooses appropriate default integration methods and constructs
the main NDSolve\`{}State\-Data data structure.
NDSolve\`{}It\-er\-ate advances the numerical solution.  The first
invocation initializes the numerical integration methods.
NDSolve\`{}Proc\-ess\-So\-lu\-tions converts numerical data into an
Interpolating\-Function to represent each solution.  More precisely,
the curve parameter range is divided into $n$ equidistant intervals
yielding $n$ points on the curve which are generally not equidistant
in space.  For our simulation the number of points $n$ is chosen
between 130 and 300.  The discretization of flow equation
(\ref{eq:flownum}) with respect to the variable $u$ needs to convert
the derivatives into finite differences.  The second-order centered
(with respect to to the set of sample points around $x(u_i)$) formula
for the first derivative is given by
\begin{equation}
  \label{eq:finitediff}
  x'(u_i)=\frac{x(u_{i+1})-x(u_{i-1})}{2h} +\order(h^2),
\end{equation}
where $h$ is the grid spacing.  Here, finite differences of sixth
order are used which are computed with Mathematica's
NDSolve\`{}Fi\-nite\-Dif\-fer\-ence\-De\-riv\-a\-tive.  In the
following, every quantity involving derivatives evaluated on a
discrete set of data points is computed using Mathematica's
NDSolve\`{}Fi\-nite\-Dif\-fer\-ence\-De\-riv\-a\-tive.  After
NDSolve\`{}Proc\-ess\-Equations is invoked the first time at $\tau=
\tau_1 =0$, the numerical solution is advanced using
NDSolve\`{}Iterate by a unit `time' step $\Delta\tau =1$ up to
$\tau_2$.  Then the computation is interrupted to compute an error
estimate that indicates whether a virtual singularity is starting to
evolve.  The error estimate exploits that $ A(\tau)= A_0 -2\pi\,\tau$
and is computed as
\begin{equation}
  \label{eq:0EE}
  10^4\times(A(\tau_2) -A(\tau_1) +2\pi(\tau_2-\tau_1)),
\end{equation}
where $A(\tau)$ is given by the discrete version of
Eq.~(\ref{eq:area}) evaluated on the point grid given by NDSolve.

Until $\tau_2$ reaches $T$, the solution is advanced step by step as
long as the error estimate does not exceed the empirically found value
of 2.  But if it does, the by then obtained solution is fitted at
$\tau_2-1$ in such a way that a new discretization yields (spatially)
well separated points to restart the procedure.  In
Fig.~\ref{fig:0EvolutionPlots}(b) such a situation is shown.  The
fitted curve is obtained as follows.  At $\tau_1$, one determines the
minimal arc length $s_{\mathrm{min}}$ which is the least of all arc length
between adjacent points on the curve.  Then, at $\tau_2 -1$, all those
points on the curve are dropped the arc length of which to their next
neighbors is less than the minimal arc length $s_{\mathrm{min}}$.  The
remaining points are fitted by trigonometric functions, where the
order of the fit is chosen to depend on $\tau$ (since the curve is
getting smoother with increasing $\tau$).  In the case of the error
estimate of the fitted curve exceeding the tolerance, the number of
grid points has to be increased or the initial curve needs to be
smoothed slightly.  In order to avoid discontinuities in the
$\tau$-evolution of $L$, $A$ and $x_{\mbox{\tiny COM}}$, and
singularities in their derivatives that occur since the fit procedure
generates piecewise defined functions, and since after the fit the
values of $A$ and $L$ slightly deviate from their former values, these
quantities are interpolated by polynomials for $0\le\tau\le T$ using
Find\-Fit.  To improve the accuracy of $L$ near the critical value
$T$, the isoperimetric ratio $\frac{L^2}{A}$ is fitted instead of $L$,
and $L$ is calculated from
$\sqrt{\left(\frac{L^2}{A}\right)_{\mbox{\tiny fitted}} \cdot A}$.

The analytical results of Sec.~\ref{sec:EC} such as the convergence of
$\frac{L^2}{A}$ to $4\pi$, the constancy of $\dot A$ or the vanishing
of $L$ and $A$ for $\tau\nearrow T$ are numerically well reproduced,
thereby confirming the validity of the simulation.

The implicit first-order differential equation
$\frac{\mathrm{d}Z}{\mathrm{d}\tau}=0$ for the coefficient is solved
using NDSolve.  If not set at will, the initial condition $c_0$ for
$c(\tau)$ was derived from Eq.~(\ref{eq:ic}) using Mathematica's
Find\-Root.  The variance of the position of COM was computed.  The
square of the coefficient $c(\tau)$ associated to the non-conformal
factor was fitted with function
\begin{equation}
  \label{eq:cfit}
  c(\tau)^2 = k(T_0 - \tau)^{\alpha},
\end{equation}
where $k$ and $\alpha$ are fit parameters.  We have determined the
critical exponent of the coefficient to $\frac{\alpha}{2}=0.5$ as
$\tau\rar T$, in accordance with the theoretical value of
Eq.~\ref{eq:limc}.  For checking purpose, we have also used $T_0$ as
fit parameter, yielding excellent agreement within the numerical
precision.

A CD-ROM containing the used Mathematica notebooks is attached to the
thesis\footnote{The results for non-selfintersecting curves were
  obtained using Mathematica version 6.0.2 or below.  Here, a remark
  concerning the used Mathematica version is in order.  Due to
  incomprehensible reasons version 6.0.3 is not capable to solve the
  implicit ODE for the coefficient $c(\tau)$, not even in the trivial
  case of an ensemble consisting of a single curve.  In the case of
  one-fold selfintersecting curves, version 6.0.3 still works and was
  used.}.

\subsection{Renormalization-group invariance of partition function}
\label{sec:0RGI}

\begin{figure}
  \centering
  \includegraphics[width=0.95\textwidth]{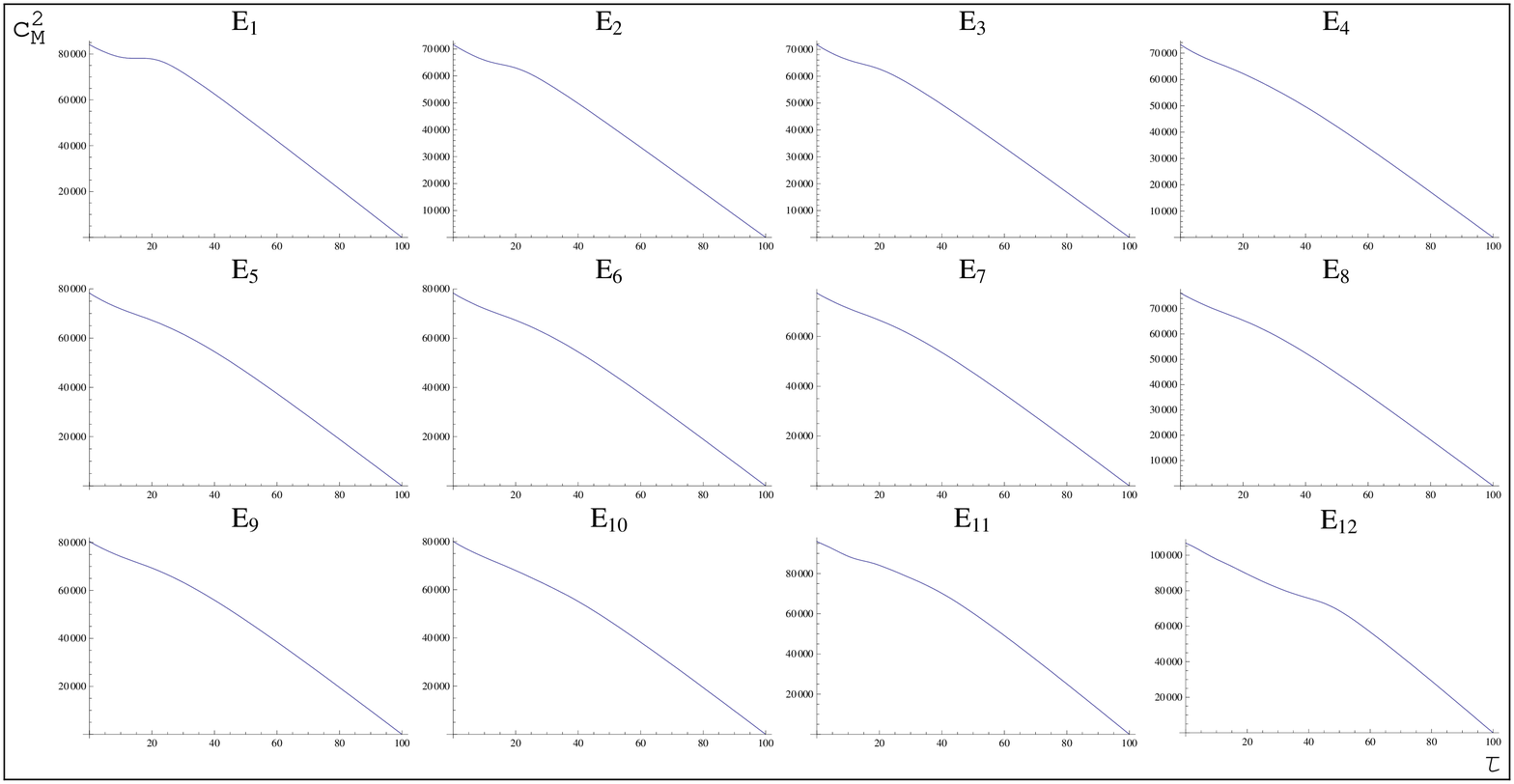}
  \caption{The square of the coefficient $c_M(\tau)$ entering the
    effective `action' $S_M =\frac{L(\tau)^2}{A(\tau)}\left(1
      +\frac{c_M(\tau)}{L(\tau)}\right)$ for various ensemble sizes
    $M=1,\dots,12$.  Notice the early onset of the linear drop of
    $c^2_M(\tau)$.  The slope of $c^2_M(\tau)$ near $\tau=T$ does not
    depend on $c_{0,M}^2\equiv c^2_M(\tau=0)$ and thus not on the
    initial choice of $\bar{L}$, but only on the specific choice of
    curves included in the ensemble.}
  \protect{\label{fig:0CoefficientSquared}}
\end{figure}

The function $c_M^2(\tau)$ is plotted in
Fig.~\ref{fig:0CoefficientSquared}.  According to
Fig.~\ref{fig:0CoefficientSquared} it seems that the larger the
ensemble the closer $c_M^2(\tau)$ is to the evolution of a single
circle of initial radius $R=\sqrt{\frac{A_0}{\pi}}$.  For growing $M$
the function $c_M^2(\tau)$ approaches the form
\begin{equation}
  \label{eq:casymp}
  c^2_{\mbox{\tiny as},M}(\tau)=k_M(T-\tau)\,, 
\end{equation}
where the slope $k_M$ depends on the strength of deviation from
circles of the representatives in the ensemble $E_M$ at $\tau=0$, that
is, on the variance $\Delta L_M$ at a given value $A_0$.  Physically
speaking, the value $\tau=0$ is associated with a certain initial
resolution of the measuring device (the strictly monotonic function
$\tau(Q)$, $Q$ being a physical scale such as energy or momentum
transfer, expresses the characteristics of the measuring device and
the measuring process), the value of $A_0$ describes the strength of
noise associated with the environment ($A_0$ determines how fast the
conformal limit of circular points is reached), and the values of
$c_{0,M}$ and $k_M$, see Eq.~(\ref{eq:casymp}), are associated with
the conditions at which the to-be-coarse-grained system is prepared.
Notice that this interpretation is valid for the `action'
\begin{equation*}
  S_M=\frac{L(\tau)^2}{A(\tau)}\left(1+\frac{c_M(\tau)}{L(\tau)}\right)  
\end{equation*}
only.
\begin{figure}
  \centering
  \includegraphics[width=0.95\textwidth]{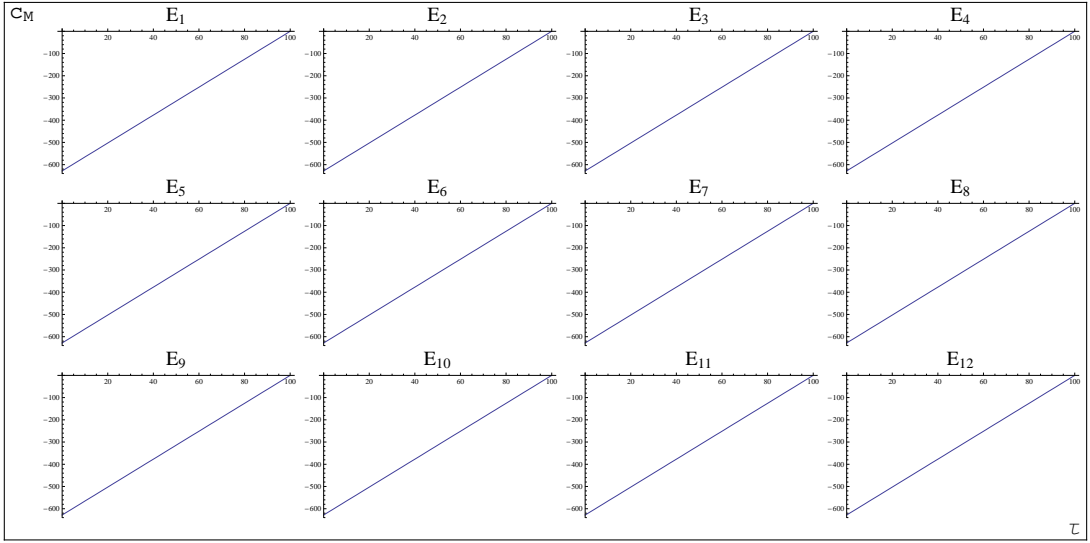}
  \caption{The coefficient $c_M(\tau)$ entering the effective `action'
    $S_M =\frac{L(\tau)^2}{A(\tau)}\left(1
      +\frac{c_M(\tau)}{A(\tau)}\right)$ for ensemble sizes
    $M=1,\dots,12$.}
  \protect{\label{fig:0Coefficient}}
\end{figure}

If we relax initial condition $\bar{L}_M(0)=\tilde{L}_M(0)$ for
$c_{0,M}$ and set the initial value for $c_{0,M}$ at will, the
coefficient starts at the given value and rapidly adapts to the
evolution depicted in Fig.~\ref{fig:0CoefficientSquared} and
respectively, Fig.~\ref{fig:0Coefficient}.  In Sec.~\ref{sec:0EA}, we
have argued that for the modified `action' and the initial condition $
\bar{L}_M(0)=\tilde{L}_M(0)$ the curves are uniformly distributed.
Relaxing this initial condition in the case of the modified `action',
means that the curves are no longer uniformly distributed for
$\tau=0$.  However, the uniform distribution is restored rapidly as
the curves evolve under the flow.

\subsection{Variance of mean `center of mass'}
\label{sec:0COM}

Having obtained the coefficient in the non-conformal factor of the
effective `action', we are now able to compute the flow of an
`observable', such as the COM position in a given ensemble and its
statistical variance.  The COM position $\x_{\mbox{\tiny{COM}}}$ of a
given curve $\x(s,\tau)$ is defined as
\begin{equation}
  \label{eq:com}
  \x_{\mbox{\tiny COM}}(\tau)
  =(x_{\mbox{\tiny COM}}(\tau),y_{\mbox{\tiny COM}}(\tau))^T
  =\frac{1}{L(\tau)}\int_0^{L(\tau)}
  \mathrm{d}s\,\x(s,\tau)\,.
\end{equation}
We will present below results on the statistical variance of the COM
position.

\begin{figure}
  \centering
  \includegraphics[totalheight=8cm,width=0.8\textwidth]{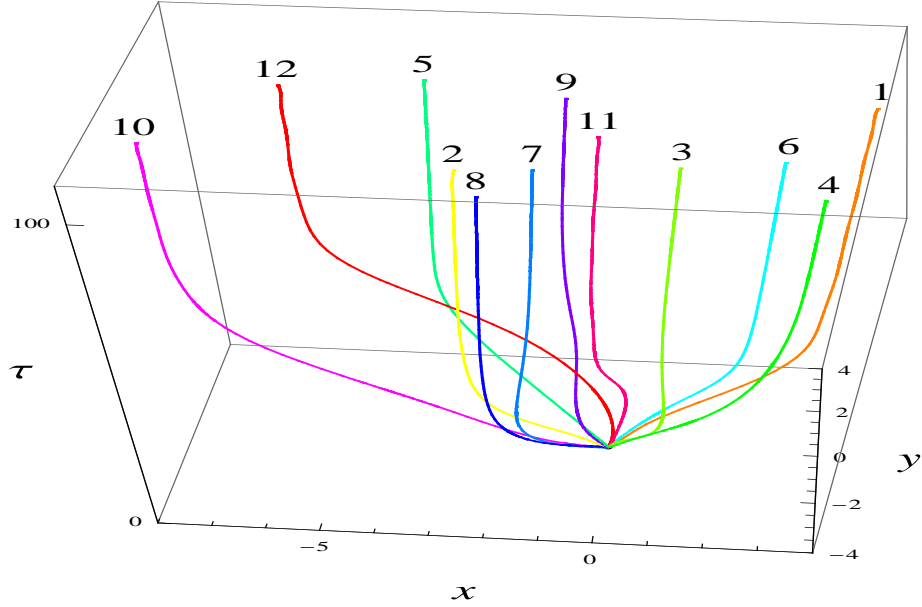}
  \caption{Flow of the positions of the `centers of masses' for the
    initial curves depicted in Fig.~\ref{fig:0InitialCurves}.}
  \protect{\label{fig:0TrajectoriesOfCOMs}}
\end{figure}

At $\tau=0$, the statistical variance in the position of the COM is
prepared to be nil, physically corresponding to an infinite resolution
applied to the system by the measuring device.  In
Fig.~\ref{fig:0TrajectoriesOfCOMs}, the flow of the COM position
corresponding to the initial curves depicted in
Fig.~\ref{fig:0InitialCurves} is shown.

\begin{figure}
  \centering
  \includegraphics[width=0.95\textwidth]{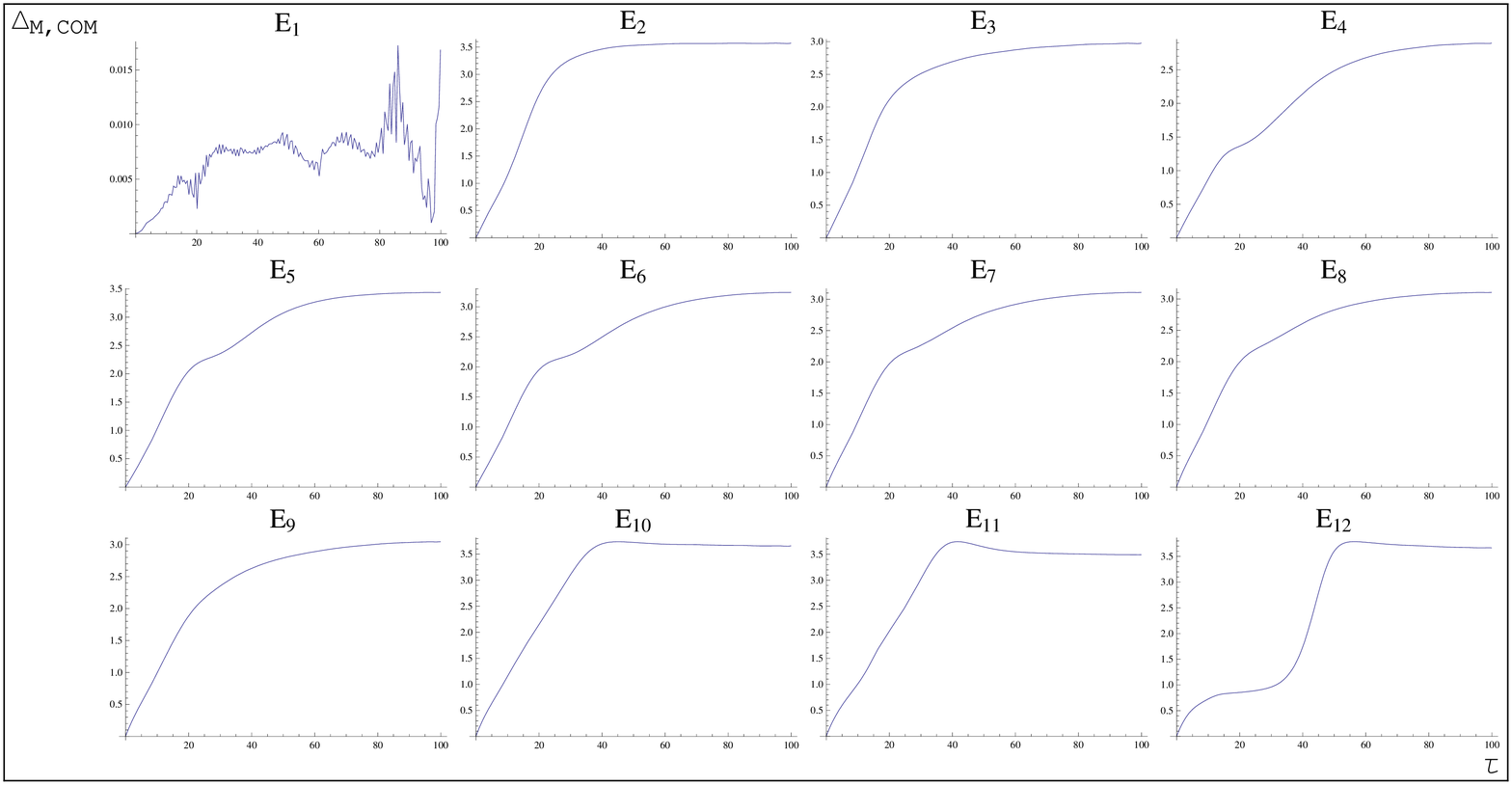}
  \caption{Plots of $\Delta_{M,\mbox{\tiny COM}}(\tau)$ for
    $M=1,\dots,12$ when evaluated with the `action' $S_M
    =\frac{L(\tau)^2}{A(\tau)}\left(1
      +\frac{c_M(\tau)}{L(\tau)}\right)$.  Notice the rapid generation
    of an uncertainty in the COM position under the flow and its
    saturation when approaching the conformal limit $\tau\nearrow
    T$. There also is a saturation of this limiting value with a
    growing ensemble size.}
  \protect{\label{fig:0VarianceOfCOMReg}}
\end{figure}

\begin{figure}
  \centering
  \includegraphics[width=0.95\textwidth]{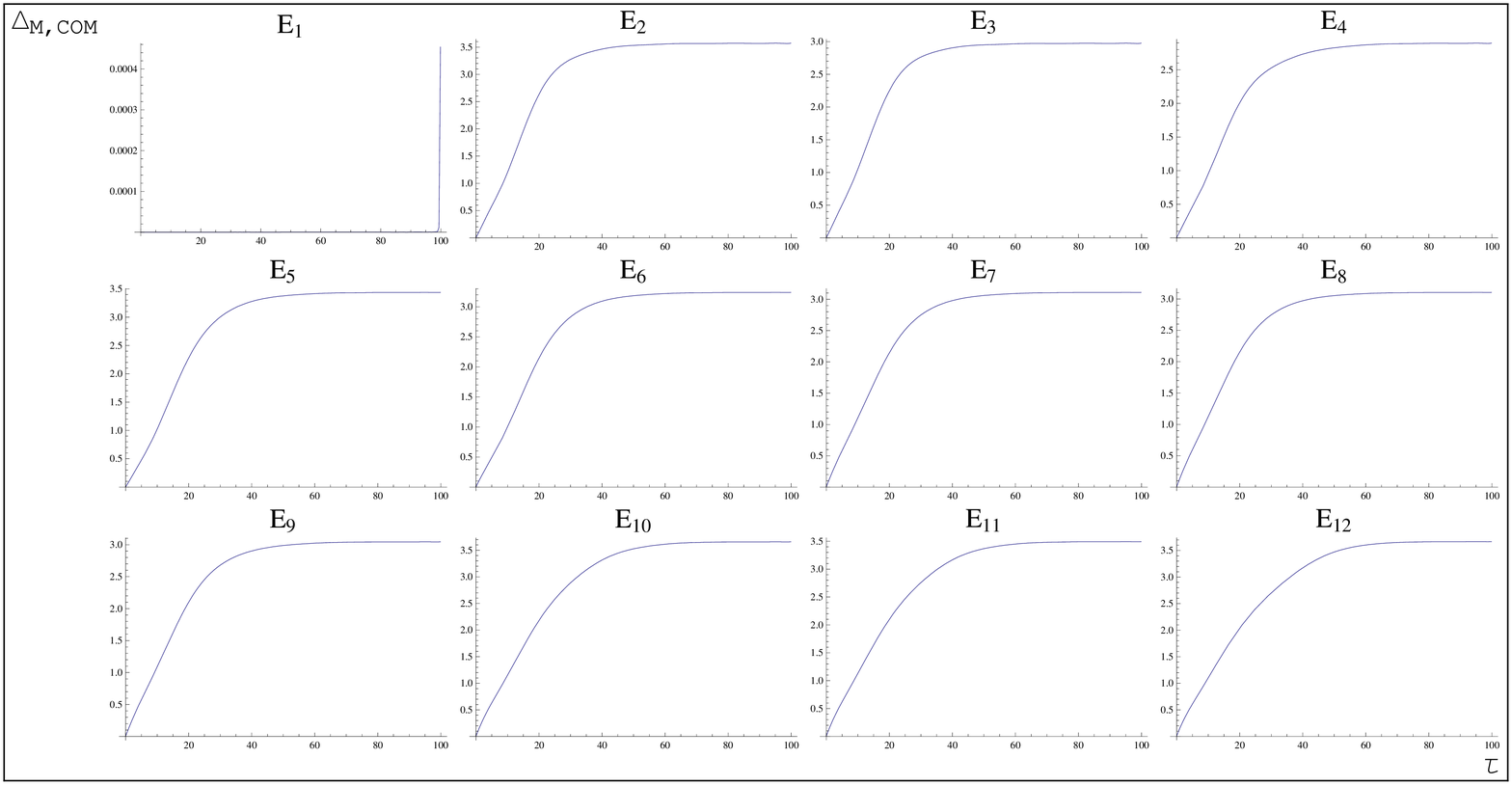}
  \caption{Plots of $\Delta_{M,\mbox{\tiny COM}}(\tau)$ for
    $M=1,\dots,12$ when evaluated with the modified `action' $S_M
    =\frac{L(\tau)^2}{A(\tau)} \left(1 + \frac{c_M(\tau)}{A(\tau)}
    \right)$.  This corresponds to a uniform distribution when
    evaluated with initial condition (\ref{eq:ic}). Notice the
    qualitative agreement with the results displayed in
    Fig.~\ref{fig:0VarianceOfCOMReg}.}
  \protect{\label{fig:0VarianceOfCOMAlt}}
\end{figure}

The mean COM position $\bar{\x}_{\mbox{\tiny COM}} $ over the ensemble
$E_M$ is defined as
\begin{equation}
  \label{eq:meancom}
  \bar{\x}_{\mbox{\tiny COM}}(\tau) 
  =(\bar{x}_{\mbox{\tiny COM}}(\tau),\bar{y}_{\mbox{\tiny COM}}(\tau))^T
  \equiv\frac{1}{Z_M}\sum_{i=1}^M \x_{\mbox{\tiny COM},i}(\tau)
  \exp\left(-S_M[\x_i(\tau)]\right)\,.
\end{equation}
The scalar statistical deviation $\Delta_{M,\mbox{\tiny COM}}$ of
$\bar{\x}_{\mbox{\tiny COM}}$ over the ensemble $E_M$ is defined as
\begin{equation}
  \label{eq:varcom}
  \Delta_{M,\mbox{\tiny COM}}(\tau)
  \equiv \sqrt{\mbox{var}_{M,\mbox{\tiny COM};x}(\tau) +
    \mbox{var}_{M,\mbox{\tiny COM};y}(\tau)}\,,
\end{equation}
where
\begin{eqnarray}
  \label{eq:varcomx}
  \mbox{var}_{M,\mbox{\tiny COM};x}&\equiv&\frac{1}{Z_M}\sum_{i=1}^M
  \left(x_{\mbox{\tiny COM},i}(\tau)
    -\bar{x}_{\mbox{\tiny COM}}(\tau)\right)^2\,
  \exp\left(-S_M[\x_i(\tau)]\right)\nonumber\\ 
  &=&-\bar{x}^2_{\mbox{\tiny COM}}(\tau)
  +\frac{1}{Z_M}\sum_{i=1}^M x^2_{\mbox{\tiny COM},i}(\tau)\,
  \exp\left(-S_M[\x_i(\tau)]\right) ,
\end{eqnarray}
and similarly for the coordinate $y$. In
Fig.~\ref{fig:0VarianceOfCOMReg}, plots of $\Delta_{M,\mbox{\tiny
    COM}}(\tau)$ are shown when $\Delta_{M,\mbox{\tiny COM}}(\tau)$ is
evaluated over the ensembles $E_1,\dots,E_{12}$ with the `action'
\begin{displaymath}
  S_M=\frac{L(\tau)^2}{A(\tau)}\left(1+\frac{c_M(\tau)}{L(\tau)}\right)
\end{displaymath}
and subject to the initial condition
$\bar{L}_M(\tau=0)=\tilde{L}_M(\tau=0)$. In
Fig.~\ref{fig:0VarianceOfCOMAlt}, the according plots of
$\Delta_{M,\mbox{\tiny COM}}(\tau)$ are depicted as obtained with the
modified `action'
\begin{displaymath}
  S_M=\frac{L(\tau)^2}{A(\tau)}\left(1+\frac{c_M(\tau)}{A(\tau)}\right)
\end{displaymath}
and subject to the initial condition
$\bar{L}_M(\tau=0)=\tilde{L}_M(\tau=0)$. In this case, one has
$c_M(\tau)=-A(\tau)$ leading to equal weights for each curve in $E_M$.
Note that the slight qualitative deviation of the last graph $E_{12}$
for small values of $\tau$ regarding to the previous graphs in
Fig.~\ref{fig:0VarianceOfCOMReg} is due to the fact that the curves
which were added to the ensemble at last are the most twisted ones.
Graph $E_{12}$ still saturates at a finite value of $\tau$,
nevertheless.  The fluctuations in graph $E_1$ of
Fig.~\ref{fig:0VarianceOfCOMReg} are within the range of the numerical
precision.

\subsection{Quantum mechanical versus statistical uncertainty}
\label{sec:0QMStat}

In view of the results obtained in the last section, we would say
that an ensemble of evolving planar CVLs in the $N=0$ sector
qualitatively resembles the Quantum Mechanics of a free point
particle\footnote{It is no relevance at this point whether this
  particle carries spin or not.} of mass $m$ in one space dimension
$x$.  Namely, an initially localized square of the wave function
$\psi$ with $|\psi(\tau=0,x)|^2\propto
\exp\left[-\frac{x^2}{a_0^2}\right]$, where $\Delta x(\tau=0)=a_0$,
according to unitary time evolution in quantum mechanics evolves as
\begin{equation}
  \label{eq:QMevolution}
  |\psi(\tau,x)|^2
  =|\exp\left[-\i\frac{H\tau}{\hbar}\right]\psi(\tau=0,x)|^2
  \propto \exp\left[-\frac{(x-\frac{p}{m}\tau)^2}{a^2(\tau)}\right], 
\end{equation}
where $H=\frac{p^2}{2m}$ is the free-particle Hamiltonian, $p$  the
spatial momentum, and $a(\tau)\equiv
a_0\sqrt{1+\left(\frac{\hbar\tau}{m a_0^2}\right)^2}$.  In agreement
with Heisenberg's uncertainty relation, one has during the evolution that
\begin{equation}
  \label{eq:Heisenberg}
  \Delta x\Delta p 
  = \frac{\hbar}{2} \sqrt{1 +\left(\frac{\tau\hbar}{m a_0^2}\right)^2} 
  \ge \frac{\hbar}{2}.
\end{equation}
The time evolution in a quantum mechanical system and the process of
lowering the resolution in a statistical system describing planar CVLs
share the same property: the `time' (resolution) evolution generates
out of a small initial position uncertainty (corresponding to a large
initial resolution $\Delta p$) a larger position uncertainty as `time'
increases (resolution decreases).  Possibly, future development will
show that interference effects in Quantum Mechanics can be traced back
to the non-local nature of the degrees of freedom (CVLs) entering a
statistical partition function.

\cleardoublepage


\chapter{Selfintersecting center-vortex loops}
\label{chap:1CVL}

Let us now turn to the case of $N=1$ CVLs.  We proceed as far as
possible in close analogy to the $N=0$ sector.

\section{Wilsonian renormalization-group flow}
\label{sec:1WRGF}

\subsection{Geometric partition function}
\label{sec:1GPF}

As in the $N=0$ sector, we interpret curve-shrinking as a Wilsonian
renormalization-group flow.  The partition function is now defined
over an ensemble of $N=1$ CVLs, and we consider it to be independent
under a change of resolution $Q$ and thus independent of $\tau$.  We
express the effective `action' in terms of integrals over local
densities in $s$, and demand the following symmetries in order to
conceive an ansatz for the effective `action':

\noindent (i) scaling symmetry $\x\to \lambda\x\,,\ 
\lambda\in{\mathbb R}_+$: for $\lambda\to\infty$, implying $\lambda
L\to\infty$ at fixed $L$, the `action' $S$ should be invariant under
further finite rescalings (decoupling of the fixed length scales
$\sigma^{-1/2}$ and $\Lambda^{-1}$).

\noindent (ii) Euclidean point symmetry of the plane: this is
sufficiently satisfied for a representation of $S$ in terms of
integrals over scalar densities with respect to these symmetries.
Thus, we can represent the `action' density as a series involving
products of Euclidean scalar products of $\frac{\pd^n}{\pd s^n}\x\,,\
 n\in\mathbb{N}_+\,,$ or constancy.

As in Sec.~\ref{sec:0GPF}, we resort to a factorization ansatz as
\begin{equation}
  \label{eq:1action}
  S=F_c\times F_{nc},
\end{equation}
where in addition to Euclidean point symmetry $F_c$ ($F_{nc}$) is (is
not) invariant under $\x\to \lambda\x$.  In principle, infinitely many
operators can be defined to contribute to $F_c$.  Since the evolution
homogenizes the curvature, except for a small vicinity of the
intersection point where one or both loops of the curve vanish,
higher derivatives of $k$ with respect to $s$ should not be of
importance. This should also hold for Euclidean scalar products
involving higher derivatives $\frac{\pd^n}{\pd s^n}\x$.  Conformally
invariant expressions are obtained from such integrals if multiplied
by powers of $\sqrt{A}$ and/or $L$ or the inverse of
integrals involving lower derivatives.  The conformal factor $F_c$ is
set equal to the isoperimetric ratio,
\begin{equation}
  \label{eq:1Fc}
  F_c(\tau)\equiv\frac{L(\tau)^2}{A(\tau)}\,.  
\end{equation}
The property of conformal invariance for $L,A\to\infty$ suggests to
express the non-conformal factor $F_{nc}$  as a formal expansion
in inverse powers of $L$ or $A\equiv A_1+A_2$.  We allow for an
explicit $\tau$ dependence of the coefficient $c$ of the lowest
nontrivial power $\frac{1}{L}$ or $\frac{1}{A}$.  In principle, this
sums up the contribution to $F_{nc}$ of certain higher-power operators
which do not exhibit an explicit $\tau$ dependence.

We restrict to the following two ans\"atze for the non-conformal
factor in Eq.~(\ref{eq:1action}),
\begin{equation}
  \label{eq:1Fnc}
  F_{nc}(\tau)=1+\frac{c(\tau)}{L(\tau)}\,. 
\end{equation}
and for the modified `action'
\begin{equation}
  \label{eq:1Fncmod}
  F_{nc}(\tau)=1+\frac{c(\tau)}{A(\tau)}\,.  
\end{equation}
The initial value $c(\tau=0)$ is determined from the physical boundary
condition such as the mean length $\bar{L}$ at $\tau=0$.  Although
the modified ansatz (\ref{eq:1Fncmod}) in $F_{nc}$ of the geometric `action' is
profoundly different physical results such as the evolution of entropy
or the variance of intersection of a given ensemble agree remarkably
well, see Sec.~\ref{sec:1Simulation}

\subsection{Effective `action'}
\label{sec:1EA}

The effective `action' $S_M[\x(\tau)]$ results from a partition
function $Z_M$ which is defined as the average
\begin{equation}
  \label{PartZM}
  Z_M=\sum_{i}^M \exp\left(-S_M[\x_i(\tau)]\right) 
\end{equation}
over the nontrivial ensemble $E_M=\{\x_1,\dots\,\x_M\}$.  The ensemble
$E_M$, consisting of $M$ curves, is obtained from $E_{M-1}$ by adding
a new curve $\x_M(\tau)$.  The effective `action' $S_M$ in
Eq.~(\ref{eq:1action}) is determined by the function $c_M(\tau)$, the
flow of which follows from the requirement of $\tau$-independence of
the partition function:
\begin{equation}
  \label{eq:1renflow}
  \frac{\mathrm{d}}{\mathrm{d}\tau}Z_M=0.
\end{equation}
As in Sec.~\ref{sec:0EA}, we obtain the initial condition
$c_{0,M}=c_M(\tau=0)$ to this implicit first-order ordinary
differential equation by the constraint that the geometric mean
coincides with the statistic mean at $\tau=0$,
\begin{equation}
  \label{eq:1ic}
  \bar{L}_M(0)=\tilde{L}_M(0).  
\end{equation}

\section{Results of simulation}
\label{sec:1Simulation}

\subsection{Preparation of ensembles}
\label{sec:1POE}

Similar to Sec.~\ref{sec:0POE}, all curves are normalized curves to
have the same initial total area $A_0=A_{0,1}+A_{0,2}$ and since we
are now interested in the position of the intersection where the
(anti)monopole is localized (see Sec.~\ref{sec:ConPhase} or
Fig.~\ref{fig:SectorTransition}), we have applied a translation to
each curve in the ensemble $E_M$ such that the location of the
intersections initially coincide with the origin.  Again, such a
transition does not alter the effective `action' due to Euclidean
point symmetry.

We order the members of the maximal-size ensemble $E_{M=16}$ into
sub-ensembles $E_{M<16}$ such that $T_{i=1}\ge T_{i=2}\ge\dots\ge
T_M$, because the critical value $T$ of the flow parameter $\tau$
varies from curve to curve.  These ensembles $E_M$ are referred to as
$T$-ordered.  We have also performed all simulations with ensembles
$E^\prime_{M<16}$ the members of which are picked randomly from
$E_{M=16}$ and have obtained similar results for ensemble averages of
`observables' using $E_{M<16}$ and $E^\prime_{M<16}$ for the $\tau$
evolution to the left of $\tau=\min\{T_i|\x_i\in E^\prime_{M<16}\}$.
The main difference is that the computation of the coefficient, and
with it the flow of `observables', terminates at a smaller $\tau$
since the ensembles $E^{\prime}_{M}$ are no longer $T$-ordered.
\begin{figure}
  \centering
  \includegraphics[width=0.95\textwidth]{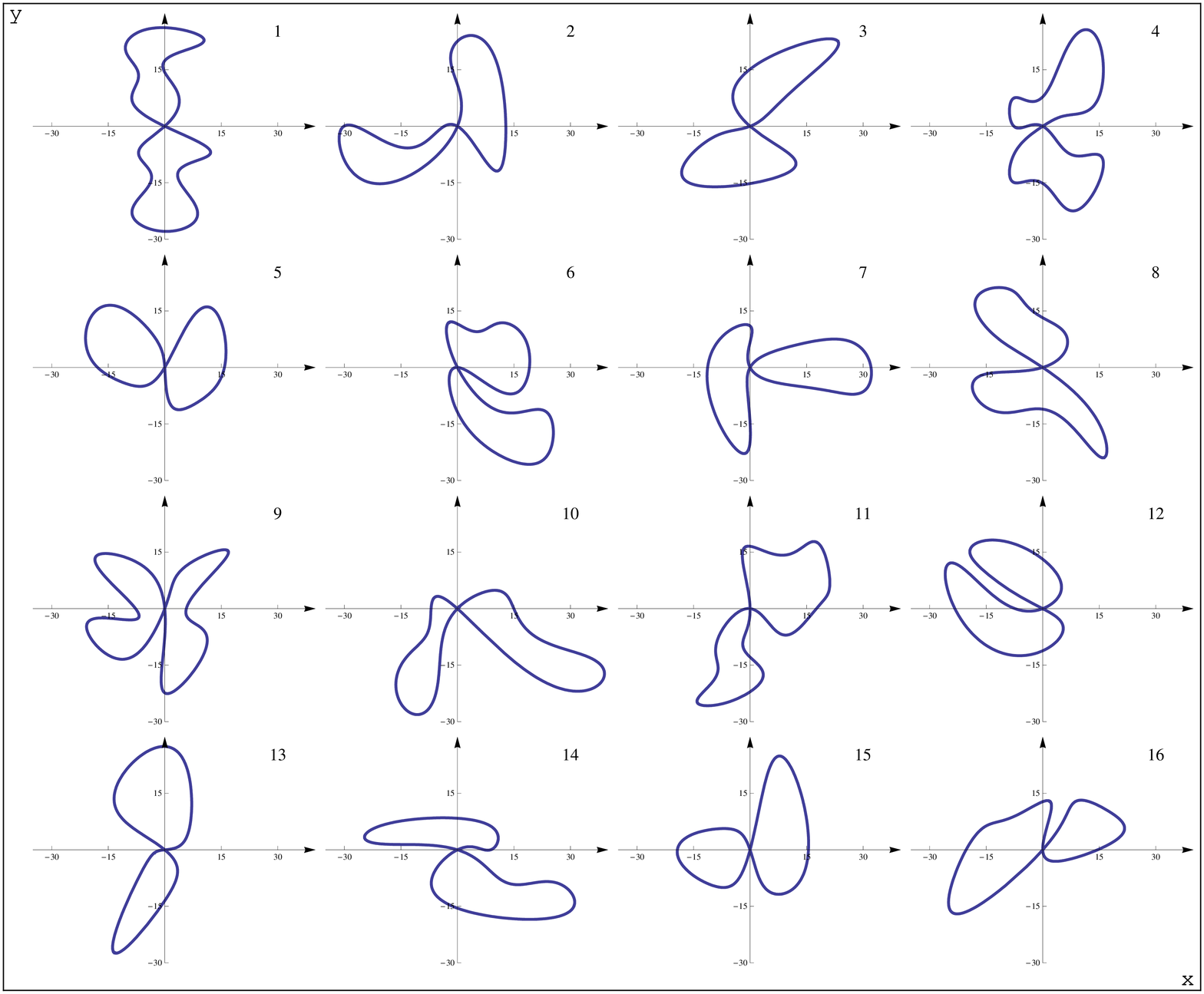}
  \caption{Initial curves $\x_i(u,\tau=0)$ contributing to the
    ensemble $E_{M=16}$. The intersection points $x_{{\mbox{\tiny
          int}},i}(\tau=0)$ coincide with the origin, and all curves
    have the same area $200\,\pi$.  By definition $E_{M=16}$ is
    $T$-ordered.}
  \protect{\label{fig:1InitialCurves}}
\end{figure}

\begin{figure}
  \centering
  \includegraphics[totalheight=9cm,width=1\textwidth]{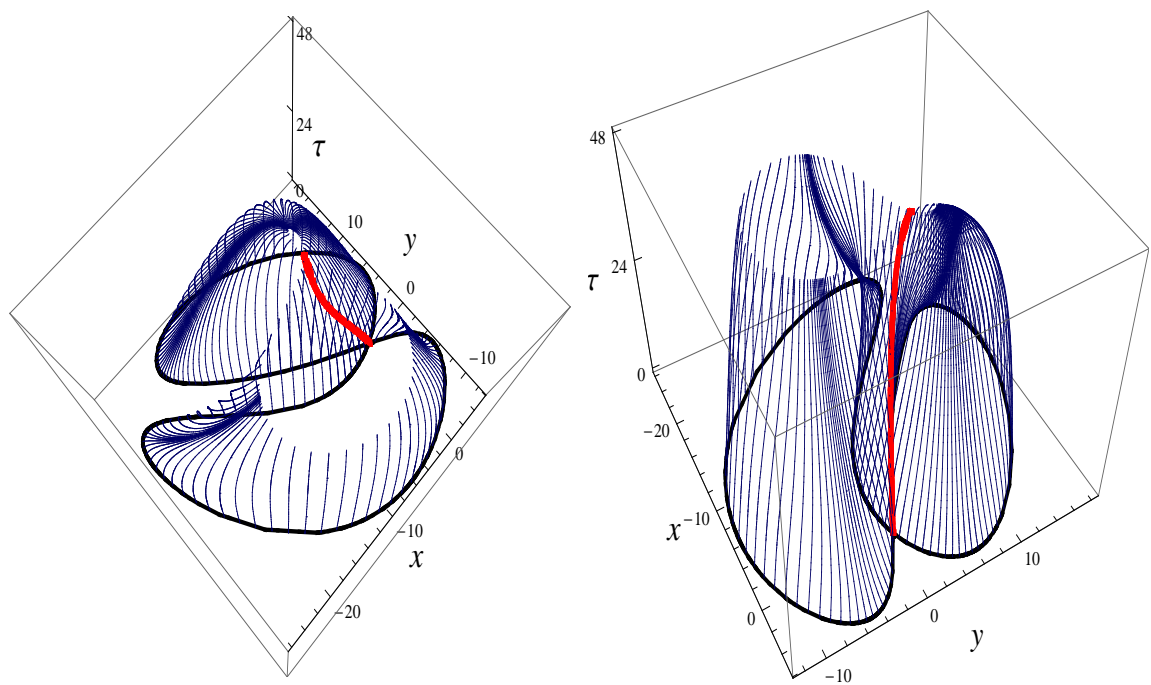}
  \caption{Plots of the evolution of an $N=1$ center-vortex loop
    (curve 12 of Fig.~\ref{fig:1InitialCurves}) under curve
    shrinking. The thick central line depicts the trajectory of the
    intersection point which coincides with the origin at $\tau=0$.}
  \protect{\label{fig:1EvolutionPlots}}
\end{figure}

\begin{figure}
  \centering
  \includegraphics[totalheight=9cm,width=0.95\textwidth]{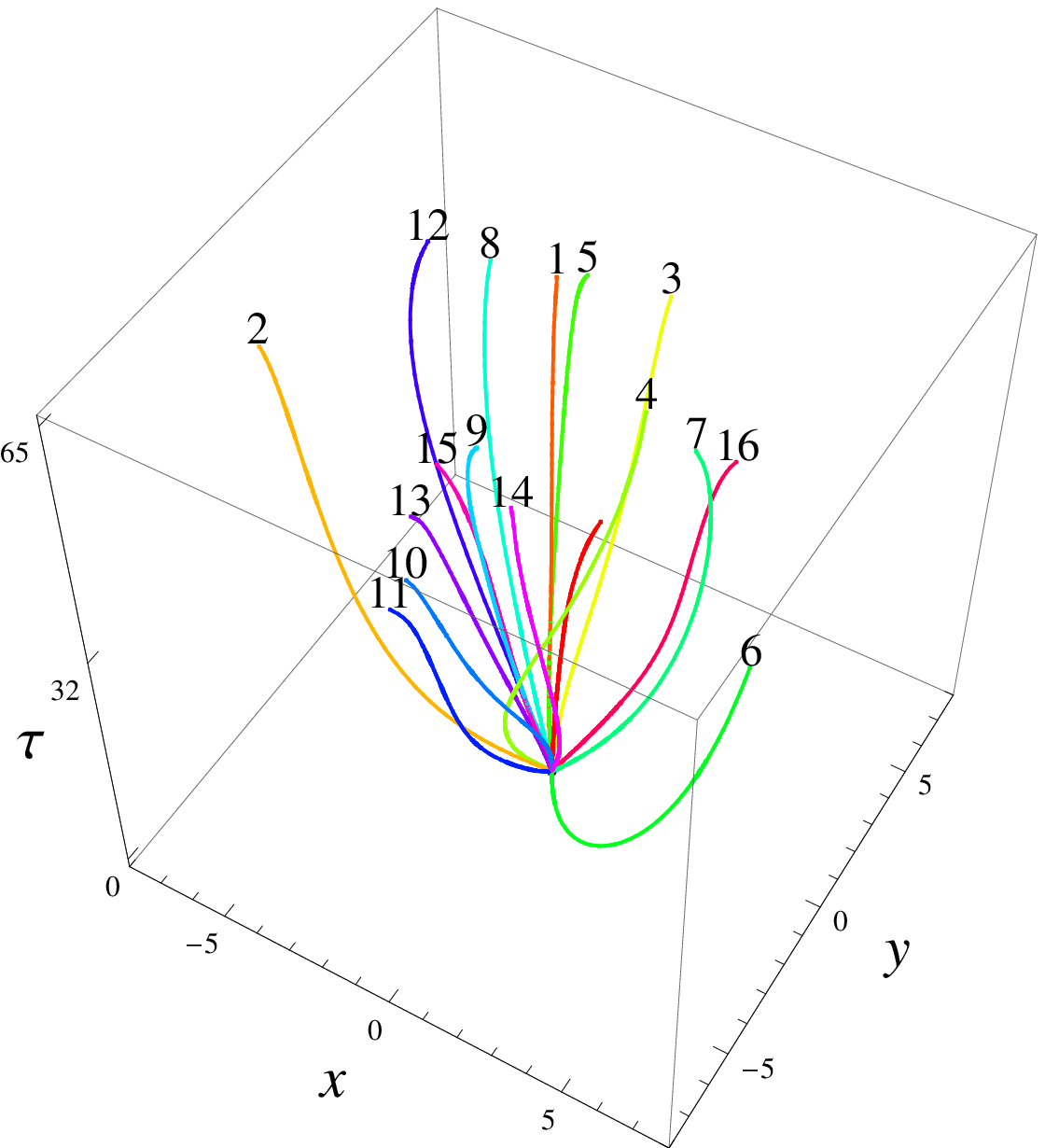}
  \caption{Flow of the intersection points
    $\x_{{\mbox{\tiny{int}}},i}(\tau)$ for the initial curves depicted
    in Fig.~\ref{fig:1InitialCurves}.}
  \protect{\label{fig:1TrajectoriesOfIntersections}}
\end{figure}

The maximal-size ensemble $E_{M=16}$ at $\tau=0$ is depicted in
Fig.~\ref{fig:1InitialCurves} with the universal choice
$A_0=200\,\pi$.  The curves in Fig.~\ref{fig:1InitialCurves} are
arranged in a $T$-ordered way.  We have $T_{i=1}=65\ge
T_{2}\ge\dots\ge T_M=43$.  In Fig.~\ref{fig:1EvolutionPlots}, the
evolution of an initial curve (number 12 of
Fig.~\ref{fig:1InitialCurves}) under curve shrinking is shown from two
points of view.  The flow is started at $\tau=0$ and stopped at a
value of $\tau$ shortly below $T$.
In Fig.~\ref{fig:1TrajectoriesOfIntersections}, the flow of the
intersection points $\x_{\mbox{\tiny int},i}(\tau)$ corresponding to
the initial curves depicted in Fig.~\ref{fig:1InitialCurves} is shown.

\subsection{Numerical investigation}
\label{sec:1NI}

In general, the procedure is in close analogy to Sec.~\ref{sec:0NI}.
Therefore, we solely mention the differences compared to
Sec.~\ref{sec:0NI}.  The choice of sample points, used to generate the
initial curves, is now done in such a way that, if interpolated one
after another by a cubic spline with coinciding initial and end point,
the curve will cross itself once.  The number of grid points $n$ is
set to 300 for all curves, except for curve 9 where 500 points are
used.  For the simplicity of the computation of the intersection of
the initial curve, the intersection point is included in the set of
sample points as a double point and is chosen to coincide with the
origin.

The search for solutions to the flow equation is proceeded as in
Sec.~\ref{sec:0NI} using the Numerical Method of Lines.  Starting from
$\tau=\tau_1=0$ the solution is advanced step by step as long as
$\tau<T$ or a virtual singularity evolves.  If the latter is the case
we start the same fitting procedure as in Sec.~\ref{sec:0NI}.  The
error estimate of Sec.~\ref{sec:0NI}, based on the constant value of
$\dot A$, is no longer applicable in the case of selfintersecting
curves, however, $\Delta A(\tau)=A_1(\tau)-A_2(\tau)
=\mathrm{constant}$ can serve to estimate a value that indicates the
validity of the numerical solution.  The error estimate is defined as
\begin{equation}
  \label{eq:1EE}
  10^6\times\left(1-\frac{\Delta A(\tau)}{\Delta A(\tau_1)}\right).
\end{equation}
In the case where both areas of the curve have almost the same value,
and the absolute of $A$ falls below the numerical precision, we have
used
\begin{equation}
  \label{eq:1EEalt}
  10^6\times(\Delta A(\tau_1)-\Delta A(\tau)).
\end{equation}
Since Eq.~(\ref{eq:diffA}) involves the two-dimensional curl the
straight forward discretization of this equation already computes the
(signed) difference $\Delta A$ of the areas enclosed by the curve.
Therefore, we do not have to keep track of the selfintersection during
the numerical evolution of the curve.  However, the computation of the
total area involves knowledge of $A_1$ and $A_2$, and this in turn of
the intersection point.  Once the error estimate exceeds a value of 2 the
curve is fitted to yield again spatially well separated points.  We also
used the error estimate to recognize the final (non-virtual) singularity
at $T$, where a further evolution in the sense of the flow equation is
impossible and does not make sense.

To compute the position of intersection point one searches, at a first
step, for those two points in the solution set at given $\tau$ which are
spatially nearest to each other, but with the restriction that they are
element of different line segments of the curve which generate the
intersection point.  Therefore, the spatial distances between the $i^{t
  h}$ and $j^{t h}$ point on the curve are determined for all pairs of
points with $|i-j|>d_{\mathrm{min}}$, where $d_{\mathrm{min}}$ depends
on the considered curve and has to be adjusted for each individually.
Then it is searched for the least of all distances to find that pair of
points which is closest to the intersection point.  The minimal distance
of indices $d_{\mathrm{min}}$ is introduced because, as the flow evolves
the curve, next neighboring points could become (spatially) closer to
each other than the points nearest to the intersection.  Once the pair
of points which is next to the intersection is found, the two curve
segments around these points are approximated by cubic splines.  Now the
intersection of these two splines is computed using Mathematica's
Find\-Root

At given $\tau$, length and area of the curves are computed with their
discrete formulas using Mathematica's Finite\-Difference\-Derivative.
For the same reasons as in Sec.~\ref{sec:0NI} the $\tau$-evolution of
$L$, $A$ and $\x_{\mbox{\tiny int}}$ are interpolated by polynomials.

Finally, the implicit first-order differential equation
$\mathrm{d}Z/\mathrm{d}\tau=0$ for the coefficient is solved using NDSolve for all
ensembles sizes and orderings and for both `actions'.  If not set at
will, the initial condition $c_0$ for $c(\tau)$ was derived from
Eq.~(\ref{eq:1ic}) using Mathematica's FindRoot.  Variance of mean
intersection and entropy were computed.

A CD-ROM containing the used Mathematica notebooks is attached to the
thesis\footnote{Mathematica version 6.0.3 was used.  Pay attention to
  the footnote in Sec.~\ref{sec:0NI}.}.

\subsection{Renormalization-group invariance of partition function}
\label{sec:1RGI}

\begin{figure}
  \centering
  \includegraphics[totalheight=8.5cm,width=0.95\textwidth]{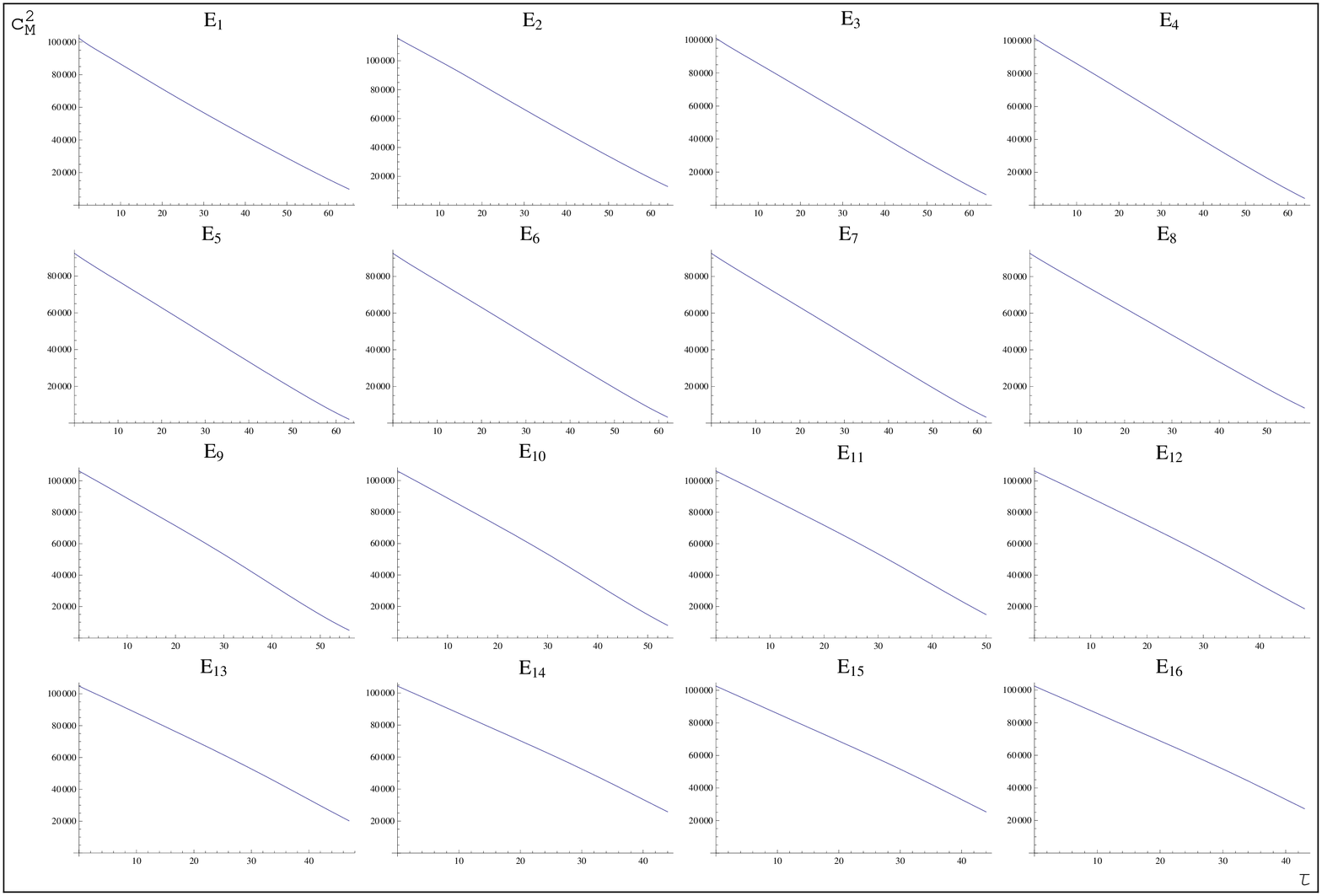}
  \caption{The squares of the coefficients $c_M(\tau)$ entering the
    ansatz for effective `action' $S_M = \frac{L(\tau)^2}{A(\tau)}
    \left(1 + \frac{c_M(\tau)}{L(\tau)}\right)$ for $T$-ordered
    ensembles up to $M=16$.}
  \protect{\label{fig:1CoefficientSquared}}
\end{figure}

\begin{figure}
  \centering
  \includegraphics[totalheight=8.5cm,width=0.95\textwidth]{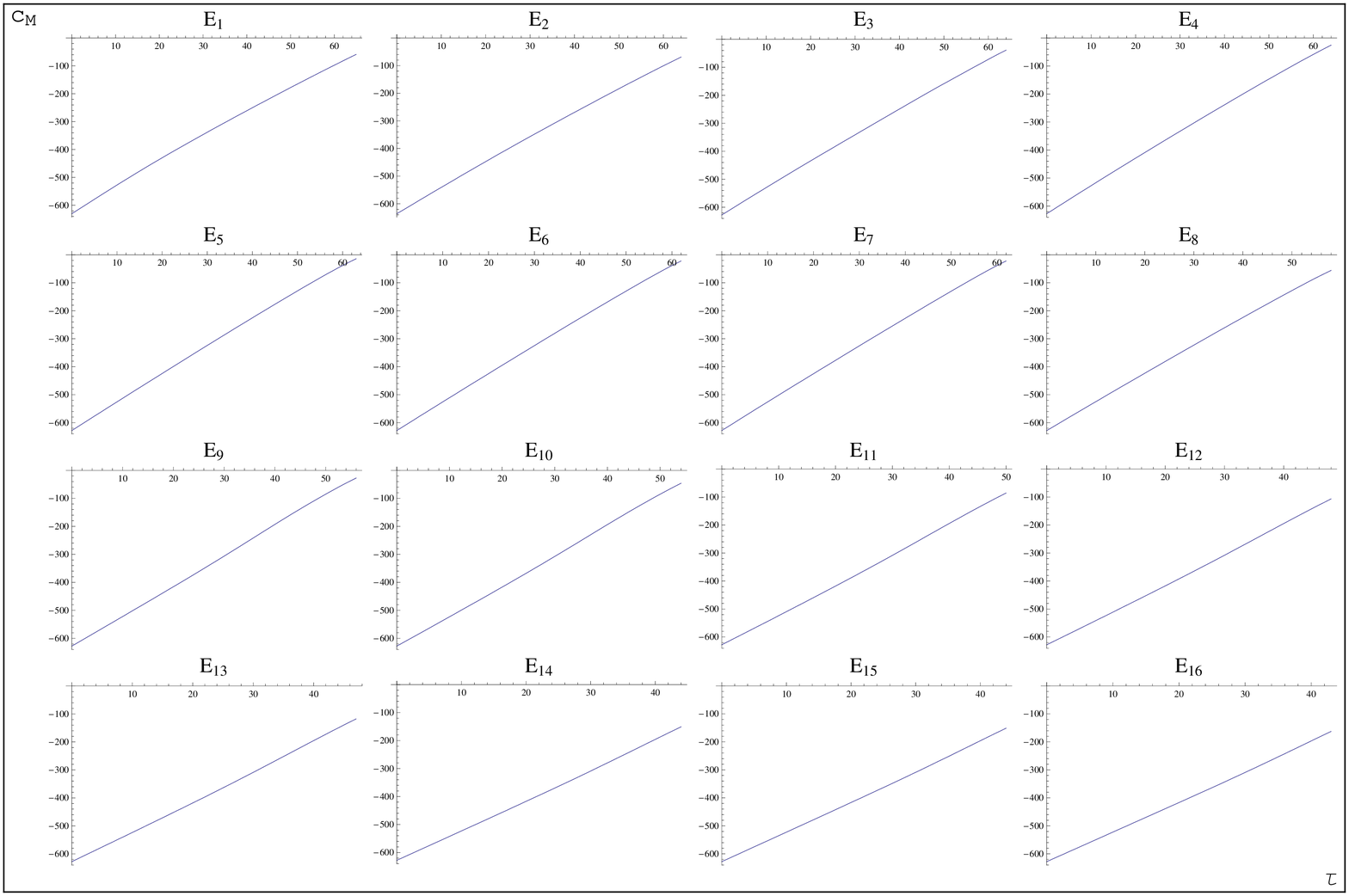}
  \caption{The coefficient $c_M(\tau)$ entering the ansatz for the
    effective `action' $S_M = \frac{L(\tau)^2}{A(\tau)} \left(1 +
      \frac{c_M(\tau)}{A(\tau)}\right)$ for $T$-ordered ensembles up
    to $M=16$.}
  \protect{\label{fig:1Coefficient}}
\end{figure}

We now present the results of the simulation.  For all ensembles
$E_M$, the $\tau$ dependence of the coefficient $c_M$ in
Eq.~(\ref{eq:1Fnc}) roughly behaves like a square root
$\propto\sqrt{T_M-\tau}$ where $T_M$ is the weakly ensemble-dependent
minimal resolution.  For the modified `action'
$S_M=\frac{L(t)^2}{A(t)}\left (1+\frac{c_M(t)}{A(t)}\right)$ the
coefficient $c_M(\tau)$ is well approximated by a linear function
$\propto T_M-\tau$.  Again, $T_M$ is a weakly ensemble-dependent
minimal resolution.  For $T$-ordered ensembles the results for
$c_M(\tau)$ for the `actions' Eq.~(\ref{eq:1Fnc}) and
Eq.~(\ref{eq:1Fncmod}) are shown in Fig.~\ref{fig:1CoefficientSquared}
and respectively, in Fig.~\ref{fig:1Coefficient}.  The results for
the ensembles $E^\prime_M$ do not differ sizably from those presented
in Fig.~\ref{fig:1CoefficientSquared} and respectively, in
Fig.~\ref{fig:1Coefficient}.

\subsection{Variance of location of selfintersection}
\label{sec:1Variance}

\begin{figure}
  \centering
  \includegraphics[totalheight=8.5cm,width=0.95\textwidth]{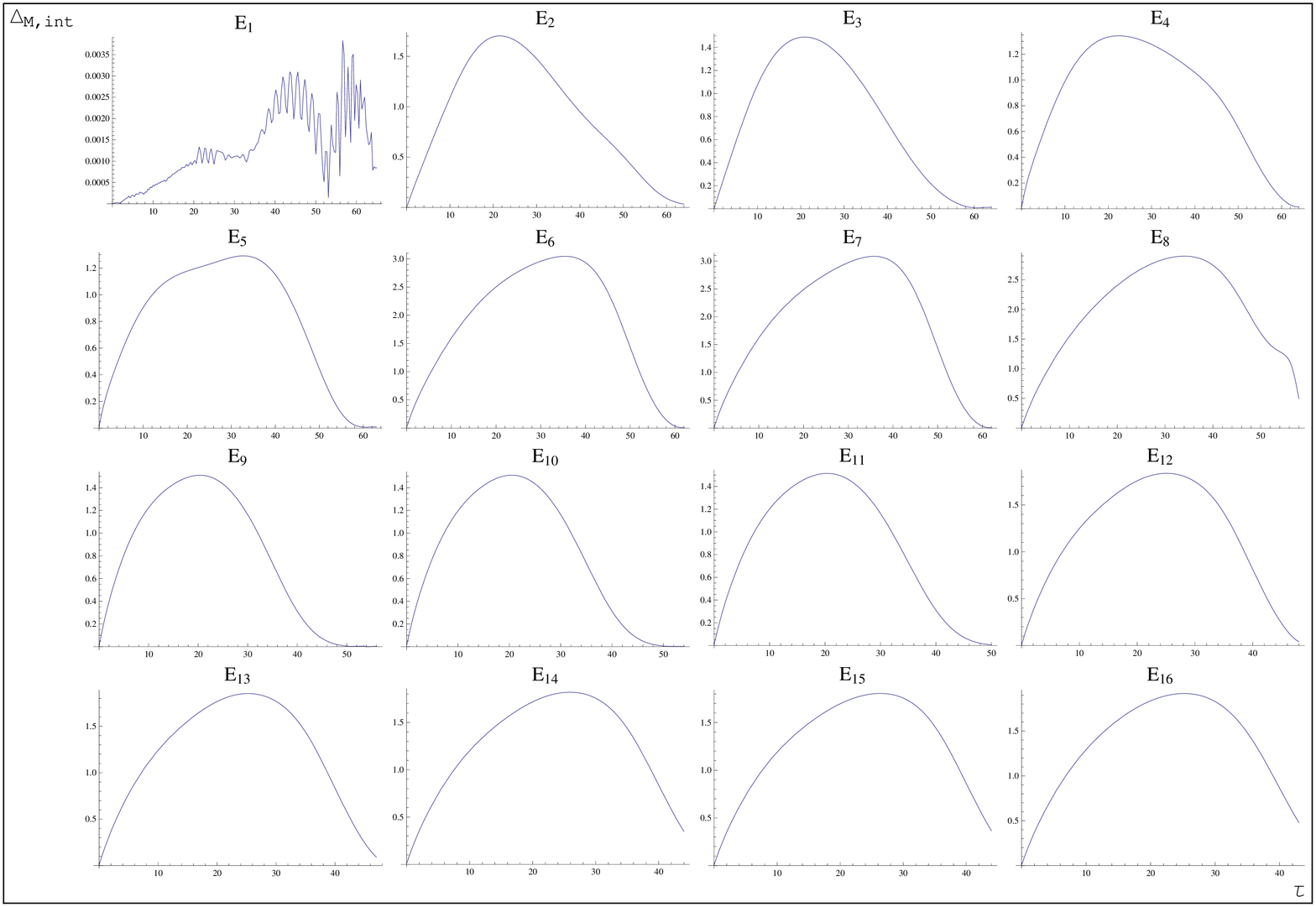}
  \caption{Plots of $\Delta_{M,\mbox{\tiny int}}(\tau)$ for the
    $T$-ordered ensembles $E_M$ with $M=1,\dots,16$.  We have employed
    the ansatz for the `action' $S_M = \frac{L(\tau)^2}{A(\tau)}
    \left(1 + \frac{c_M(\tau)}{L(\tau)} \right)$.}
  \protect{\label{fig:1VarianceOfMeanIntersectionReg}}
\end{figure}

\begin{figure}
  \centering
  \includegraphics[totalheight=8.5cm,width=0.95\textwidth]{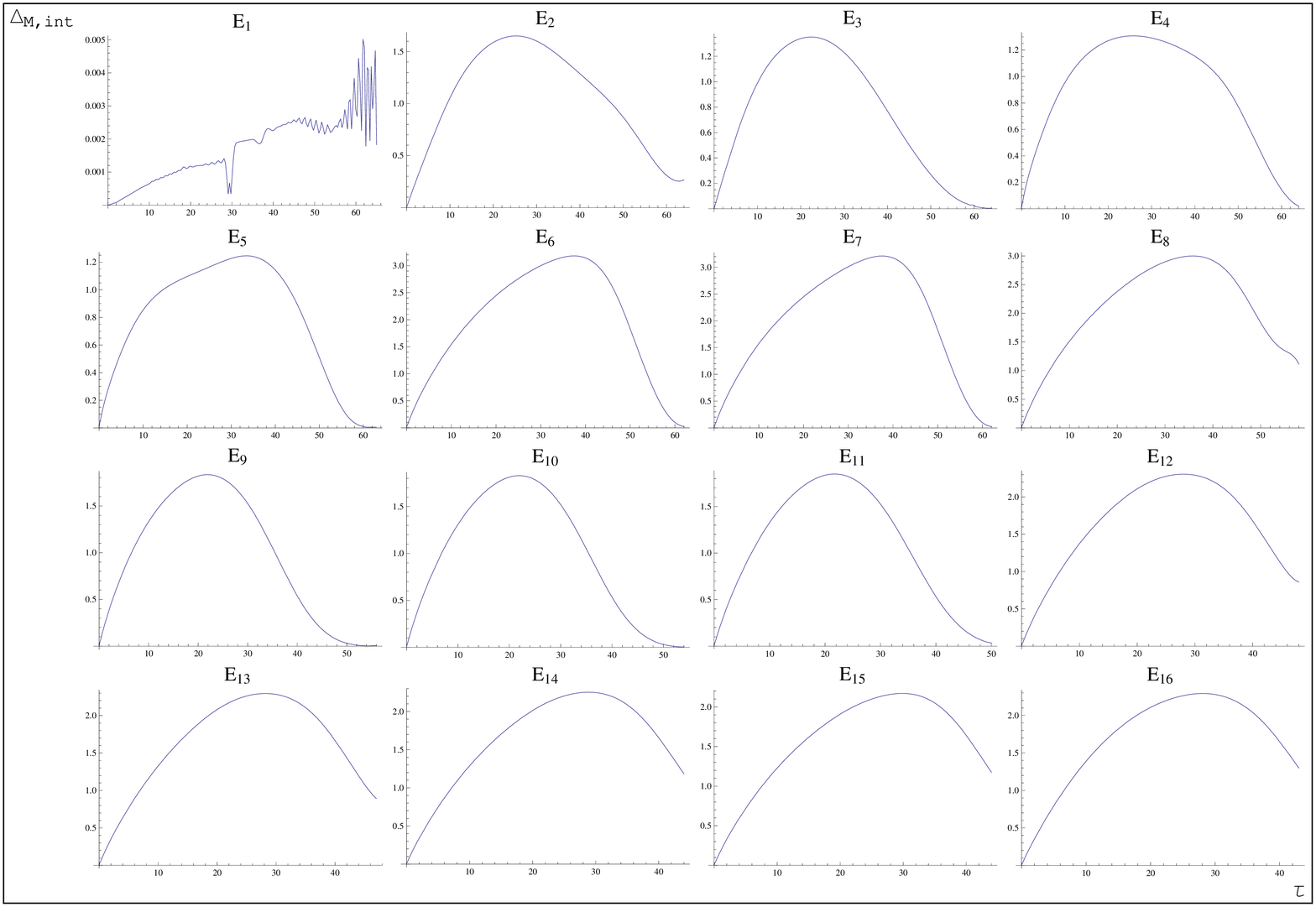}
  \caption{Plots of $\Delta_{M,\mbox{\tiny int}}(\tau)$ for the
    $T$-ordered ensembles $E_M$ with $M=1,\dots,16$.  We have employed
    the ansatz for the modified `action' $S_M =
    \frac{L(\tau)^2}{A(\tau)} \left(1 + \frac{c_M(\tau)}{A(\tau)}
    \right)$.}
  \protect{\label{fig:1VarianceOfMeanIntersectionAlt}}
\end{figure}

The mean intersection $\bar{\x}_{\mbox{\tiny int}}(\tau)$ over the
ensemble $E_M$ is defined as
\begin{equation}
  \label{eq:meanint}
  \bar{\x}_{\mbox{\tiny int}}(\tau) 
=(\bar{x}_{\mbox{\tiny int}}(\tau),\bar{y}_{\mbox{\tiny int}}(\tau))^T
\equiv \frac{1}{Z_M}\sum_{i=1}^M
  \x_{\mbox{\tiny int},i}(\tau)\exp\left(-S_M[\x_i(\tau)]\right)\,,
\end{equation}
where $\x_{\mbox{\tiny int},i}(\tau)=(x_{\mbox{\tiny int}}(\tau),
y_{\mbox{\tiny int}}(\tau))^T$ is the location of the point of
selfintersection of curve $\x_i$ at $\tau$.  The scalar statistical
deviation $\Delta_{M,\mbox{\tiny int}}$ of $\bar{\x}_{\mbox{\tiny
    int}}$ over the ensemble $E_M$ is defined as
\begin{equation}
  \label{eq:varint}
  \Delta_{M,\mbox{\tiny int}}(\tau)\equiv
  \sqrt{\mbox{var}_{M,\mbox{\tiny int};x}(\tau)+
    \mbox{var}_{M,\mbox{\tiny int};y}(\tau)}\,,
\end{equation}
where
\begin{eqnarray}
  \label{eq:variance}
  \mbox{var}_{M,\mbox{\tiny int};x} &\equiv &\frac{1}{Z_M}\sum_{i=1}^M
  \left(x_{\mbox{\tiny int},i}(\tau)
    -\bar{x}_{\mbox{\tiny int}}(\tau)\right)^2\, 
  \exp\left(-S_M[\x_i(\tau)]\right)
  \nonumber\\ &=& -\bar{x}^2_{\mbox{\tiny int}}(\tau) +\frac{1}{Z_M}\sum_{i=1}^M
  x^2_{\mbox{\tiny int},i}(\tau)\, \exp\left(-S_M[\x_i(\tau)]\right)  
\end{eqnarray}
and similarly for the coordinate $y$. In
Fig.~\ref{fig:1VarianceOfMeanIntersectionReg}, plots of
$\Delta_{M,\mbox{\tiny int}}(\tau)$ are shown when evaluated over the
ensembles $E_1,\dots,E_{16}$ subject to the `action'
\begin{displaymath}
  S_M=\frac{L(\tau)^2}{A(\tau)}\left(1+\frac{c_M(\tau)}{L(\tau)}\right)
\end{displaymath}
and the initial condition $\bar{L}_M(\tau=0)=\tilde{L}_M(\tau=0)$. In
Fig.~\ref{fig:1VarianceOfMeanIntersectionAlt}, the according plots of
$\Delta_{M,\mbox{\tiny int}}(\tau)$ are depicted as obtained with the
`action'
\begin{displaymath}
  S_M=\frac{L(\tau)^2}{A(\tau)}\left(1+\frac{c_M(\tau)}{A(\tau)}\right)
\end{displaymath}
and subject to the initial condition
$\bar{L}_M(\tau=0)=\tilde{L}_M(\tau=0)$.
Relaxing the constraint of $T$-ordering ($E_M\to E^\prime_M$) does not
entail a qualitative change of the results.  The fluctuation in the
first graph of Fig.~\ref{fig:1VarianceOfMeanIntersectionReg} and in
Fig.~\ref{fig:1VarianceOfMeanIntersectionAlt}, representing the
trivial ensemble $E_{M=1}$ is within the range of the numerical
precision.  The results presented in
Fig.~\ref{fig:1VarianceOfMeanIntersectionReg} and
Fig.~\ref{fig:1VarianceOfMeanIntersectionAlt} are unexpected since in
the $N=0$ sector the variance of the `center of mass' saturates
rapidly to finite values.  In contrast, for the $N=1$ sector, the
variance of the location of selfintersection initially increases,
reaches a maximum, and decreases to zero at a \emph{finite} value of
$\tau$.  This is readily confirmed by the evaluation of the entropy,
see next section.

\subsection{Evolution of entropy}
\label{sec:1Entropy}

\begin{figure}
  \centering
  \includegraphics[totalheight=8.5cm,width=0.95\textwidth]{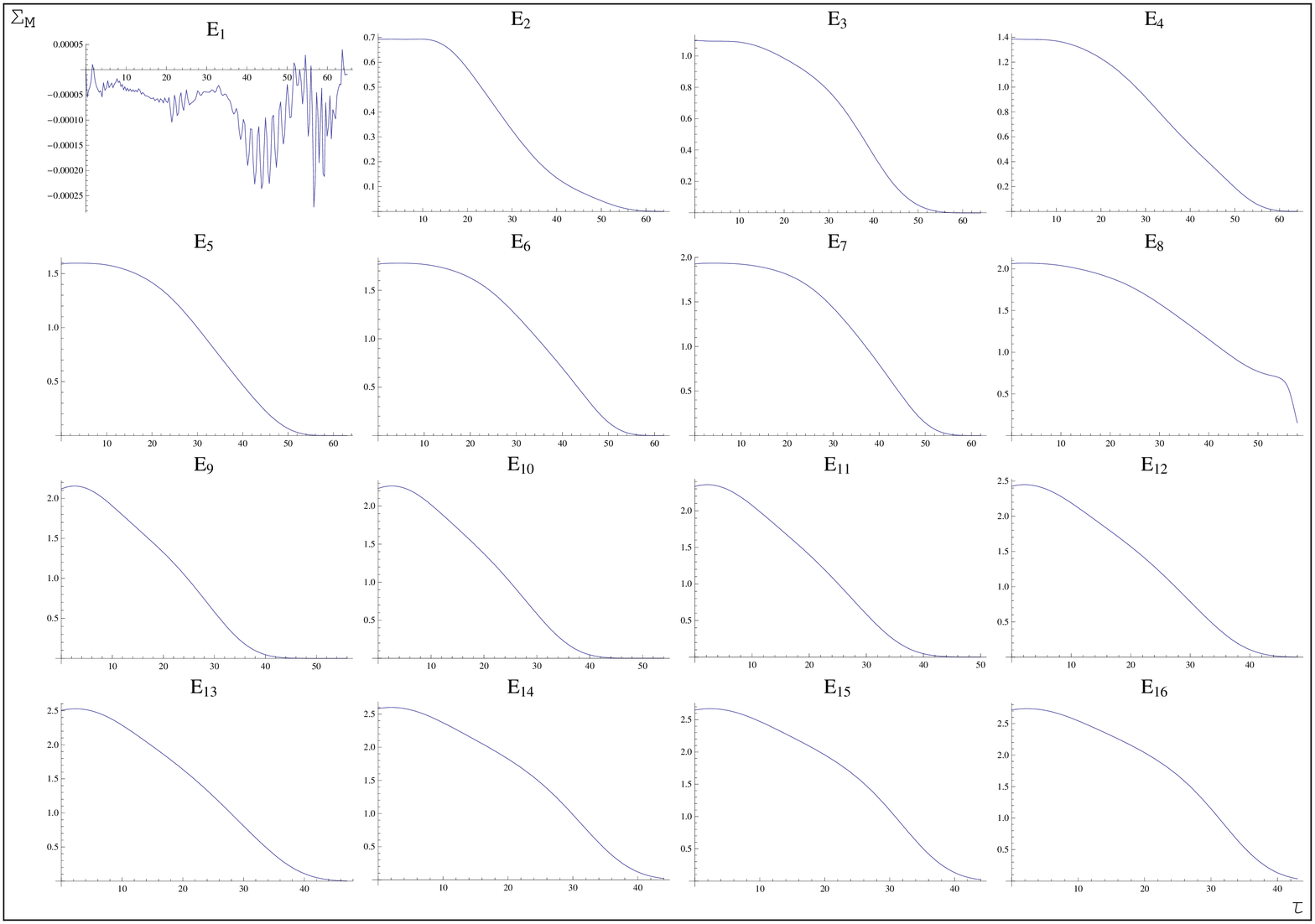}
  \caption{Flow of the entropies $\Sigma_M$ for $T$-ordered ensembles
    of size $M=1,\dots,16$ when evaluated with the `action' $S_M
    =\frac{L(\tau)^2}{A(\tau)}\left(1
      +\frac{c_M(\tau)}{L(\tau)}\right)$.}
  \protect{\label{fig:1EntropyReg}}
\end{figure}

\begin{figure}
  \centering
  \includegraphics[totalheight=8.5cm,width=0.95\textwidth]{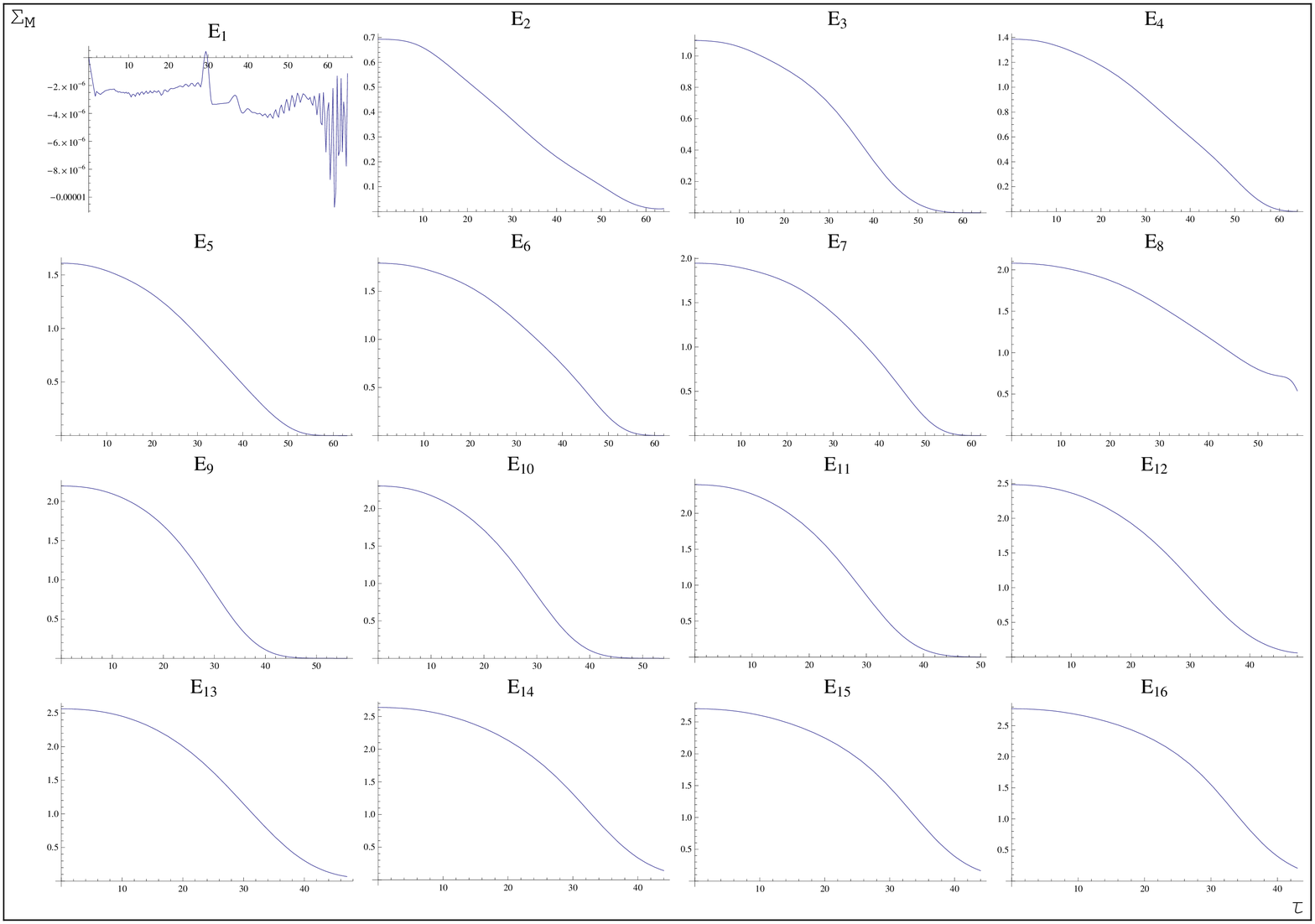}
  \caption{Flow of the entropies $\Sigma_M$ for $T$-ordered ensembles
    of size $M=1,\dots,16$ when evaluated with the modified `action'
    $S_M =\frac{L(\tau)^2}{A(\tau)}\left(1
      +\frac{c_M(\tau)}{A(\tau)}\right)$.}
  \protect{\label{fig:1EntropyAlt}}
\end{figure}

\begin{figure}[t!]
  \centering
  \includegraphics[width=\textwidth]{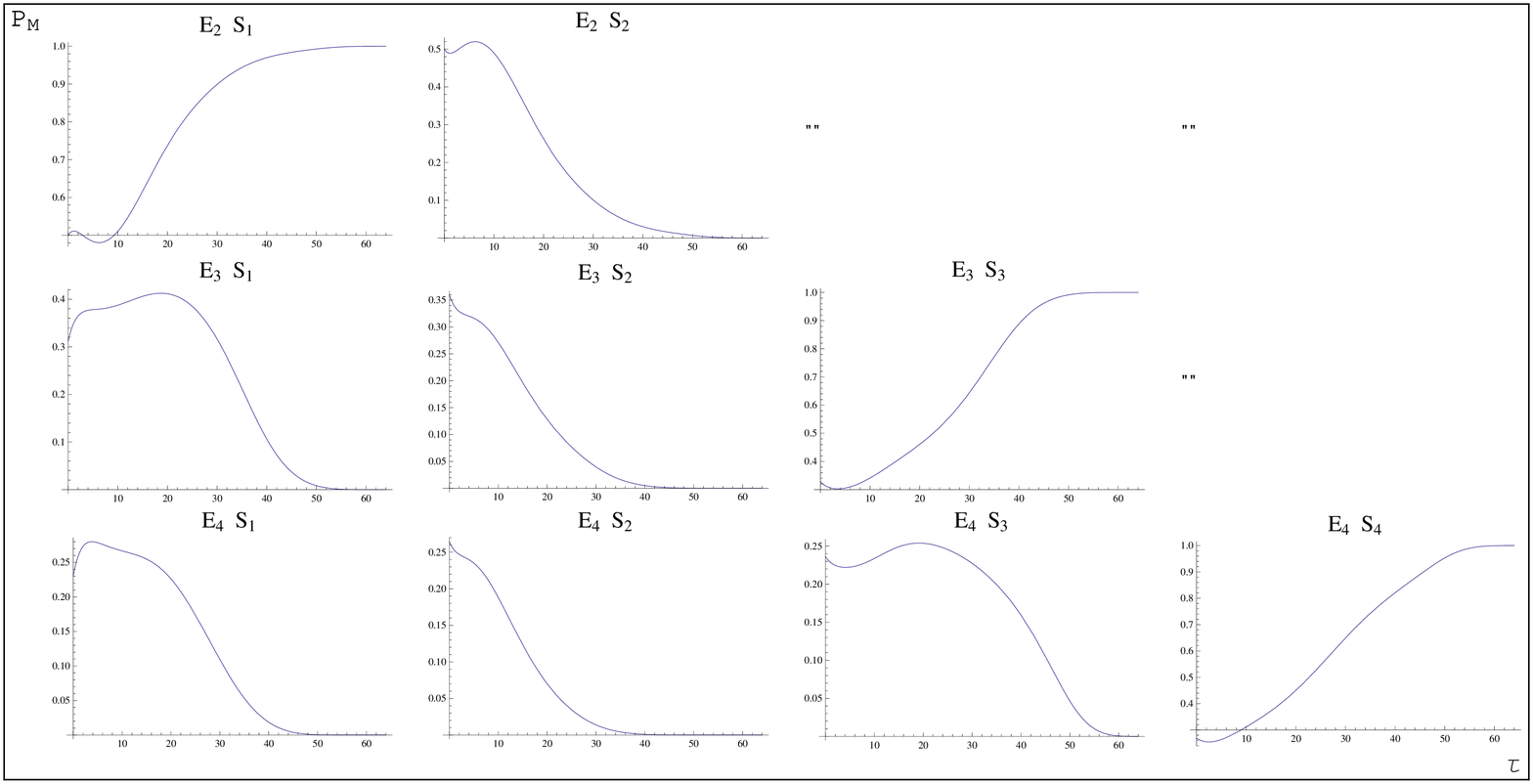}
  \caption{Weights $\frac{1}{Z_M} e^{-S_M[\x_i(\tau)]}$ of the curves
    $\x_i$ for $i=1,\dots,M$ and for $T$-ordered ensembles of size
    $M=2,\dots,4$ when evaluated with the `action' $S_M
    =\frac{L(\tau)^2}{A(\tau)}\left(1+
      \frac{c_M(\tau)}{L(\tau)}\right)$.}
  \protect{\label{fig:1WeightsReg}}
\end{figure}

Let us now investigate the flow of entropy.  The weight-functional
$P_M$ is defined as
\begin{equation}
  \label{eq:weights}
  P_M(\tau) = P_M[\x_{{\mbox{\tiny int}},i}(\tau)] 
  \equiv \frac{1}{Z_M} \exp({-S_M[\x_i(\tau)]}),
\end{equation}
and the entropy $\Sigma_M$ as
\begin{eqnarray}
  \label{eq:entropy}
  \Sigma_M(\tau) = \Sigma_M[\x_{{\mbox{\tiny int}},i}(\tau)] 
  & \equiv & \sum_{i=1}^{M} P_M[\x_{{\mbox{\tiny int}},i}(\tau)]
  \log \left( P_M[\x_{{\mbox{\tiny int}},i}(\tau)]\right)\\
  & = & \log Z_M + \frac{1}{Z_M}\sum_{i=1}^{M}S_M[\x_i(\tau)]\,
  \exp\left(-S_M[\x_i(\tau)]\right)
\end{eqnarray}
where $S_M[\x_i(\tau)]$ is given by Eq.~(\ref{eq:1action}).  In
Figures~\ref{fig:1EntropyReg} and \ref{fig:1EntropyAlt}, plots are
shown for $\Sigma_M(\tau)$ when evaluated with the `action'
\begin{equation*}
  S_M  =\frac{L(\tau)^2}{A(\tau)}\left(1 
+\frac{c_M(\tau)}{L(\tau)}\right)  
\end{equation*}
and respectively, when evaluated with the modified `action' 
\begin{equation*}
  S_M
  =\frac{L(\tau)^2}{A(\tau)}\left(1 +\frac{c_M(\tau)}{A(\tau)}\right)  
\end{equation*}
for $T$-ordered ensembles of size $M=1,\dots,16$.
The continuous approach of entropy to zero at finite values of $\tau$
implies the spontaneous emergence of order in the system as the
resolution decreases: starting at a finite value of $\tau$, a
particular member of $E_M$ is singled out by its weight approaching
unity.  This is validated by Fig.~\ref{fig:1WeightsReg} where the
weight-functionals $P_M$ are shown for $T$-ordered ensembles of size
$M=2,\dots,4$.  The pattern that a curve is singled out by its
weight-functional as $\tau$ increases continues for all ensemble sizes
$M$.  In view of Chapter~\ref{chap:0CVL}, this behavior is highly
unexpected and we conclude that the nontrivial topology of the $N=1$
sector induces qualitative differences into the coarse-graining
process.

\clearpage \cleardoublepage


\chapter{Applications}
\label{chap:Applications}

Here, it should be recalled that a magnetic charge emerging as a
result of the dynamically broken gauge-symmetry SU(2) $\rightarrow$
U(1) in the deconfining phase is interpreted as an electric charge
with respect to the U(1)$_Y$ subgroup of the electroweak sector.  In
view of Sec.~\ref{sec:HTS}, recall that the magnetic center flux of
the confining SU(2)$_e$ is dually interpreted as electric flux.

\section{Solitonic fermions}
\label{sec:Solitons}

The notion of a point-like electron has always been plagued by its
diverging self-energy.  Even in classical field theory it is present
in the shape of the infinite self-energy of a point charge.  If the
electron is considered to be a little sphere of radius $R$ and mass
$m_e$ with electric charge $e$ attached to the sphere, then the
electric field energy $U$ is given by
\begin{equation}
  \label{eq:ClassSelfE}
  U=\frac{e^2}{8\pi R}.
\end{equation}
Sending $R$ to zero, as we have to if we think of the electron as a
point particle, the self-energy contribution to the mass of the
electron diverges.  In Quantum Electrodynamics (QED), the problem
persists: the correction to the electron mass is still infinite,
although it has a much softer logarithmic divergence $(c=\hbar=1)$,
\begin{equation}
  \label{eq:QEDSelfE}
  \delta m_{\mbox{\tiny QED}} = 
  3\frac{e^2}{8\pi^2}\ln(m_{e}R).
\end{equation}
Therefore, one needs to employ renormalization theory to cope with the
emerging divergences that are to a large extent a direct consequence
of locality: the point-particle like nature of the electron.  ``...,
and despite the comparative success of renormalisation theory the
feeling remains that there ought to be a more satisfactory way of
doing things.'' as Lewis Ryder put in \cite{Ryder}.

In the Standard Model, the electron is represented by the famous Dirac
equation.  Though it successfully predicts the electron's
antiparticle, the positron, and the magnetic moment with an $g$-factor
of 2 it has to introduce the concept of the Dirac sea to make sense of
the infinite number of negative-energy eigenstates.  The Dirac sea
leads to an infinite contribution to the energy density of the
`vacuum' which has to be canceled, somehow.  Furthermore, the Standard
Model does not provide for a deeper explanation of the value of the
magnetic dipole moment other than that following from the Dirac
equation and small radiative corrections. Moreover, the electron mass
enters the QED Lagrangian as a free parameter, and the running of
which with resolution needs additional experimental input.

The excitations in the confining phase of SU(2) Yang-Mills
thermodynamics are single and selfintersecting center-vortex loops.
The mass of each intersection point in a selfintersecting
center-vortex loop is given by the Yang-Mills scale $\Lambda_c$.
Since a monopole (antimonopole) is located at the intersection, it
carries one unit of electric charge.  Recall that a magnetically
charged object in the defining gauge theory has to be interpreted as
an electrically charged object in the Standard Model - and vice versa.
In a given segment of a flux tube, the monopoles (antimonopoles) can
move in both directions: there is a two-fold degeneracy of direction
of the center flux that is analogue to the two-fold degeneracy of the
spin projection.  Moreover, for each center-vortex loop it is possible
to move along the entire flux system on a closed curve.  Thus, the
projection of the dipole moment generated by the current of monopoles
and antimonopoles inside the vortex core onto a given direction in
space is two-fold degenerated as well.  Therefore, we identify each
soliton with a spin-1/2 fermion.  Setting the Yang-Mills scale
$\Lambda_C$ equal to the electron mass $m_e$, we are led to interpret
$N=1$ center-vortex loops as electrons or positrons
\cite{Hofmann2005,ZPinch,Axion}.

\begin{figure}
  \centering
  \includegraphics[width=0.90\textwidth]{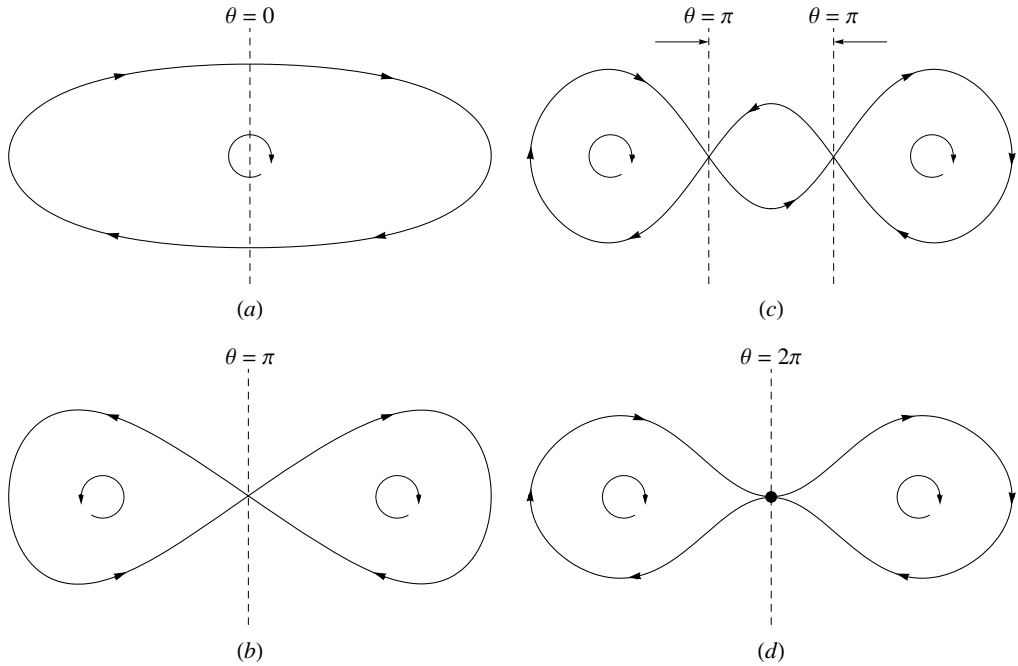}
  \caption{(Topological) transition from the $N=0$ sector (a), (b),
    (c) to the $N=1$ sector (d) by twisting and subsequent capture of
    a magnetic (anti)monopole in the core of the final intersection.
    Arrows indicate the direction of center flux.}
  \protect{\label{fig:SectorTransition}}
\end{figure}

Let us consider the process of twisting and charge localization more
closely.  The transition from a non-selfintersecting to
selfintersecting center-vortex loop is by twisting of a
non-selfintersecting curve.  The emergence of a localized
(anti)monopole in the process is due to its capture by oppositely
directed center fluxes in the core of the intersection (eye of the
storm).  By a rotation of the left half-plane in
Fig.~\ref{fig:SectorTransition}(a) by an angle of $\pi$, see
Fig.~\ref{fig:SectorTransition}(b), each wing of the center-vortex
loop forms a closed flux loop by itself, thereby introducing equally
directed center fluxes at the intersection point.
This does not allow for an isolation of a single spinning
(anti)monopole in the core of the intersection and thus is
topologically equivalent to the untwisted case
Fig.~\ref{fig:SectorTransition}(a).  However, another rotation of the
left-most half-plane in Fig.~\ref{fig:SectorTransition}(c) introduces
an intermediate loop which by shrinking is capable of isolating a
spinning (anti)monopole due to oppositely directed center fluxes.
Notice that in the last stage of such a shrinking process (short
distances between the cores of the flux lines), where propagating dual
gauge modes are available\footnote{On large distances these modes are
  infinitely massive which is characteristic of the confining phase.},
there is repulsion due to Biot-Savart which needs to be overcome.
This necessitates an investment of energy manifesting itself in terms
of the mass of the isolated (anti)monopole (eye of the storm).
Alternatively, the emergence of an isolated (anti)monopole is possible
by a simple pinching of the untwisted curve, again having to overcome
local repulsion in the final stage of this process.

In the analysis performed in Chapter~\ref{chap:1CVL}, we have solely
regarded the situation depicted in Fig.~\ref{fig:SectorTransition}(d),
since the direction of center flux within a given curve segment is
irrelevant for the process of a spatial coarse-graining
microscopically described by the same curve-shrinking flow as applied
to $N=0$ center-vortex loops in Chapter~\ref{chap:0CVL}.

There are also phenomenological reasons that argue for a non-local
nature of the electron.  Recall that the imaginary part of the
pressure in the confining phase starts to dominate when approaching
the Hagedorn transition, thereby inducing microturbulences in the
plasma (see Sec.~\ref{sec:ConPhase}).  Such a nonthermal behaviour is
likely to be related to the observed but poorly understood
microturbulences and internal transport barriers in tokamak
experiments with magnetically confined plasma.
This presumes to identify the neutrino and the electron with the
non-selfintersecting and the one-fold selfintersecting center-vortex
loop of the confining phase of SU(2)$_e$ with Yang-Mills scale
$\Lambda_c=m_e=511$~keV.  Here, it should be noticed that due to the
absence of an antiparticle in the case of a $N=0$ soliton neutrinos
need to be of Majorana type which is in compliance with the successful
search for the neutrinoless double $\beta$ decay \cite{KKG}.

This interpretation is also supported by recent high-temperature
$Z$-Pinch experiments at Sandia National Laboratories detecting an
unexpected powerful contained explosion.  There, an electric current
rises in a wire array up to $\sim 20$~MA within $\sim 100$~ns, thereby
the wire is transformed into a plasma column.  The strong magnetic
field induced by the current results in an inward directed (magnetic)
pressure $P_m$ which compresses the plasma until it collapses.  In the
course of the implosion, the ions and electrons are accelerated
towards the plasma axis.  The radiated soft x-ray energy is as much as
four times the kinetic energy that is expected to be released by the
intersection of ions and electrons. However, before the plasma
explodes it stabilizes for about 5~ns (stagnation).  The measured
electron temperature $T_e$ is found to be $\sim 3$~keV at stagnation.
Preceding the explosion, an ion temperature $T_i$ about $300$~keV is
sustained shortly after the plasma has stagnated \cite{Sandia}.  The
outward directed plasma pressure $P_p$ needs to be equal in magnitude
to $P_m=-1.8\times 10^{-12}$~MeV in order that the implosion
stagnates.  The measured electron temperature is a factor 1/8.5 too
low if it is asserted that the plasma pressure is carried by electrons
only; this would correspond to $T_e\sim 31.55$~keV.  In \cite{Haines},
the observed imbalance between the energy in- and output was addressed
to the rapid ($\sim1\dots2$~ns) conversion of magnetic field energy to
a very-high-ion-temperature plasma by the unexpected forming of short
wavelength $m=0$ magnetohydrodynamics (MHD) instabilities at
stagnation which subsequently provide associated viscous ion heating.
At stagnation, the ions reach much more rapidly than the electrons
($\sim 1\dots2$~s) a temperature of $\sim 300$~keV, and subsequently
heat the electrons up to $\sim 300$~keV, at least locally.  By
equipartition the ion energy is transferred to the electrons, leading
to the soft x-ray radiation.

According to Sec.~\ref{sec:ConPhase} and the discussion in
\cite{ZPinch}
this will involve center-vortex loops with a higher number of
selfintersections.  These accelerate the transit of thermal energy
from ions to electrons and generate a larger energy density and
pressure than expected from electron dynamics only.  As a consequence,
the electron temperature rises rapidly after stagnation.  After the
ion-induced heating the Debye screening mass $m_D$ of a conventional
electron-photon plasma is comparable to $T_e$.  Thus, at about $\sim
5$~ns after stagnation, the plasma is absolutely opaque and no
radiation is released.

For $T_e<0.6\,m_e$ and $N_C=6$, the truncated sums $\bar
P_{N_c}=\sum_{N=1}^{N_c}P_{C,N}$ and
$\bar\rho_{N_c}=\sum_{N=1}^{N_c}\rho_{C,N}$ of the pressure
$P_C=\sum_{N=0}^{\infty}P_{C,N}$ and the energy density
$\rho_C=\sum_{N=0}^{\infty}\rho_{C,N}$ in the electronic system are in
the regime of asymptotic convergence \cite{ZPinch}. The case $i=1$
corresponds to a contribution of electrons and positrons only.  While
for $T_e\ll m_e$ center-vortex loops with higher mass (higher number
of intersections) are strongly suppressed, these do significantly
contribute to the pressure and energy density for $T_e\gtrsim
0.1$~MeV.  At $T_e=0.25$~MeV, the relative partial pressure and the
relative partial energy density is $\bar P_{6}\sim 3\,\bar P_{1}$ and
respectively, $\bar \rho_{6}\sim 4.8\,\bar \rho_{1}$, and at
$T_e=0.3$~MeV, already $\bar P_{6}\sim 5\,\bar P_{1}$ and
respectively, $\bar \rho_{6}\sim 9.4\,\bar \rho_{1}$.  Notice that at
$T_e=0.25$~MeV, the ratio of $\bar P_{1}$ to the magnetic pressure
$P_m$ at stagnation is: $\frac{\bar P_{1}}{P_m}\sim -4.4\times 10^8$.
The existence of center-vortex loops with higher mass facilitates the
rapid increase of $T_e$ to $T_i\sim 0.3$~MeV and eventually initiates
the powerful explosion.

When the electron temperature approaches a value of about 0.5~MeV the
Hagedorn transition towards the preconfining phase is expected to take
place where all charges condense densely packed into a new ground
state.  The $Z$ vector boson of the Standard Model is identified with
the decoupled dual gauge mode in the magnetic phase of SU(2)$_e$.

The electron appears to be structureless for (nearly) all external
momenta that are used to probe the system because of the existence of
a Hagedorn-like density of states: the invested energy deposited into
the vertex is converted into entropy associated with the excitations
of a large number of unstable and heavy resonances (see
Fig.~\ref{fig:Excitations} for the excitations with up to $N=3$
selfintersections).  Only for momenta comparable to the Yang-Mills
scale $\Lambda_c$, the BPS monopole located at the intersection
becomes excited and reveals a part of its structure.  For momenta
sizeably below $\Lambda_c$, there is nothing to be excited in BPS
monopole. 

\section{High-temperature superconductivity}
\label{sec:HTS}

Let us now sketch an alternative approach to high-temperature (high
$T_c$) superconductivity.  Recall, that the magnetic center flux,
dually interpreted as electric center flux, is two-fold degenerated.
Now the electric charges that travel along the flux lines in the
vortex core produce a magnetic dipole moment.  The projection of which
onto a given direction in space is either parallel or antiparallel and
represents the two-fold degeneracy of the spin projection.  Here, it
should be recalled that a shift of the intersection point of an
isolated $N=1$ center-vortex loop leaves the mass of this soliton
invariant.

Coincidentally, there are quantum systems in nature the unconventional
behaviour of which seems to be closely related to the restriction of
electron dynamics to two space dimensions.  In particular, these
include high $T_c$ superconductors such as the family of
superconducting materials largely containing (rare-earth) doped
cooper-oxide (cuprates) planes as well as the recently discovered new
class of layered oxypnictide superconductors.  Let us consider the
former at first.  In both cases, superconducting layers of magnetic
moments are interspersed with layers of nonmagnetic material.  This
nonmagnetic material  also serves as an reservoir that provides,
by doping, for the electrons and screens the Coulomb repulsion in the
superconducting layer between them.  Now the question arises how
long-range interactions of magnetic moments at given optimal doping
and sufficiently low temperature lead to superconductivity in the
cuprate layers.

Since, at small enough temperatures, copper-oxide planes are Mott
insulators with long-range antiferromagnetic order of spins, the
conventional Hubbard model must be used.  A canonical transformation
involving a Gutzwiller projection
leads to the `$t-J$' model, where $t$ describes the hopping of
electrons from site to site and $J$ the superexchange $J =4t^2/U$ with
$U$ describing the Coulomb repulsion.  Here, the Gutzwiller
projection, which removes most of the phonon pairing interaction, is
mandatory.  Variation of the electronic degrees of freedom results in
a set of gap equations for the ground state that give the predicted
$d$-wave gap and the superconducting order parameter (related to the
critical temperature $T_c$) as a function of doping \cite{Anderson}.

\begin{figure}[t]
  \centering
  \includegraphics[width=0.90\textwidth]{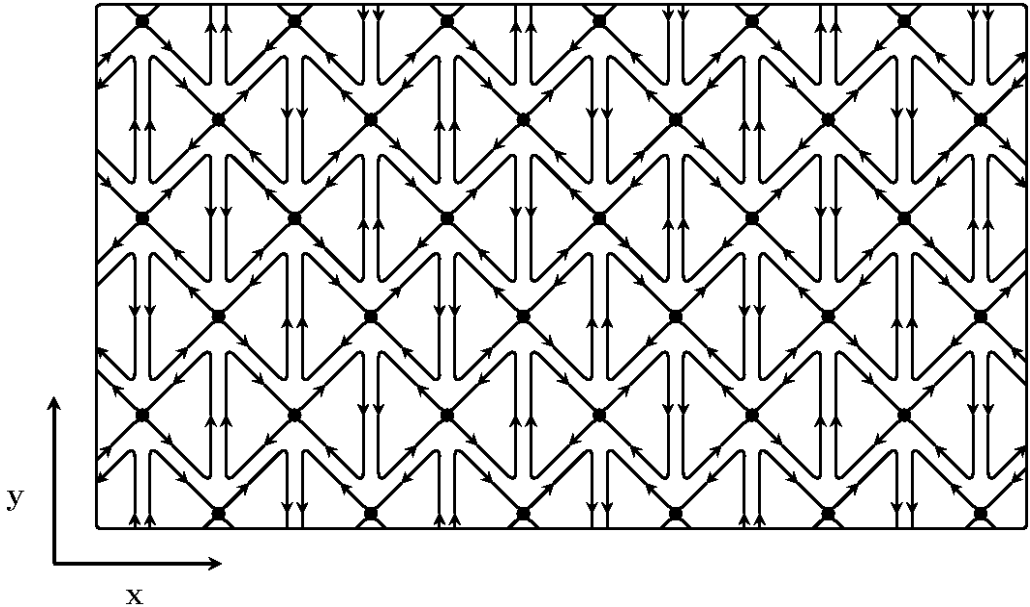}
  \caption{The figure, possibly related to the superconducting state
    in a cuprate, shows an array of strongly correlated center-vortex
    loops tiling the two-dimensional plane.  If optimal screening of
    the electron charge located at the intersection point is provided
    by doping such that an attractive interaction between
    center-vortex loops due to Amp\`ere's Law becomes important, then
    the attractive force between equally directed center flux segments
    could lead to the indicated equilibrium configuration.  For a
    given electron there are six neighbouring line segments two of
    which experience repulsion while the other four experience
    attraction.  An overlap of flux lines would create new
    intersection points, each of mass $\propto m_e$ , which is
    topologically forbidden, thus leading to a repulsion at short
    distances.}
  \protect{\label{fig:Cuprate}}
\end{figure}

Let us now sketch a somewhat speculative approach to high $T_c$
superconductivity being well aware of our lacking theoretical knowledge
on details in this field of research.  The key idea is already encoded
in Fig.~\ref{fig:SectorTransition}(d).  According to SU(2) Yang-Mills
theory, the electron represented as a selfintersecting center-vortex
loop is a non-local object the magnetic dipole moment of which is only
loosely related to the localization of its charge: the magnetic moment,
carried by the vortex core of the flux lines, receives contributions
from line segments which are spatially far separated (on the scale of
the diameter of the intersection) from the location of the electric
charge.  This suggests a system of planar center-vortex loops trapped in
a two-dimensional layer where the interaction between vortex lines
becomes important due to an effective screening of the electron charge
leading to an ordering effect.  In view of the reported strong
correlations between electrons in two-dimensional superconducting
systems \cite{Nature2008}, we imagine a situation as it is depicted in
Fig.~\ref{fig:Cuprate}.

Due to Amp\`ere's law equally directed electric flux lines attract
each other, whereas oppositely directed flux lines experience a
repulsive force.  So, for a given center-vortex loop, there is an
attractive interaction of four out of six line segments defined by the
neighbouring electrons while the other two repulse each other.  The
existence of interactions between flux lines that are mediated by the
photon is a consequence of the mixing between the gauge groups
SU(2)$_e$ and $\cmb$, the latter pertinent to the existence of
propagating photons, see \cite{Hofmann2005,Axion}.  It should be
noticed that the spin projection of a given electron is equally
directed for two of its neighbours while the other four have
oppositely directed spins.  This supports the observation that high
$T_c$ superconductivity is an effect not related to $s$-wave pairing
\cite{Nature2008}.  An overlap of flux lines would lead to additional
intersection points which require an extra amount of energy
$\Lambda_C$ for each intersection and is therefore energetically
forbidden because the fluctuations in energy density of the system
will not allow for the creation of an intersection of mass
$m_e=511$~keV.  This leads to a repulsive force as the spatial
distance between adjacent vortex segments vanishes.  In order that an
equilibrium between attraction and repulsion where the intersection
point is fixed with respect to its neighbours occurs, as it is depicted
in Fig.~\ref{fig:Cuprate}, one needs a sufficiently low temperature
(related to resolution) and an optimal screening of the Coulomb
repulsion by the surrounding reservoir layers.  If the temperature
(resolution) falls below a critical value then the fluctuations of the
intersection points relative to one another will vanish.  Applying an
external electric field parallel to the plane would set the stiff
system of locked electrons in a collective motion with zero electric
resistance.  Macroscopically, this situation is illustrated by a stiff
table cloth being pulled over the table in a frictionless way.  The
measured pseudo-gap phase in high $T_c$ superconductors will be
addressed to local distortions in this highly ordered state.  These
distortions require an finite amount of energy and are due to
insufficient screening and/or to much of a thermal noise.

Let us now turn to the recently discovered, new class of high $T_c$
superconductors that are based on oxypnictides (a class of materials
including oxygen, an element of the nitrogen group (pnictogen), and one
or more other elements).  These do not seem to exhibit strong
correlations between the electrons contained in the two-dimensional
(FeAs) layers where the electron dynamics takes place, see
\cite{KlaussBuchner2008}.  If the behaviour of a two-dimensional system
of non-interacting electrons, which are subjected to an environment
represented by a parameter $\tau$, effectively is describable by a
coarse-graining process in a statistical ensemble, as it is investigated
in Chapter~\ref{chap:1CVL}, then we should likely address the
observation that the entropy vanishes at \emph{finite} $\tau$ to this
particular kind of high $T_c$ superconductivity.  Namely, the
observation that no variance of the ensemble average of the position of
the electric charge is allowed for at a finite resolution is crucial
for the statement that the two-dimensional system of free quasiparticles
is void of any electric resistance.  Free quasiparticles in the sense
that explicit interactions between the electrons in the superconducting
layer are absent, but the distortions induced by the noise of the
environment is fully taken into account.  Again, $\tau$ should be a
monotonic function of temperature.

\cleardoublepage


\chapter{Summary}
\label{chap:Summary}

In this thesis, we have investigated center-vortex loops with and
without selfintersection, as they emerge in the confining phase of SU(2)
Yang-Mills thermodynamics.  In a noisy environment, center-vortex loops
are subject to a spatial coarse-graining due to a motion by curvature
that is described by a curve shortening flow.  In a statistical
description of ensembles of center-vortex loops which are (locally)
embedded into a two-dimensional flat plane, we have defined an effective
`action' in purely geometric terms that is governed by a
renormalization-group flow driven by the curve shrinking.  The `action'
possesses a natural decomposition into a conformal and a non-conformal
factor.  `Observables', such as the position of `center of mass'
($N=0$), or the intersection point ($N=1$), are computed as ensemble
averages of local or non-local operators on the curves.

We have made the observation that $N=0$ center-vortex loops exhibit a
second-order transition to the conformal limit of vanishing curve length
with a critical mean-field exponent of the coefficient: on average,
center-vortex loops disappear from the spectrum of confining SU(2)
Yang-Mills theory, thus generating an asymptotic mass gap.  The
evolution of the variance of the initially sharp position of `center of
mass' saturates at finite value within a finite decrease of resolution
$Q$, the latter related to the resolving power used to probe to the
system.  These findings bear a strong family resemblance with the
unitary time evolution of a free particle in quantum field theory.

Since we believe that $N=0$ center-vortex loops play the role of
Majorana neutrinos \cite{KKG}, the concept of a neutrino rest mass is no
longer applicable.  Its mass is the result of the distortions induced by
the environment it is embedded in and depends on the resolution.  The
disappearance of $N=0$ center-vortex loops from the excitation spectrum
and the absence of a corresponding antiparticle would be manifestations
of lepton-number violation forbidden in the Standard Model of Particle
Physics.

In the case of one-fold intersecting center-vortex loops, we have
obtained the unexpected result that a statistical ensemble of initial
curves evolves into a highly ordered state.  That is, only a particular
member of the ensemble survives the process of two-dimensional spatial
coarse-graining.  As a consequence, the entropy attributed to the
ensemble moves to a zero value for a sufficient decrease of resolution.

We have sketched an alternative approach to high-tem\-per\-a\-ture
superconductivity based on cuprates.  The central observation that these
depend highly on strong correlations between electrons trapped in a flat
two-dimensional layer is attributed to an array of bow-tie-like
simplices ($N=1$ center-vortex loops) tiling the plane.  We have also
speculated that the spontaneous emergence of order in an ensemble of
planar $N=1$ center-vortex loops could be relevant for the recently
discovered new class of oxypnictide layered high-temperature
superconductors that do not seem to exhibit explicit, strong
correlations between electrons within the superconducting (FeAs) planes.

In a sense, we have reversed the usage of the renormalization group.
Instead of starting with a `physical' action, from which an equation of
motion follows, and demanding the system to be invariant under
renormalization-group transformations, we have defined a geometric
effective `action' the coefficient of which is determined by a
renormalization-group flow driven by a coarse-graining process (curve
shrinking).  Afterwards, physical `observables', such as the center of
mass or the localization of electric charge, were computed as mean
values 
in a statistical ensemble.  In a manner of speaking, we have `derived' a
statistically averaged `equation of motion'.  In this context, we regard
resolution over time (or temperature) as the more fundamental quantity
describing a quantum mechanical or statistical system.  However, this
requires the introduction of a model that relates resolution to time (or
temperature).

An obvious extension of our statistical approach would be the account of
interactions between center-vortex loops; especially in view of
two-dimensional systems exhibiting high-temperature superconductivity.
At a first stage, this would include Coulomb interactions between the
charges of $N=1$ center-vortex loops localized at the intersection
point, and a delta-function-like repulsion due the topologically
forbidden overlap of center-vortex loops (contact interaction).  This
could be done by adding interaction terms to the effective `action' in
the partition function that are weighted accordingly.  Considering a
multitude of configurations of initial curves of center-vortex loops,
those configurations will be singled out the curves of which are most
likely to survive the process of coarse-graining.  However, the
nontrivial issue arises how to gain a weight which relates the purely
geometric `action' to the one stemming from the electromagnetic
interaction between center-vortex loops.

\cleardoublepage


\end{document}